\documentclass{article}
\usepackage{jheppub}
\usepackage{float}
\usepackage{amsmath, amssymb}
\usepackage{graphicx}
\numberwithin{equation}{section}

\makeatletter
\newcommand\fake@math{}
\def\fake@math#1\){{#1}}
\makeatother

\newcommand{\const}{\mathop{\mathrm{const}}\nolimits}

\begin{document}

\title{Semiclassical correlators in Jackiw--Teitelboim gravity}

\author[a]{Ksenia Bulycheva\footnote{On leave of absence from IITP, Bolshoy Karetny per. 19, Moscow, 127051 Russia}}
\abstract{
  In the semiclassical approximation to JT gravity, we find two-point and four-point correlators of heavy operators.
  To do so, we introduce a massive particle in the bulk and compute its action with gravitational backreaction.
  In Euclidean signature, the two-point function has a finite limit at large distances.
  In real time, we find that the thermal two-point function approaches an exponentially small value $\sim \exp(-N)$ at long time.
  We also find that after a period of exponential decay, the out of time ordered four-point function approaches an exponentially small value as well.
}

\affiliation[a]{Department of Physics, Princeton University, Princeton, NJ 08544}
\emailAdd{kseniab@princeton.edu}

\dedicated{Dedicated to the memory of Sergey Guts}

\maketitle

\section{Introduction}

The SYK model is a useful one-dimensional laboratory, where many effects of gravity have been observed.
In particular, it shows high degeneracy of low-lying states, finite entropy at zero temperature and chaotic growth of four-point correlation functions \cite{Maldacena:2016hyu} \cite{Kitaev:2015}.
The model is relatively easy to study numerically, and it is also solvable in the large $N$ limit.

However, the full holographic dual of the SYK is yet to be built and explored.
As an approximately dual theory, the Schwarzian action is often studied \cite{Engelsoy:2016xyb} \cite{Mertens:2017mtv} \cite{Jevicki:2016bwu} \cite{Fu:2016vas}.
It gives the first correction to the conformal description of the SYK.
Although it is not the full answer for the correlators, it shows (among other things) the exponential growth of the four-point function associated with quantum chaos.

The Schwarzian action can be found directly from the SYK \cite{Maldacena:2016hyu}.
Alternatively, we can view it as a limit of two-dimensional gravity.
The latter view allows to extend the $AdS/CFT$ correspondence to near-$AdS$/ near-$CFT$ \cite{Maldacena:2016upp}. (We write $NAdS$ and $NCFT$ for the
near-$AdS$ space and near-$CFT$.)
On the left-hand side we have an $AdS_2$ space with the position of the boundary defined by equations of motion for the dilaton, and not being at infinity as in the conventional holographic prescription.
On the right-hand side, we study a theory that is conformal in some limit, but receives corrections that break conformal invariance.

In most of the existing literature, one considers operators with fixed dimension in the $NCFT$ which lives on the boundary of the $NAdS$ space. 
Then in the weak coupling limit, these operators cause negligible back-reaction on the geometry, but they produce interesting quantum fluctuations.
In this paper, instead, we will consider operators whose dimension scales as the inverse of the coupling parameter.    
This means that in the weak coupling limit, these operators produce significant back-reaction on the geometry, or more precisely on the dilaton field, since the geometry of the $NAdS$ is really fixed by the equations of motion.
For small dimension of the operator, we reproduce the Schwarzian corrections to the two-point function, but for large dimension we find novel behavior.
We also analytically continue the two-point function from Euclidean signature in order to compute a real time thermal correlator.

It turns out that the two-point function computed in this way either for large Euclidean distances or large real time in the case of a thermal correlator has unexpected properties whose origin is not clear. 
In the Euclidean case, when the length scale in the $NCFT$ goes to infinity, the length of the relevant geodesic that appears in the correlator
approaches a finite limit, and the two-point function approaches a limit as well. 
It suggests that the large dimension operator $\mathcal O$ that we consider has a nonzero vacuum expectation value.
The interpretation of this fact is obscure.

In the thermal case, the large time limit of a two-point function is small, but finite number.
In \cite{Maldacena:2001kr}, it was argued that a useful test of the information paradox in black hole physics is to see whether the two-point function decays exponentially for large
real time or there are some finite residual correlations even after long time.
The size of these fluctuations is expected to be of order $\sim \exp(-cS)$, where $S$ is the entropy and $c$ is a constant.
The two-point function after long time is expected to fluctuate wildly, with a characteristic amplitude $\exp\left( -cS \right)$.
When these fluctuations are averaged, the two-point function becomes an exponentially small constant.

A test of this behavior  was performed for the SYK model in \cite{Cotler:2016fpe} and it was shown that the (real part of the) correlator indeed approaches a plateau after a period of exponential descent, reaching a minimum and then rising back to a constant value.
Some of these features have been later explored and partly explained in \cite{Saad:2018bqo}.
The plateau is very low, and its height is proportional to $\sim \exp(-N)$.

This effect is of order $O\left( e^{-1/G_N} \right)$ and is non-perturbative in gravity.
However, it may be accessible in a semiclassical treatment \cite{Cotler:2016fpe}, \cite{Barbon:2003aq}, and we find a similar phenomenon in our setup.
For a two-point function, the exponential decay eventually slows down, and it approaches a plateau at long (real) time. 
When we identify the parameters of our model with the parameters of the SYK, we find that the long-time limit of the two-point function is also $\sim \exp(-N)$.
However, the plateau is reached much sooner than expected on general grounds, which points out that the physics behind it may be different.
It is also worth noting that the two-point function found as the exponentiated geodesic length has a shape very similar to what was found in  \cite{Cotler:2016fpe} (see fig.~\ref{fig:sff}), although this is most likely accidental.

Our setup also allows us to study four-point functions.
In particular, we find the out of time ordered thermal four-point function.
It is related to a double commutator and serves as a measure of chaotic behavior \cite{Maldacena:2015waa}.
We find that the four-point function decays exponentially at first (as can be found from Schwarzian action), then after a relatively short time this decay stops.
This can be identified with Ruelle behavior, showing how the system approaches the thermal equilibrium.
However, we also find that that at long times the four-point function approaches a small but finite value, similarly to the real-time two-point function.
To the best of our knowledge, this has not been tested on the SYK side.

We should note that the question of $NAdS$ correlators with back-reaction included has been mentioned in  \cite{Gu:2017njx} and also  in \cite{Kourkoulou:2017zaj}.

{\bf Acknowledgments.} Author thanks Edward Witten for extensive discussions and Sergey Khilkov for his help with numerical calculations.

\section{Setup}

To set the scene, we introduce the action of two-dimensional gravity in the formulation of Jackiw and Teitelboim \cite{Jackiw:1984je}, \cite{Teitelboim:1983ux}, \cite{Almheiri:2014cka}.
The pure gravity action with the boundary term reads:
\begin{equation}
  I = -\frac{1}{16 \pi G} \int d^2x \phi \sqrt{g}\left( R+2 \right) - \frac{1}{8\pi G} \int_\partial \phi_b \sqrt{h} K.
  \label{I_JT}
\end{equation}
(We omit the term defining the extremal entropy and higher-order terms in $\phi$.)
The equations of motion for the dilaton set a constant negative curvature:
\begin{equation}
  R+2=0.
  \label{R=-2}
\end{equation}
In a pure anti--de Sitter space this action is topological.
However, if we restrict it to a region of $AdS_2$, the position of the boundary becomes a non-trivial degree of freedom.
Classically, it is set by the boundary condition for the dilaton:
\begin{equation}
  \left.\phi\right|_\partial=\phi_b=\const.
  \label{phi_b}
\end{equation}
We consider massive particles moving in the Jackiw--Teitelboim gravity.
With the condition (\ref{R=-2}) the action becomes:
\begin{equation}
  I = - \frac{ \phi_b }{8\pi G} \int_{\partial NAdS}\sqrt{h} K - m \int_{NAdS} ds.
  \label{I_our}
\end{equation}
The second integral means that we consider only the part of the worldline lying inside our near--$AdS$ space.
Also, from now on we absorb the gravitational constant $G$ in the definition of $\phi_b$.

As has been discussed in \cite{Maldacena:2016upp}, we can think of this action as a low-energy limit of some unknown theory.
The position of the boundary is then a UV cutoff, and we choose to make this cutoff consistent with the equations of motion following from the JT action (\ref{I_JT}). 

The boundary of the $NAdS$ space is at finite distance from the ``center'' of the true $AdS_2$, and has a finite length. 
This allows us to study the boundary theory using the gravitational action only.
The symmetries of the boundary theory are generated by isometries of $AdS_2$.
Therefore the theory does not possess the full conformal symmetry, but, keeping the boundary ``close'' to the boundary of the true $AdS_2$ space, we can hope to see a nearly-conformal theory.
The ``closeness'' is measured by the value of the boundary dilaton.
In particular, the true conformal theory corresponds to the boundary dilaton being infinitely large.
To be more specific and following \cite{Maldacena:2016upp}, we define:
\begin{equation}
  \phi_b = \frac{\phi_r}{\epsilon},
  \label{phi_r_def}
\end{equation}
where $\phi_r$ is the renormalized value of the dilaton, and $\epsilon$ is a small number measuring how close the $NAdS$ boundary is to the true boundary of $AdS_2$. 
We want the boundary lengths to be finite as $\epsilon \to 0$, and therefore we rescale quantum mechanical distances as:
\begin{equation}
  du_{\text{QM}} \equiv \epsilon \cdot du_{AdS}.
  \label{duQM_def}
\end{equation}

When $\epsilon$ is small, the extrinsic curvature term in (\ref{I_our}) reduces to a Schwarzian derivative:
\begin{equation}
  \phi_b \int \sqrt{h} K \to \phi_r \int \mathrm{Sch}(t,u) du.
  \label{K_Schw}
\end{equation}
The same Schwarzian term appears in the effective action of the SYK model as the first correction to the conformal answer.
This allows us to tentatively identify the parameters of the two theories as:
\begin{equation}
  {\phi_r}^{\text{(JT)}} \sim {\left(\frac{N}{J}  \right)}^{\text{(SYK)}}.
  \label{phi_r_SYK}
\end{equation}
The factor between the parameters is the function of $q$ in SYK which we are not discussing here.
So, we expect the boundary theory to be close to conformal when $\phi_r$ is large, and the $1/N$ corrections to SYK to correspond to $1/\phi_r$ corrections on the boundary of JT gravity.

We want to study this $NAdS$/$NCFT$ correspondence in a semiclassical regime with $G \ll 1$, or in our notation $\phi_b \gg 1$.
In particular, we use the conventional holographic prescription \cite{Witten:1998qj} to find the two-point function of boundary operators.

\begin{figure}
  \centering
  \includegraphics[width=.4\textwidth]{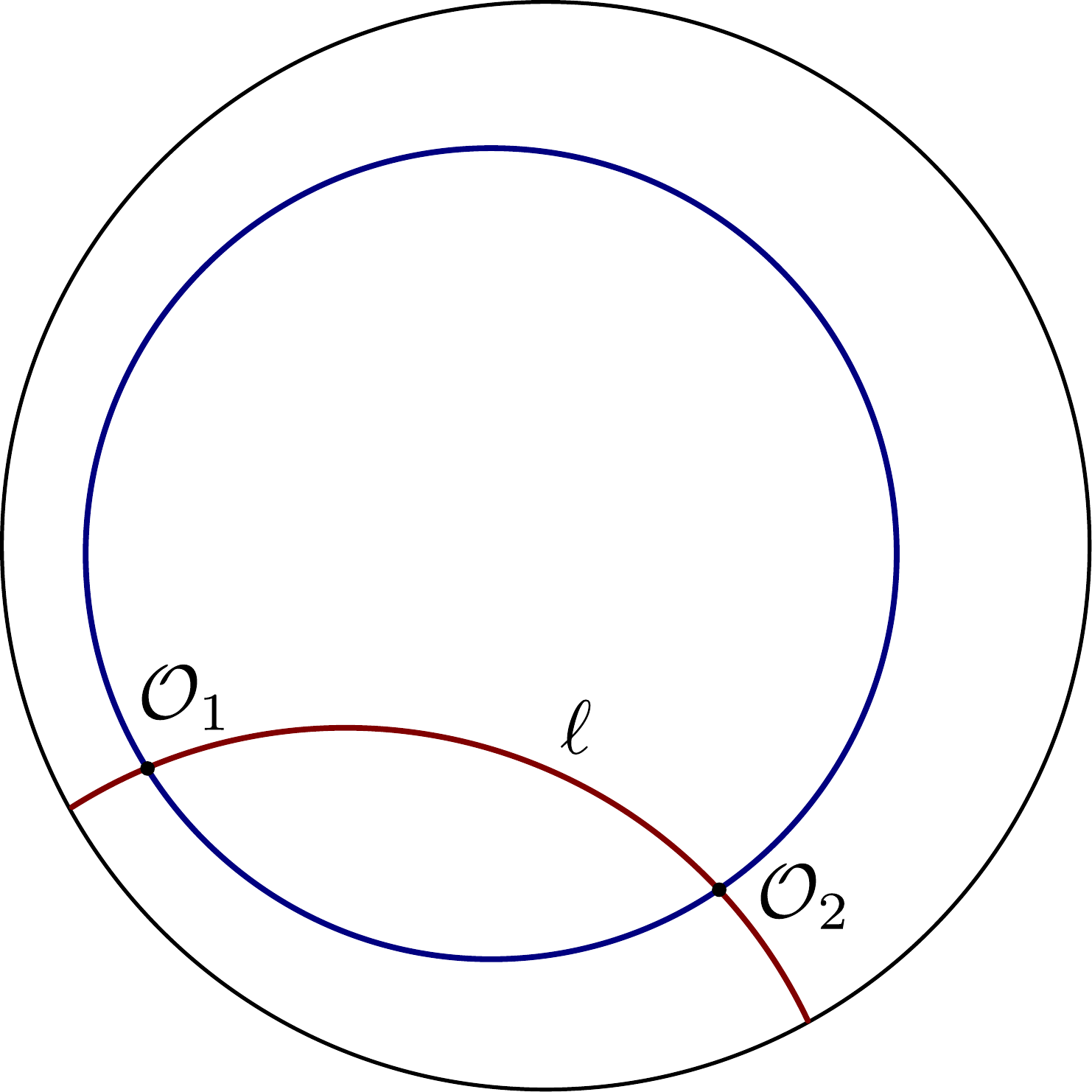}
  \caption{Poincar\'e disc (black) with near--$AdS$ space inside (blue), crossed by a massive particle (red). Two operators $\mathcal O_1, \mathcal O_2$ belong to a near-conformal theory on the boundary of $NAdS$ space. 
  The mass is small, and the back-reaction absent, therefore the $NAdS$ boundary looks like a full circle.}
  \label{fig:disc}
\end{figure}

The two-point function $G$ satisfies the Laplace equation written in terms of the geodesic length (we ignore the angular part of the Laplacian):
\begin{equation}
  \left(-\frac{1}{\sinh \ell}\frac{d}{d\ell} \sinh \ell \frac{d}{d\ell} + m^2\right) G = 0, \qquad G \equiv \left \langle \mathcal O_1 \mathcal O_2 \right \rangle. 
  \label{laplace}
\end{equation}
Here $\ell$ is the length of the geodesic connecting the two operators (see fig.~\ref{fig:disc}).
This equation typically has an exponentially growing and an exponentially decaying solution, with only the latter making physical sense.
For large $\ell$, the two-point function is:
\begin{equation}
  G\left( \ell \right) \sim e^{-\Delta \cdot \ell} \sim \exp\left( -m \int ds \right), \qquad \Delta\left( \Delta - 1 \right) = m^2.
  \label{G_semi}
\end{equation}
We are interested in particles with large mass, and will assume:
\begin{equation}
  \Delta \sim m.
  \label{Delta_def}
\end{equation}
In this prescription takes into account only the second part of the action (\ref{I_our}).
The action of extrinsic curvature provides a correction to this result.
In Section~\ref{sec:K}, we find that this correction is (numerically) small in Euclidean signature, but is significant for real-time correlation functions. 
But first, we find the two-point function as the exponentiated geodesic length.

In the absence of back-reaction (see fig.~\ref{fig:disc}), or with an extremely small mass, this prescription gives usual conformal answer:
\begin{equation}
  \left \langle \mathcal O_1(x) \mathcal O_2(0) \right \rangle \sim \frac{1}{\sin^{2\Delta} x}.
  \label{OO_sin}
\end{equation}
However, the massive particle creates a jump in the dilaton and distorts the boundary, therefore introducing corrections to this result.
This has to be taken into account to compute the length of the geodesic;
one cannot compute the length of the geodesic as if one were in an undistorted $NAdS_2$.
Our goal is to find the full semiclassical answer for the two-point function, taking this back-reaction into account.

\section{Near--\(AdS\) boundary}
\label{sec:geo}

In this Section, we set up the geometry we are working in.
We find the boundary of the $NAdS$ space, consistent with the Dirichlet condition for the dilaton:
\begin{equation}
  \left. \phi \right|_{\text{bdry}} = \phi_b.
  \label{phi=const}
\end{equation}
The classical solution for the dilaton field is found from the pure Jackiw--Teitelboim action (\ref{I_JT}).
The equations of motion for the metric define the energy-momentum tensor for the dilaton:
\begin{equation}
  T_{\mu\nu}^\phi = \frac{1}{8\pi G} \left(\nabla_\mu \nabla_\nu \phi - g_{\mu\nu} \nabla^2 \phi + g_{\mu\nu} \phi \right).
  \label{T_phi}
\end{equation}
In the absence of matter, this energy-momentum tensor vanishes:
\begin{equation}
  T_{\mu\nu}^\phi = 0.
  \label{T=0}
\end{equation}
This condition is conveniently solved in the embedding coordinates $Y_i$, which define the $AdS_2$ inside flat three-dimensional space as:
\begin{equation}
  Y = \left( Y_{0}, Y_1,  Y_2  \right), \qquad Y^2 = Y_0^2-Y_{1}^2-Y_2^2 = 1.
  \label{Y_def}
\end{equation}
In these coordinates, the solution to (\ref{T=0}) is linear:
\begin{equation}
  \phi = Z \cdot Y = Z_0 Y_0 - Z_1 Y_1 - Z_2 Y_2.
  \label{phi_Z}
\end{equation}
The $Z_i$ constants are the $SO\left( 1,2 \right)$ charges of the solution.
In what follows, we will sometimes use the word ``dilaton'' to mean this vector of charges.

\begin{figure}
  \centering
  \includegraphics[width=.4\textwidth]{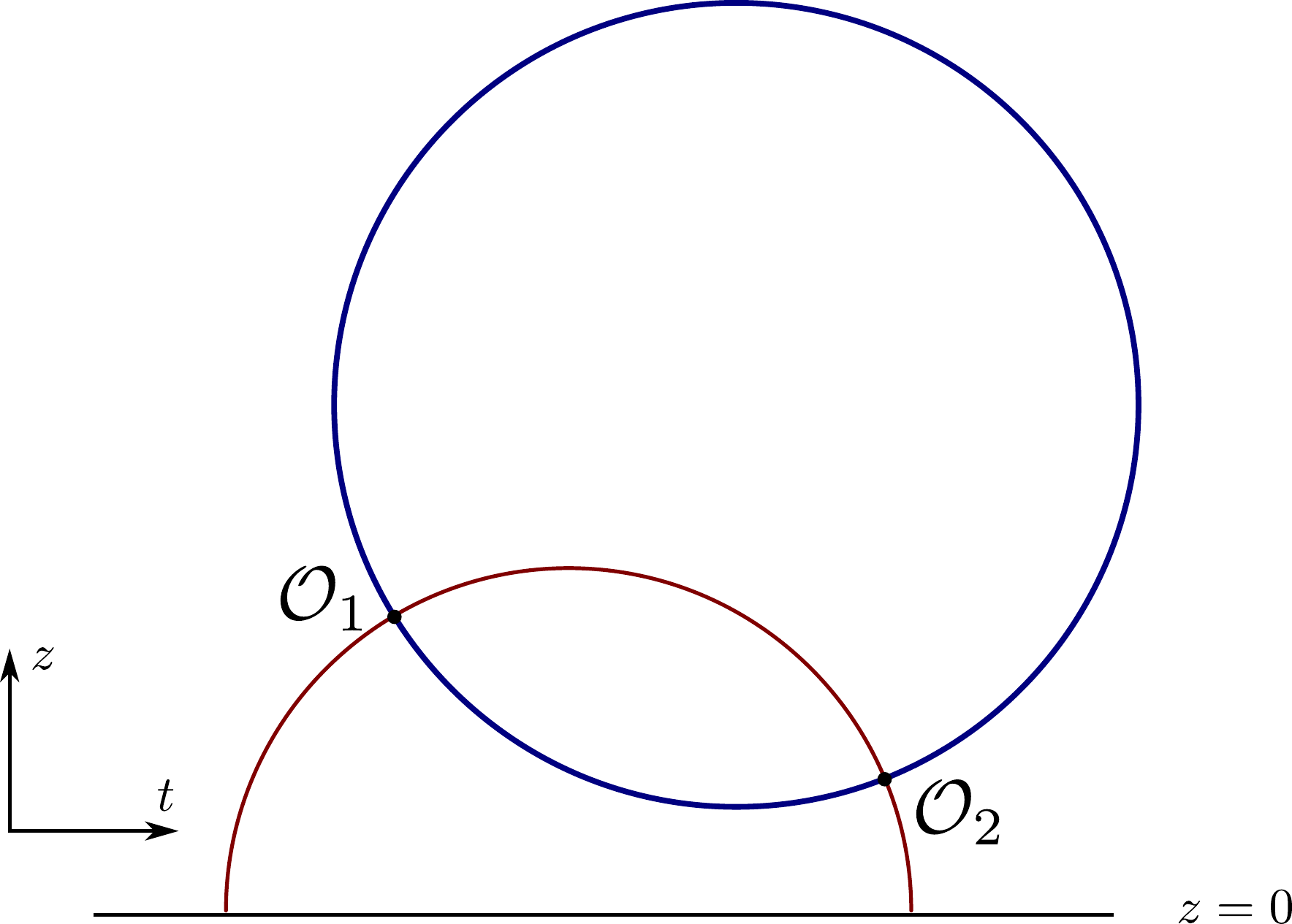}
  \caption{When mass is small, the $NAdS$ boundary is a perfect circle.}
  \label{fig:m0}
\end{figure}

We could work in the embedding coordinates, but find it more convenient to switch to the two-dimensional space.
There are several conventional representation of the (Euclidean) $AdS_2$ space.
One (perhaps more intuitive) is a Poincar\'e disc, as on fig.~\ref{fig:disc}.
We will use a hyperbolic half-plane instead (see fig. \ref{fig:m0}).

We choose the coordinates on the half-plane to be $(t,z)$, with the boundary at $z = 0$ and the $t$ axis along the boundary.
The Euclidean hyperbolic metric is:
\begin{equation}
  g = \frac{dt^2+dz^2}{z^2}.
  \label{g_hyp}
\end{equation}
The embedding coordinates map to $(t,z)$ as:
\begin{equation}
  Y_0 = \frac{1+t^2+z^2}{2z}, \quad Y_1 =- \frac{t}{z}, \quad Y_2 = \frac{1-t^2-z^2}{2z}. 
  \label{Y_tz}
\end{equation}
In coordinates $\left( t,z \right)$ on the half-plane, the classical solution for the dilaton is:
\begin{equation}
  \phi = \frac{1}{2z}\left( \left( Z_0-Z_2 \right)\left( t^2+z^2 \right)-2 Z_1t+\left( Z_0+Z_2 \right) \right). 
  \label{phi(t,z)}
\end{equation}
The boundary of the near--$AdS_2$ space is fixed by the Dirichlet boundary condition on the dilaton (\ref{phi=const}), which looks  a circle on the half-plane (see fig. \ref{fig:m0}):
\begin{equation}
  \phi = \phi_b \qquad \Rightarrow \qquad \left( t - \frac{Z_1}{Z_0-Z_2} \right)^2 + \left( z - \frac{\phi_b}{Z_0-Z_2} \right)^2 = \frac{\phi_b^2-Z^2}{\left( Z_0-Z_2 \right)^2}.
  \label{phi_circle}
\end{equation}

There are some restrictions on the parameters of the dilaton.
First, to make the Schwarzian action positive, the boundary dilaton has to be positive:
\begin{equation}
  \phi_b>0.
  \label{phi>0}
\end{equation}
The $NAdS_2$ space makes sense only if it lies completely inside the hyperbolic plane, that is, if its boundary does not intersect the true boundary at $z=0$.
This amounts to the requirement that the square of the charge vector for the dilaton is positive:
\begin{equation}
  Z^2>0,
  \label{Z_spacelike}
\end{equation}
and that the center of the circle (\ref{phi_circle}) is above the $z=0$ boundary: 
\begin{equation}
 Z_0 - Z_2 > 0.
  \label{Z_+}
\end{equation}

\begin{figure}
  \centering
\includegraphics[width = \textwidth]{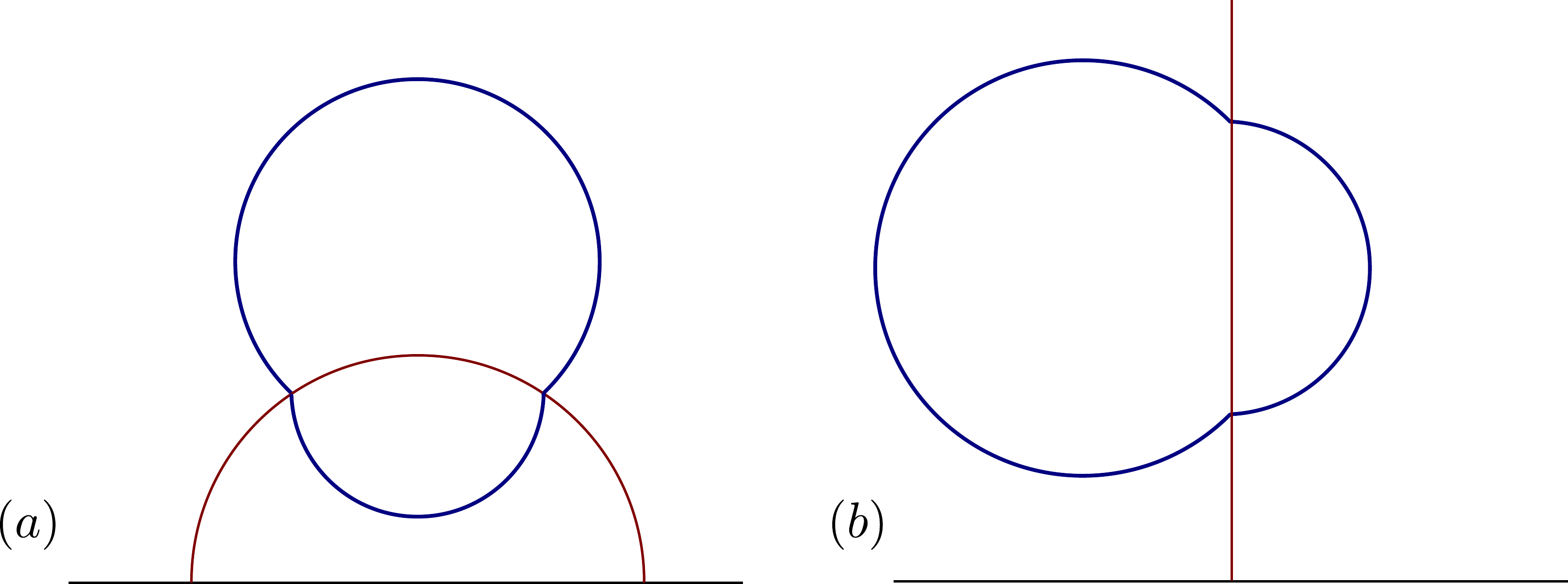}
\caption{Two different gauge choices.}
  \label{fig:sunset}
\end{figure}

To find correlators in the semi-classical approximation, we introduce a massive particle.
On a half plane, its trajectory is generally a half-circle, intersecting the boundary at a right angle (see fig.~\ref{fig:m0}). 
The parameters of the trajectory form another vector of $SO(1,2)$ charges $A$.
In the embedding coordinates, the trajectory is also given by a linear condition:
\begin{equation}
  A \cdot Y =0. 
  \label{geo_Y}
\end{equation}

In the coordinates on the half-plane, this condition reads:
\begin{equation}
  \left( A_0 - A_2 \right)\left( t^2 + z^2 \right) - 2A_1 t + \left( A_0 + A_2 \right) = 0.
  \label{geo_circle}
\end{equation}
Generally, the radius of the circle is:
\begin{equation}
  r^2 = \frac{-A^2}{\left( A_0-A_2 \right)^2}.
  \label{r2_def}
\end{equation}
From this we see that $A^2<0$. 
The $SO\left( 1,2 \right)$ transformations allow us to rotate the $A$ vector, keeping its square invariant.
This invariant fixes the mass of the particle:
\begin{equation}
  A^2 = -m^2.
  \label{A_m}
\end{equation}

\begin{figure}
  \centering
  \includegraphics[width = \textwidth]{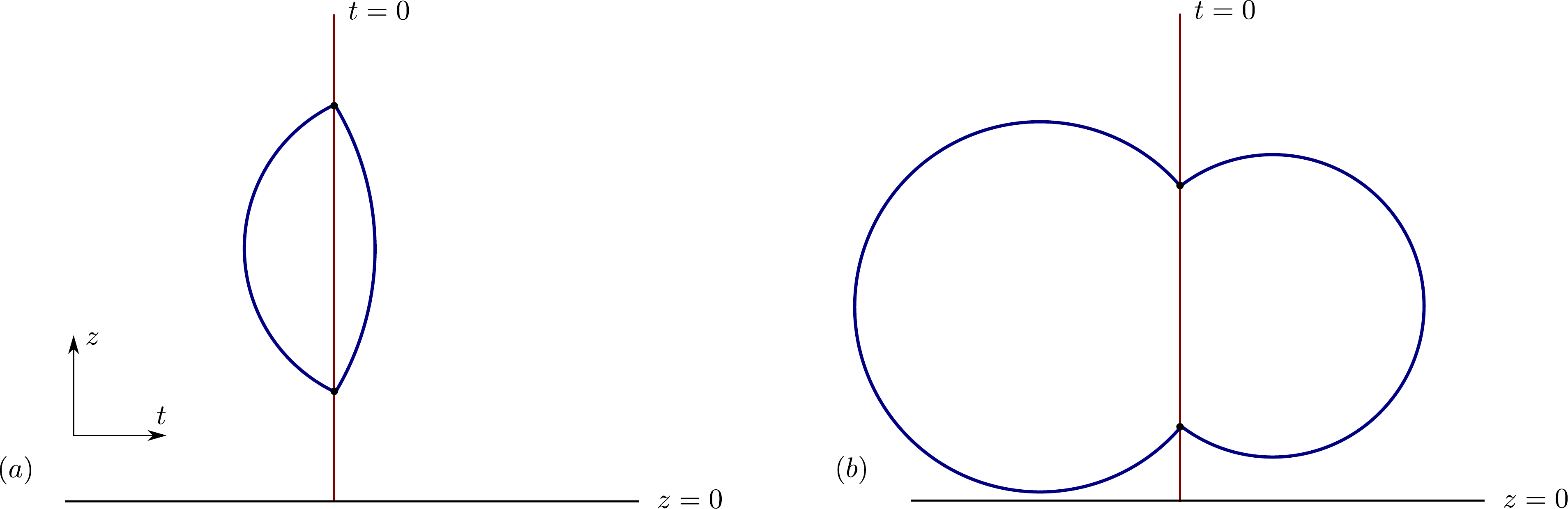}
  \caption{Positive mass makes the cusps turn inward (b), negative mass turns them outward (a).}
  \label{fig:ears}
\end{figure}

The massive particle creates a jump in the parameters of the dilaton. 
That is, on the other side of the geodesic the dilaton changes to:
\begin{equation}
  \phi = Z\cdot Y \qquad \Rightarrow \qquad \phi = \left( Z+A \right)\cdot Y.
  \label{phi_jump}
\end{equation}
This means that the boundary of the $NAdS_2$ space with massive particle inside consists of two arcs meeting at an angle (see fig. \ref{fig:sunset}).
This angle is conformally invariant, and is zero for massless particles, corresponding to a fully conformal theory on the boundary.
For a finite mass, there is cusp.
A particle with positive mass draws together pieces of the boundary, creating an inward cusp as on part $(b)$ of fig.~\ref{fig:ears}.
A particle with negative mass would pull the boundary apart, and creates an outward cusp, as can be seen on part $(a)$.
We discuss negative mass in Appendix \ref{app:m_neg}.

We can use an $SO(1,2)$ transformation to choose a convenient gauge.
With it, we can either make a picture look symmetric, or choose the degenerate case with the geodesic being a straight line, as on fig. \ref{fig:sunset}:
\begin{equation}
  (a): A_1 = 0, \qquad (b): A_0 = A_2 = 0.
  \label{sunset_grain}
\end{equation}
We will find it easier to use the $(b)$ choice.

Since the particle creates a jump in the parameters of the dilaton, we have to make sure once again that the whole boundary of the near--$AdS$ lies above the true boundary at $z=0$.
To do that, we can require:
\begin{equation}
  Z^2>0, \qquad \left( Z+A \right)^2>0,
  \label{y>R}
\end{equation}
together with the condition that the centers of the circle segments lie above the boundary:
\begin{equation}
  Z_0-Z_2>0, \qquad \left( Z+A \right)_0 -\left( Z+A \right)_2>0.
  \label{y>0}
\end{equation}

\section{Euclidean two-point function}
\label{sec:2pt}

\begin{figure}
  \centering
  \includegraphics[width = .5\textwidth]{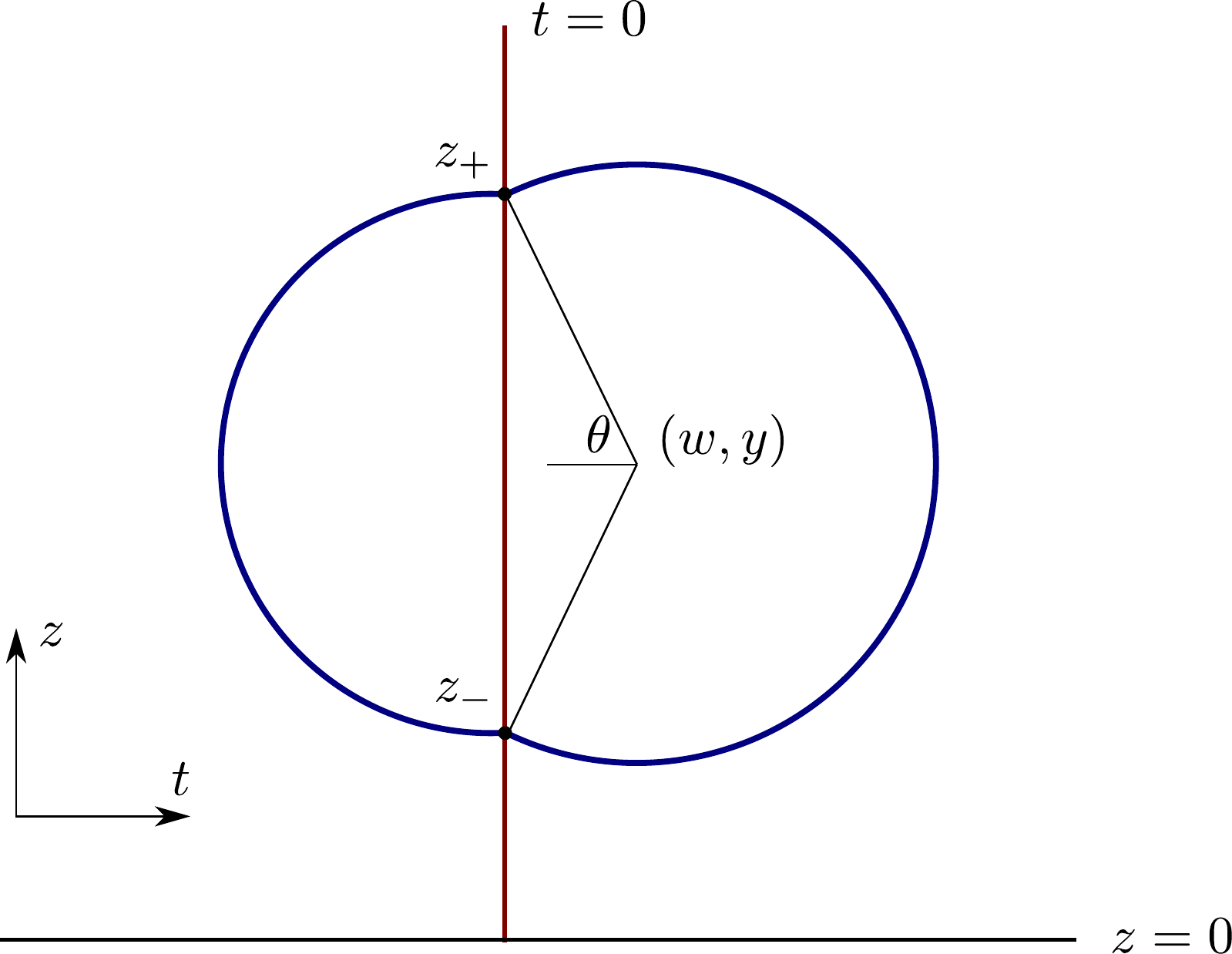}
  \caption{Massive particle distorting the boundary of a near-$AdS_2$ space. Red line is the trajectory of the particle, blue line is two arcs of the $NAdS_2$ boundary, and solid black line is the boundary of the true $AdS$ space.}
  \label{fig:grain}
\end{figure}

From the bulk point of view and in the semi-classical picture, the two-point function depends on the length of the trajectory of the massive particle.
Since we work in a part of the anti--de Sitter space, we are interested in the part of the geodesic cut out by the requirement $\phi = \phi_b$ (the part of the dark-red line in fig. \ref{fig:grain} inside the blue boundary of the $NAdS$ space).
This geodesic length is:
\begin{equation}
  \ell = \ln \frac{z_+}{z_-}.
  \label{l_log}
\end{equation}

However, the boundary quantum mechanics does not know about the trajectory of a particle in the bulk.
So from the boundary point of view, the two-point function should be expressed in terms of the distance {\it along the boundary}, which by definition must be longer than the geodesic distance.
In fig. \ref{fig:grain} it is the length of the blue segment of the circle:
\begin{equation}
  u_{12} \sim \int_{-}^{+}\frac{\sqrt{dz^2+dt^2}}{z} \qquad \text{(along the $NAdS$ boundary)}.
  \label{u12_def_1}
\end{equation}
The metric in the boundary theory can differ from the metric inherited from the bulk of $AdS_2$ by a constant factor.
We choose this factor so that when we come close to the boundary of the true $AdS$, the quantum mechanical length remains finite.
Using the $\epsilon$ parameter defined in (\ref{phi_r_def}), we define the distance in the boundary theory as:
\begin{equation}
  u_{12} \equiv \epsilon \int_{-}^{+}\frac{\sqrt{dz^2+dt^2}}{z} \qquad \text{(along the $NAdS$ boundary)}.
  \label{u12_def_2}
\end{equation}
Accordingly, we rescale the two-point function so that the limit of large dilaton is meaningful.
We can do it since the Laplace equation (\ref{laplace}) defines the two-point function up to a constant factor.
We want the two-point function to be consistent with OPE at small distances:
\begin{equation}
  G=\frac{1}{\left|u_{12}\right|^{2\Delta}}, \qquad u_{12} \to 0.
  \label{OPE}
\end{equation}
In what follows, we see that it is so if we rescale $G$ as:
\begin{equation}
  G =  \frac{1}{\epsilon^{2\Delta}} \exp\left( -\Delta \cdot \ell  \right).
  \label{G_rescaled}
\end{equation}
In what follows, it will sometimes be convenient to define the exponentiated distance $\gamma \equiv  \epsilon \cdot \exp \left( \ell/2 \right)$,  so that the two-point function is:
\begin{equation}
  G = \frac{1}{\gamma^{2\Delta}}. 
  \label{gamma_def}
\end{equation}

Our goal is to express the two-point function (\ref{gamma_def}) in terms of the boundary length (\ref{u12_def_2}).
In our choice of the gauge, if the center of this circle is at $\left( t,z \right) = \left( w, y \right)$ and the radius of the circle is $R$, this length is:

\begin{equation}
  u_{12} = \epsilon \int_{\theta_-}^{\theta_+}\frac{R d\theta}{ y - R \sin \theta} =  \epsilon \cdot \frac{2R}{\sqrt{1-w^2}}\arccos\left( \frac{wy}{R} \right),
  \label{u_wy}
\end{equation}
with a similar formula for the length of the other segment:
\begin{equation}
  u_{21} =  \epsilon \cdot \frac{2R}{\sqrt{1-w^2}}\left(\pi -  \arccos\left( \frac{wy}{R} \right) \right).
  \label{u21_wy}
\end{equation}
Here and in what follows $u_{12}$ is the length of the segment on the left-hand side of the geodesic on fig.~\ref{fig:grain}, and $u_{21}$ is the length on the right-hand side.

The sum of the lengths of these two segments is the length of the boundary:
\begin{equation}
  L \equiv u_{12} + u_{21}.
  \label{L_def}
\end{equation}

The geodesic length in terms of the same variables is given by:
\begin{equation}
  \ell = \ln \frac{z_+}{z_-} = \ln \frac{y + \sqrt{R^2-w^2}}{y - \sqrt{R^2-w^2}}.
  \label{ell_wy}
\end{equation}

The parameters $(y,w,R)$ of the circle are defined by the $SO\left( 1,2 \right)$ charges of the dilaton.
From (\ref{phi_circle}), we see that:
\begin{equation}
  w = \frac{Z_1}{Z_0 - Z_2}, \qquad y = \frac{\phi_b}{Z_0 - Z_2}, \qquad R = \frac{\sqrt{\phi_b^2-Z^2}}{Z_0-Z_2}.
  \label{wy_Z}
\end{equation}
The charge vectors are different on the different sides of the particle's worldline.
The worldline is specified by the vector of charges $A$.
We take the worldline to be vertical, therefore $A$ is fixed to be:
\begin{equation}
  A = \begin{pmatrix}
    0 \\ m \\ 0
  \end{pmatrix}.
  \label{A_vert}
\end{equation}
The dilaton charges get shifted by this vector when crossing the worldline:
\begin{equation}
  Z \mapsto Z+A.
  \label{Z_jump}
\end{equation}
This means that the $Z_1$ component of the dilaton is fixed.
However, we still have the freedom of boosting $Z_0$ and $Z_2$.
The boost acts as a rescaling of the boundary:
\begin{equation}
  w \mapsto e^\rho w, \qquad y \mapsto e^\rho y, \qquad R \mapsto e^\rho R.
  \label{boost_wy}
\end{equation}
We choose the boost in such a way that the exchange of the two operator insertions acts as an inversion:
\begin{equation}
  t \rightarrow \frac{-t}{t^2+z^2}, \qquad z \rightarrow \frac{z}{t^2+z^2},
  \label{inv_def}
\end{equation}
that is, such that:
\begin{equation}
  z_+ z_- = 1.
  \label{z+z-}
\end{equation}
This choice corresponds to $Z_2 = 0$.
It allows us to connect the radius of the boundary to the coordinates of the center:
\begin{equation}
  R^2 = y^2 + w^2 - 1.
  \label{R_wy}
\end{equation}

In the semiclassical approximation to gravity, the boundary value of the dilaton is a large parameter. 
This allows us to treat $y$ as a large parameter as well:
\begin{equation}
  \phi_b \gg 1 \qquad \Rightarrow \qquad y = \frac{\phi_b}{Z_0} \gg 1,
  \label{y>>1}
\end{equation}
In this choice of gauge and this approximation, the geodesic length is:
\begin{equation}
  \ell = 2 \cosh^{-1} y \sim 2 \ln \left( 2y \right),
  \label{l(y)}
\end{equation}
and the two-point function becomes:
\begin{equation}
  G = \frac{1}{\gamma^{2\Delta}}, \qquad \gamma =2\epsilon  y. 
  \label{G(y)}
\end{equation}
The $y \gg 1$ approximation simplifies our expression for the length of the segment $u_{12}$:
\begin{equation}
  u_{12} = \epsilon \frac{2y}{\sqrt{1-w^2}} \arccos w,
  \label{u12_y_large}
\end{equation}
which asks for the following parameterization:
\begin{equation}
  w \equiv \cos \alpha.
  \label{alpha_def}
\end{equation}
We express everything in terms of the angle $\alpha$, and in what follows call (\ref{u12_y_large}) the small $\epsilon$ approximation.

There is an important difference between working in the small $\epsilon$ approximation (\ref{u12_y_large}) and using the precise answer (\ref{u_wy}).
For the integral in (\ref{u_wy}) to make sense when $w<0$, we need to require that $y>1$.
This means that very small distances are not available to us.
The boundary dilaton $\phi_b$, or rather the $\epsilon$ parameter, plays the role of a UV cutoff.
This makes sense since we are working in the semiclassical approximation to gravity, and the underlying ultraviolet theory is beyond our scope.
However, the small $\epsilon$ approximation formally continues to small distances and, in particular, it gives the correct OPE for the operators.
We interpret the small $\epsilon$ approximation as a certain limit where $\epsilon$ is extremely small; however, we need to keep in mind that at any given value of $\epsilon$ at some point in the UV this approximation, together with OPE, is bound to break down.

We discuss the implications of the small $\epsilon$ approximation in more detail in Section~\ref{sec:angular}.
Before that, we find the two-point function in terms of the boundary length.
First we do it with back-reaction absent, then consider the symmetric case where the distance between the operators is exactly half the length of the boundary, and finally we look how the two-point function depends on the separation between the operators when the boundary length is fixed.
In doing so, we work in the small $\epsilon$ approximation and rely on numerical methods.

\subsection{No back-reaction}
\label{sec:conf}

\begin{figure}
  \centering
  \includegraphics[width = .5\textwidth]{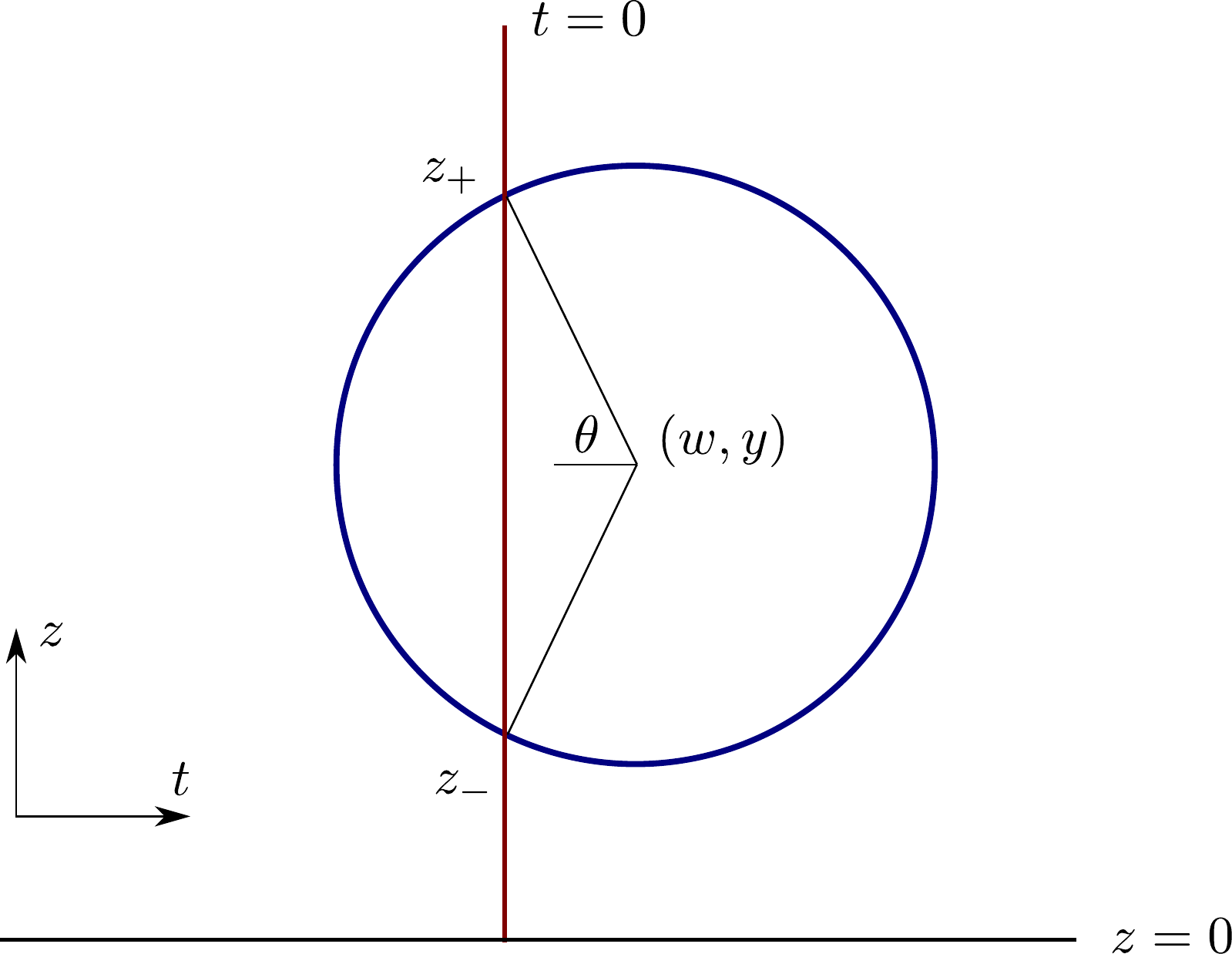}
  \caption{Near--$AdS$ space without backreaction.}
  \label{fig:conf}
\end{figure}

Without the back-reaction, the boundary of the $NAdS$ remains a perfect circle (see fig.~\ref{fig:conf}).
In particular, it is invariant under a part of the conformal group, namely rescaling of the boundary (\ref{boost_wy}).
Therefore, in the expression for the two-point function we should recover the familiar conformal answer.
Without back-reaction, there is no jump in the parameters of the dilaton, and it remains the same on both side of the particle's trajectory:
\begin{equation}
  \phi = Z \cdot Y.
  \label{Z_conf}
\end{equation}

The parameters $(y, w, R)$ are the same on both sides of the circles, and the boundary distances between the operators in the small $\epsilon$ approximation are:
\begin{equation}
  \begin{aligned}  
  u_{12} &= {2y}{\epsilon} \frac{\alpha}{\sin \alpha}, \\
  u_{21} &= {2y}{\epsilon} \frac{\pi-\alpha}{\sin \alpha}.
 \end{aligned}
  \label{u12_alpha}
\end{equation}
Then the length of the boundary is:
\begin{equation}
  L = {y}\frac{2\pi \epsilon}{\sin \alpha}.
  \label{L_alpha}
\end{equation}
From here, we see that $2\alpha$ is the arc angles between the two operators on  fig.~\ref{fig:conf}:
\begin{equation}
  u_{12} = \frac{L}{\pi}\alpha.
  \label{u_arc}
\end{equation}
The conformal (or no-backreaction) limit is when these arc angles add up to precisely a full circle.
In Section~\ref{sec:small_m} we consider small deviations from this, recovering the perturbative answer for the two-point function.

The two-point function depends only on the coordinate of the center of the circle $y$.
From (\ref{L_alpha}), we find:
\begin{equation}
  y = \frac{L}{2\pi \epsilon}\sin \alpha.
  \label{y_u12}
\end{equation}
Then the two-point function has the form we would expect in a conformal theory:
\begin{equation}
  G = \left(\frac{L}{\pi} \sin \frac{\pi u_{12}}{L} \right)^{-2\Delta}.
  \label{G_conf_ans}
\end{equation}
Notice that because of rescaling $G$ in (\ref{G_rescaled}), we in particular recover the usual OPE expected in a conformal theory:
\begin{equation}
  G \sim \frac{1}{{u_{12}}^{2\Delta}}, \qquad u_{12} \ll L.
  \label{G_OPE_conf}
\end{equation}

\subsection{Two-point function with back-reaction: symmetric case}
\label{sec:sym}

\begin{figure}
  \centering
  \includegraphics[width = .5\textwidth]{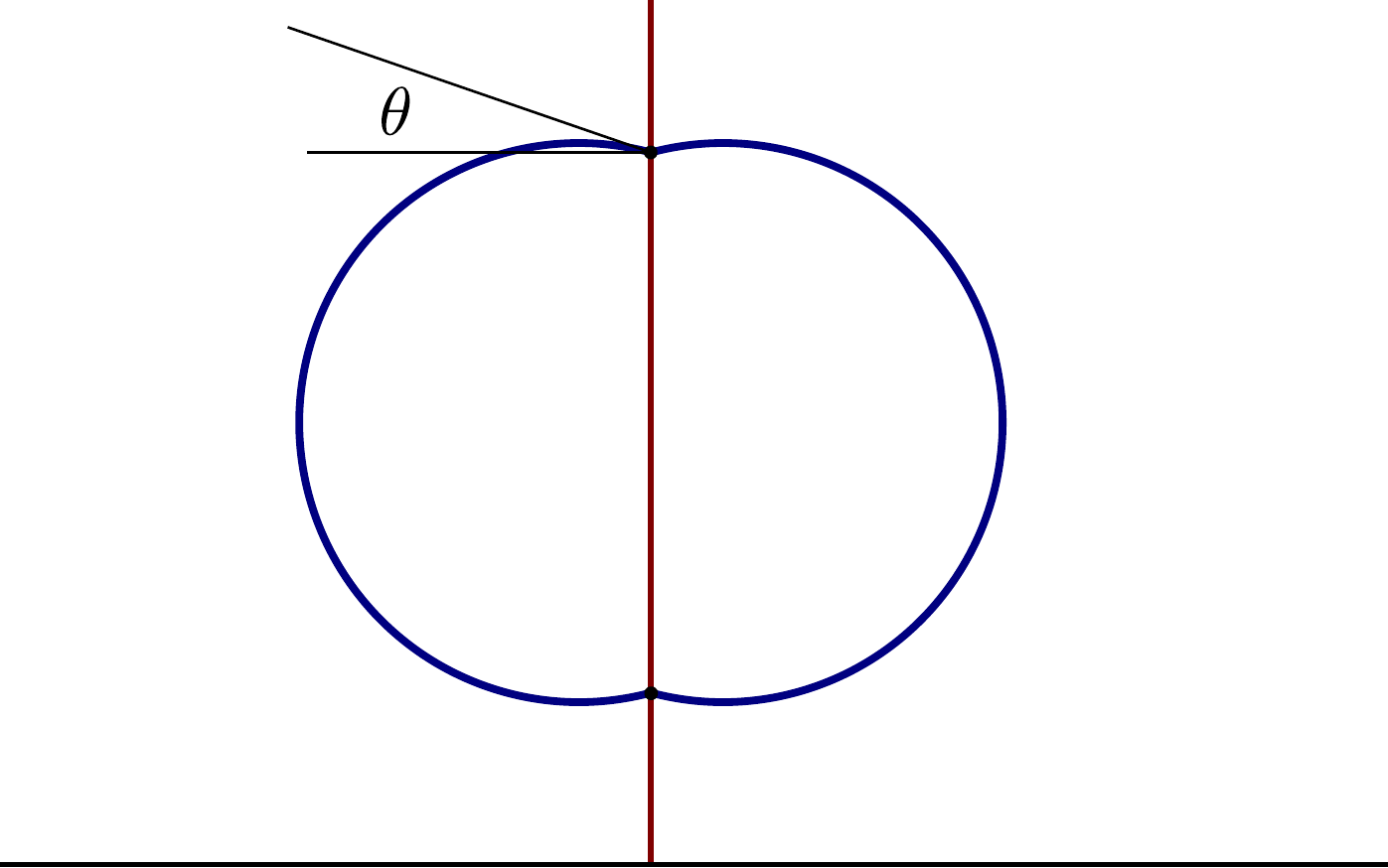}
  \caption{The near-$AdS$ boundary for the symmetric case.}
  \label{fig:sym}
\end{figure}

Having reproduced the conformal result, let us turn to discussing the two-point function with back-reaction.
To simplify, we first consider the symmetric case, when $u_{12} = u_{21} = L/2$ (see fig. \ref{fig:sym}).
The conformal answer gives:
\begin{equation}
  G = \left( \frac{L}{\pi} \sin \frac{\pi}{2} \right)^{-2\Delta}  = \left(\frac{L}{\pi} \right)^{-2\Delta}.
  \label{G_conf_sym}
\end{equation}

The parameters of the dilaton jump by $\left( 0, m, 0 \right)$ when crossing the trajectory of the particle.
Left-right symmetry of the picture corresponds to $Z_1 \leftrightarrow -Z_1$.
Since $Z_1$ has to jump by $m$ when crossing the worldline of the particle, in the symmetric case we can choose the parameters of the dilaton to be:
\begin{equation}
  Z = \left( Z_0, \pm \frac{m}{2}, 0 \right).
  \label{Z_sym}
\end{equation}
The angle $\theta$ in fig. \ref{fig:sym} is conformally invariant, and it shows how large the back-reaction is:
\begin{equation}
  \sin \theta = \frac{m}{2\phi_b}. 
  \label{theta_sym}
\end{equation}
In the massless, or conformal, case, $\theta=0$ and the cusp on the boundary is absent.
The combination $\phi_b/m$ governs how close our theory is to conformal.
It roughly tells the $AdS$ distance at which the conformal symmetry breaks down in the infrared.
We want this distance to be small:
\begin{equation}
  \frac{m}{ \phi_b} \ll 1.
  \label{l0_def}
\end{equation}

In the symmetric case, the $w$ parameters have opposite signs at the opposite sides of the worldline:
\begin{equation}
  w_{\text{left}} = -w_{\text{right}}<0.
  \label{w_sym}
\end{equation}
The boundary of $NAdS$ touches the true $AdS$ boundary when $w=-1$. 
It is convenient to use the $\alpha$ angles, as before:
\begin{equation}
  w = \cos \alpha, \qquad \pi/2 \le \alpha \le \pi.
  \label{w_alpha_2}
\end{equation}
These angles add up to $\pi$ on the opposite sides:
\begin{equation}
  \alpha_{\text{left}} = \pi - \alpha_{\text{right}}.
  \label{alpha_sym}
\end{equation}

From the jump in the parameters of the dilaton, we find:
\begin{equation}
  w_{\text{left}} = -\frac{m}{Z_0}.
  \label{w_mz}
\end{equation}
We are still working in the gauge $Z_2 = 0$ where inversion acts as the exchange of the two operators, so the radius is given by (\ref{R_wy}):
\begin{equation}
  R^2 = y^2 - \sin^2 \alpha.
  \label{R_sym}
\end{equation}
This time, $y$ is also expressed using $\alpha$:
\begin{equation}
  y = \frac{\phi_b}{Z_0} = -\frac{2\phi_b}{m} \cos \alpha, \qquad \gamma =- \frac{4 \phi_r}{m}\cos \alpha.
  \label{y_sym_alpha}
\end{equation}
The length of the boundary is found from (\ref{u12_alpha}):
\begin{equation}
  L = 2u_{12} = 4 \epsilon R \frac{\alpha}{\sin \alpha},
  \label{L_sym}
\end{equation}
\begin{figure}
  \centering
  \includegraphics[width=.5\textwidth]{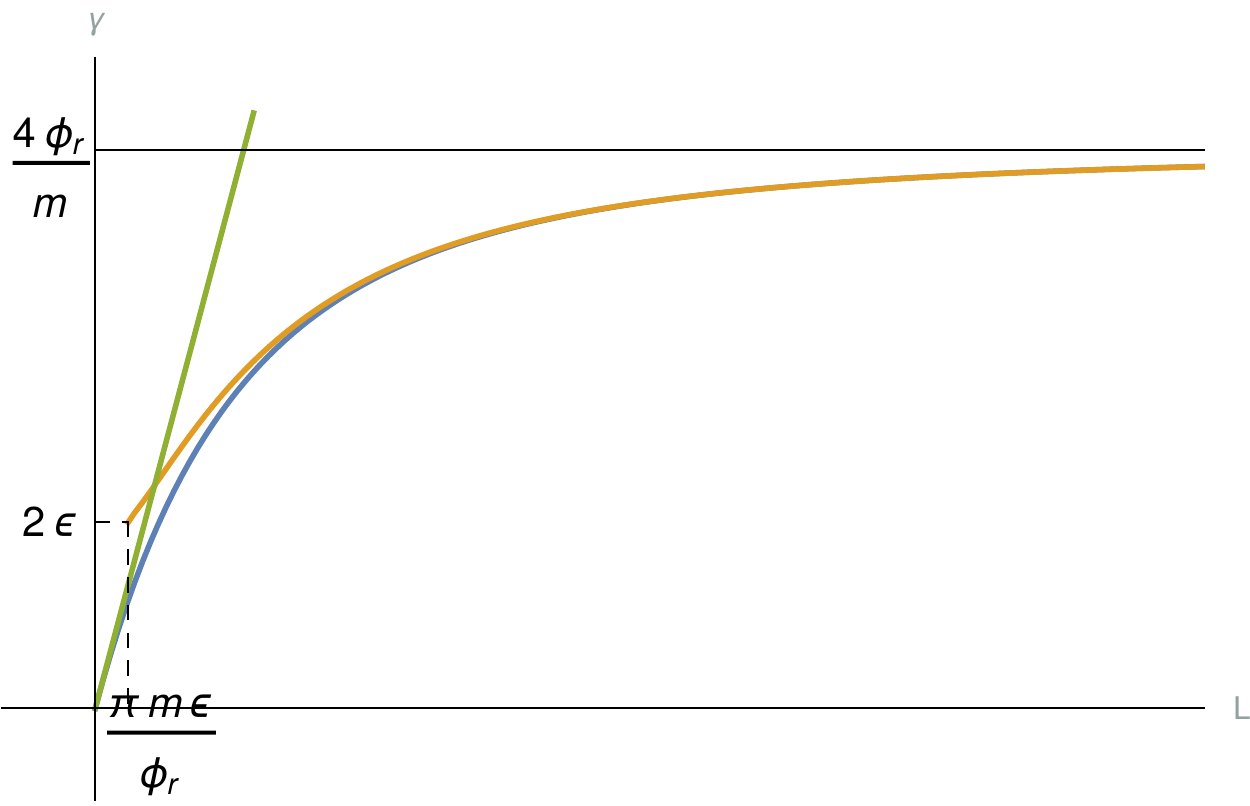}
  \caption{Numerical solution to (\ref{L_sym}) and (\ref{y_sym_alpha}) when mass is positive.
  The blue line is the conformal result, the orange line is the full solution, and the green line is the solution with $R \sim y$, or $\phi_b/m$ taken to be large.}
  \label{fig:yL_m+}
\end{figure}

Using (\ref{L_sym}) and (\ref{y_sym_alpha}), we cannot find the two-point function of the boundary theory in a closed form, as we did in Section \ref{sec:conf}.
However, we can solve these two conditions numerically, the result shown in fig.~\ref{fig:yL_m+}.

The limits of small and large $L$ can also be studied numerically.
If $\alpha$ is close to $\pi/2$,
\begin{equation}
  \alpha = \frac{\pi}{2}+\delta, \qquad \delta \gg \frac{m}{\phi_b},
  \label{alpha_conf_+}
\end{equation}
then
\begin{equation}
  \gamma = \frac{4 \phi_r}{m} \delta, \qquad L \sim \pi \gamma.
  \label{y_alpha_pi2+}
\end{equation}
This is the conformal limit.
Note that in this limit, only the approximate formulas (\ref{L_sym}, \ref{y_sym_alpha}) make sense.
The precise distance (\ref{u_wy}) stops working when $L \sim \pi m \epsilon^2 /\phi_r$. 
This is when $y$ goes to 1, so the circles forming the boundary barely touch the worldline.
$y=1$ is the physical limit, restricting our insight into the UV physics.
This reminds us once again that our theory is cut off in the ultraviolet, much like the SYK model.
Our $\alpha$ parameterization allows us to continue past the cutoff and in fact make theory nearly conformal in the UV.
But this continuation is formal, and we should keep in mind that the true ultraviolet theory is not accessible for us.

Also, from fig.~\ref{fig:yL_m+} we see that there is a lower limit on $\phi_b$ parameter:
\begin{equation}
  \phi_b > m/2.
  \label{l>.5}
\end{equation}

In the opposite limit, when $\alpha$ is near $\pi$,
\begin{equation}
  \alpha = \pi - \delta,
  \label{alpha_pi}
\end{equation}
the length of the boundary becomes infinite, and $\gamma$ saturates:
\begin{equation}
  \gamma = \frac{4 \phi_r}{m}, \qquad L \sim \frac{1}{\delta}.
  \label{yl_pi}
\end{equation}
The two-point function depends on the $y$ parameter and also becomes constant in this limit:
\begin{equation}
  G = \gamma^{-2\Delta} \sim \left( \frac{4m}{\phi_r} \right)^{2m}.
  \label{G_a_pi}
\end{equation}
So we see that the two-point function approaches a non-zero constant when boundary length is large.

\begin{figure}
  \centering
  \includegraphics[width = .6\textwidth]{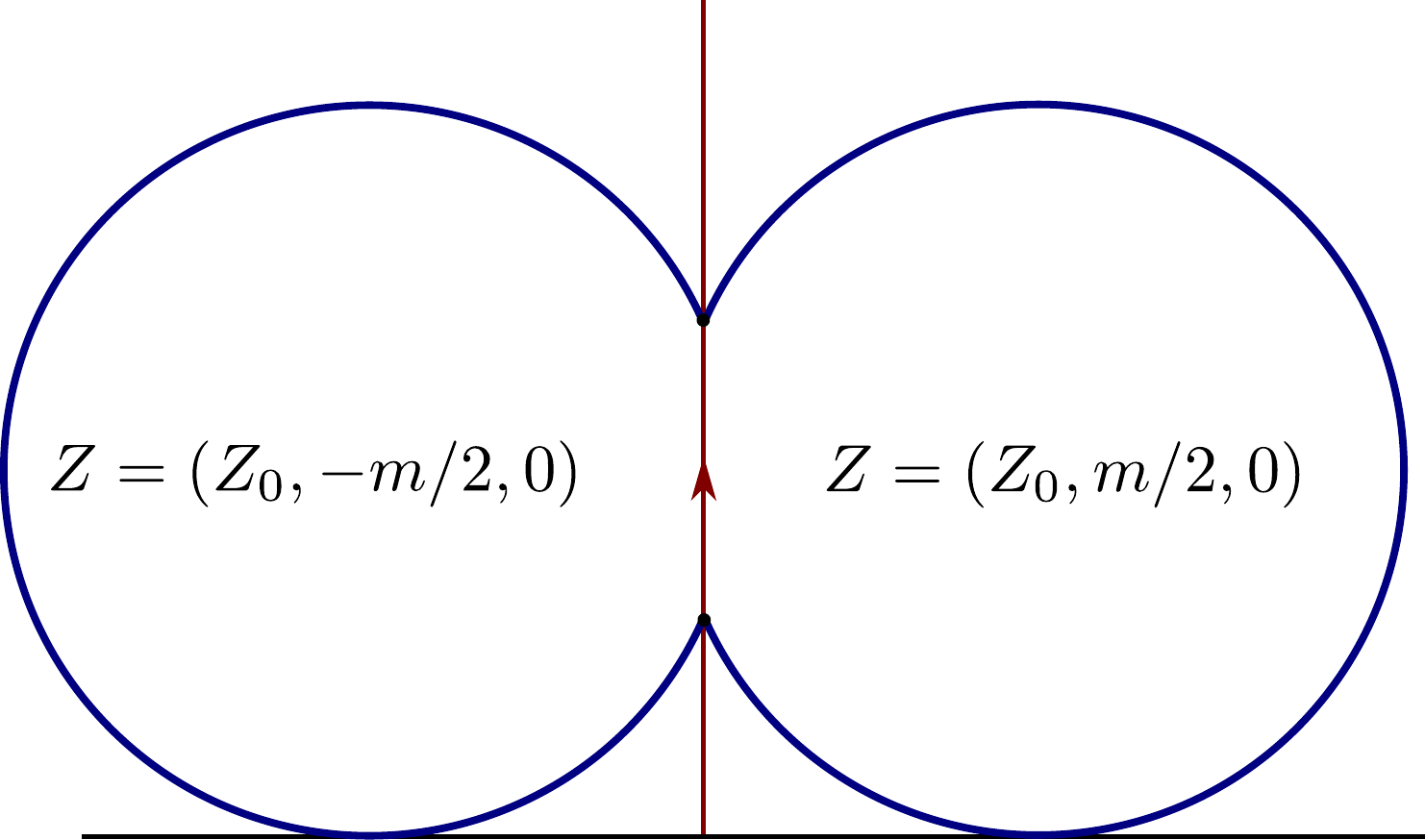}
  \caption{The two-point function stays finite as the $NAdS$ boundary touches the boundary of the true $AdS$, making the boundary length diverge.}
  \label{fig:sym_diverge}
\end{figure}

Although we have found it in the small $\epsilon$ approximation, the full answer for the two-point function gives the same result.
Geometrically, this happens because we can easily bring the length of the boundary to diverge, just making it approach the boundary of the true $AdS$ space.
If $\phi_b$ is finite, the bulk distance between the two operators can be kept finite in this limit, as on fig.~\ref{fig:sym_diverge}.

There is a comment to be made about the exchange symmetry of the operators.
So far, we treated the operator insertions as identical, and naively the picture on fig.~\ref{fig:sym_diverge} is symmetric under inversion.
However, a closer look shows that this is not so.
The parameters of the dilaton jump by $A_1>0$ when we cross the worldline of the particle from left to right. 
(We need $A_1$ to be positive, so that the cusps at the operator insertions point inward.)
This is a matter of convention, but this convention breaks the exchange symmetry between the operators.
To make ``left'' and ``right'' well-defined, we draw an arrow pointing from one operator insertion to the other on fig.~\ref{fig:sym_diverge}. 
Inversion reverses the direction of this arrow and at the same time exchanges left and right sides of the picture.
In what follows, we implicitly assume that all the trajectories are directional.

\subsection{Two-point function: generic case}
\label{sec:2pt_gen}
Having discussed the symmetric case, we move on to consider $u_{12} \neq u_{21}$.
From what we have seen above, we expect the two-point function to be approximately conformal at short distances, and saturate at a constant when the operators are far away. 
We also expect to approach the conformal result when mass goes to zero.

The parameters of the dilaton are:
\begin{equation}
  Z = \left( Z_0, Z_1 \pm \frac{m}{2}, 0 \right),
  \label{Z_gen}
\end{equation}
where we have again fixed the gauge so that the inversion exchanges the two operators.
The horizontal displacements for the centers of the circles are:
\begin{equation}
  w_{1,2} = \frac{Z_1 \pm \frac{m}{2}}{Z_0}.
  \label{w_gen}
\end{equation}
The radii of the circular segments are now different:
\begin{equation}
  R_{1,2}^2 = y^2+w_{1,2}^2-1. 
  \label{R_i}
\end{equation}
Here $y$ is the vertical coordinate of the center, and it is the same for both segments.
We can express it via $w_{1,2}$:
\begin{equation}
  y = \frac{\phi_b}{Z_0} = \frac{\phi_b}{m} \left( w_2-w_1 \right).
  \label{y_w}
\end{equation}

The lengths of the boundary segments are given by:
\begin{equation}
  \begin{aligned}
  u_{12} &=\epsilon \frac{2 R_1}{\sqrt{1-w_1^2}} \arccos \left( \frac{w_1 y}{R_1}\right),\\
  u_{21} &=\epsilon \frac{2 R_2}{\sqrt{1-w_2^2}}\left(\pi -   \arccos \left( \frac{w_2 y}{R_2}\right) \right).
  \end{aligned}
  \label{u_generic}
\end{equation}
For these lengths to be real, the argument of the $\arccos$ has to be greater than -1.
It implies that:

\begin{equation}
  y>1.
  \label{y>1}
\end{equation}
This is our condition for the UV cutoff.
For notational simplicity, we switch signs in (\ref{u_generic}), so that the boundary distances are positive when mass is positive.

If we take $y \gg 1$, we can assume $y \sim R$ and use the angular ansatz, as above,
\begin{equation}
  w_{1,2} = \cos \alpha_{1,2},
  \label{w_gen_alpha}
\end{equation}
and find a relatively simple expression for the boundary distances:
\begin{equation}
  \begin{aligned}
  u_{12} =& \frac{2 \phi_r}{m} \frac{\alpha_1}{\sin \alpha_1} \left( \cos \alpha_2 - \cos \alpha_1 \right), \\
  u_{21} =& \frac{2 \phi_r}{m} \frac{\pi- \alpha_2}{\sin \alpha_2} \left( \cos \alpha_2 - \cos \alpha_1 \right).
  \end{aligned}
  \label{u_12_21}
\end{equation}
The $\gamma$ parameter then is given by:
\begin{equation}
  \gamma = \frac{2 \phi_r}{m} \left( \cos \alpha_2 - \cos \alpha_1 \right).
  \label{y_alpha}
\end{equation}
An important property of the small $\epsilon$ approximation is that neither distances nor the two-point function depends on $\epsilon$.
In a way, we can treat $\epsilon$ as a parameter saying how close the small $\epsilon$ approximation is to the true answer.
We will see how the two-point function depends on this parameter in Section~\ref{sec:angular}.

\begin{figure}
  \centering
\includegraphics[width=.4\textwidth]{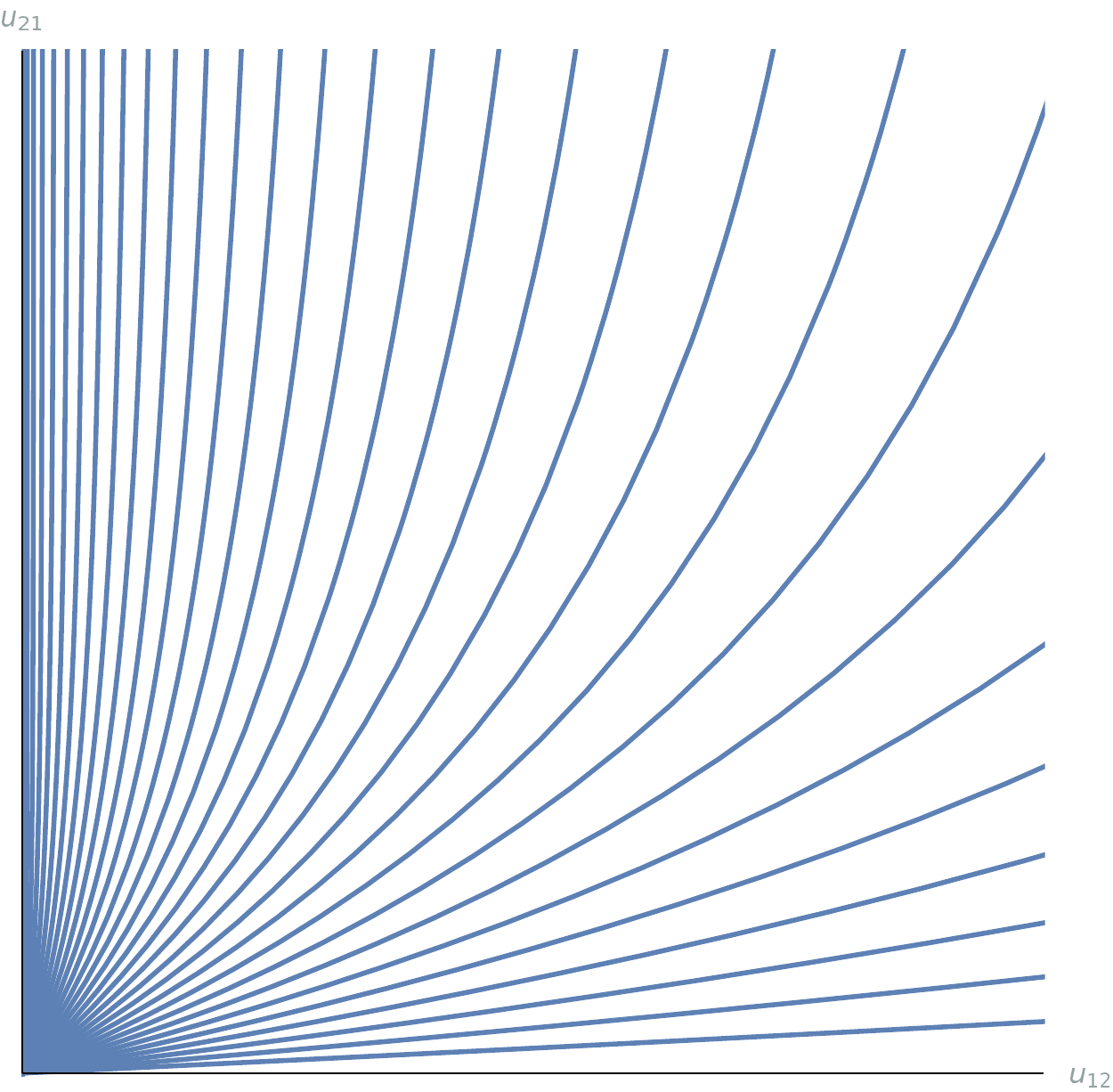}
\caption{Every line on the $u$ plane corresponds to fixed $\alpha_1$ and varying $\alpha_2$.}
  \label{fig:uu_minus}
\end{figure}

A natural question is whether the angular parameters $\alpha_{1,2}$ are in one-to-one correspondence with the boundary distances.
On fig.~\ref{fig:uu_minus} we vary the angles and find that the quarter of the $u$ plane with both positive distance gets covered exactly once.
From here, we conclude that we can indeed change variables to $\alpha_{1,2}$ without introducing any additional singularities.

In the small $\epsilon$ approximation, the UV theory is nearly conformal.
Indeed, if we take:
\begin{equation}
  \alpha_1 = \delta_1, \qquad \alpha_2 = \delta_2, \qquad \delta_1^2-\delta_2^2 = \frac{mL}{\pi \phi_r} \cdot \delta_2, \qquad \delta_{1,2} \ll 1,
  \label{2pt_OPE}
\end{equation}
we get for the distances:
\begin{equation}
  u_{21}\sim L, \qquad u_{12} \sim \frac{L}{\pi} \delta_2,
  \label{u_OPE}
\end{equation}
and the two-point function is:
\begin{equation}
  G =\frac{1}{\gamma^{2\Delta}} = \frac{1}{|u_{12}|^{2\Delta}}.
  \label{G_OPE}
\end{equation}
This is the behavior one expects from the OPE.
We found in the small $\epsilon$ approximation, and cannot take it literally.
However, for every $u_{12}$ we can find $\epsilon$ small enough so that at that distance, the two-point function looks like (\ref{G_OPE}). 
For that same $\epsilon$, the OPE will break down as we go to shorter distances.

\begin{figure}
  \centering
  \includegraphics[width=.6\textwidth]{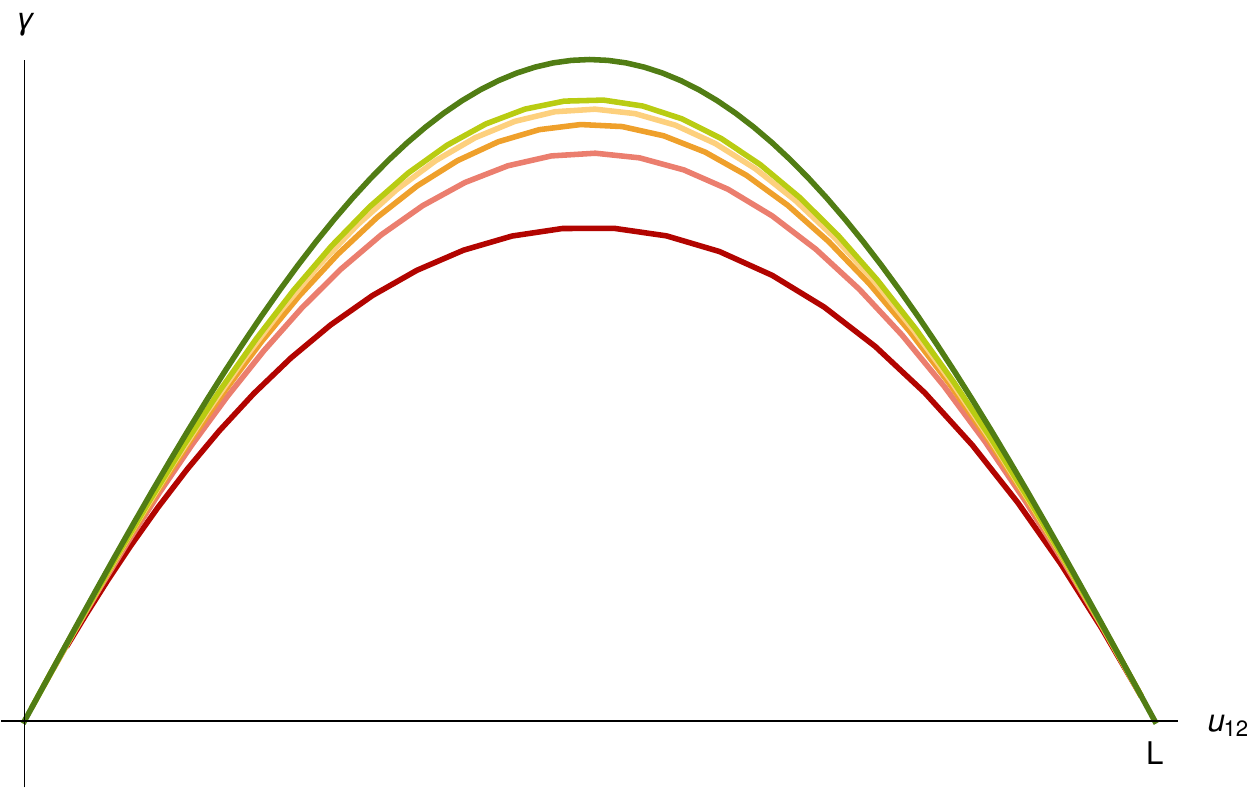}
  \caption{$\gamma\left( u_{12} \right)$ for various masses, the curves becoming greener as mass decreases. The dark green line is the conformal result $\gamma=L/\pi \cdot \sin \left( \pi  u_{12}/ L\right)$. 
  }
  \label{fig:yu+}
\end{figure}

Let us also look in more detail how the two-point function depends on mass.
To do that, we fix the full length of the boundary $L$ and find $\gamma$ as a function of $u_{12}$.
The two-point function, as before, is $G=\gamma^{-2\Delta}$.
There is no analytical solution, however the equations can be solved numerically with good convergence.

The solutions are plotted on fig.~\ref{fig:yu+}.
We see that $\gamma\left( u_{12} \right)$ starts linear for small distances.
For every mass, when distance is small enough, the two-point function is indistinguishable from the conformal one, as we have seen in (\ref{G_OPE}).
As distance grows, $\gamma$ deviates from the conformal answer, however having a similar general shape.
As mass decreases, $\gamma$ comes closer to conformal, as we expect on general grounds.

\subsection{Small \texorpdfstring{$\epsilon $}{epsilon}  approximation}
\label{sec:angular}

\begin{figure}
  \centering
  \includegraphics[width=.5\textwidth]{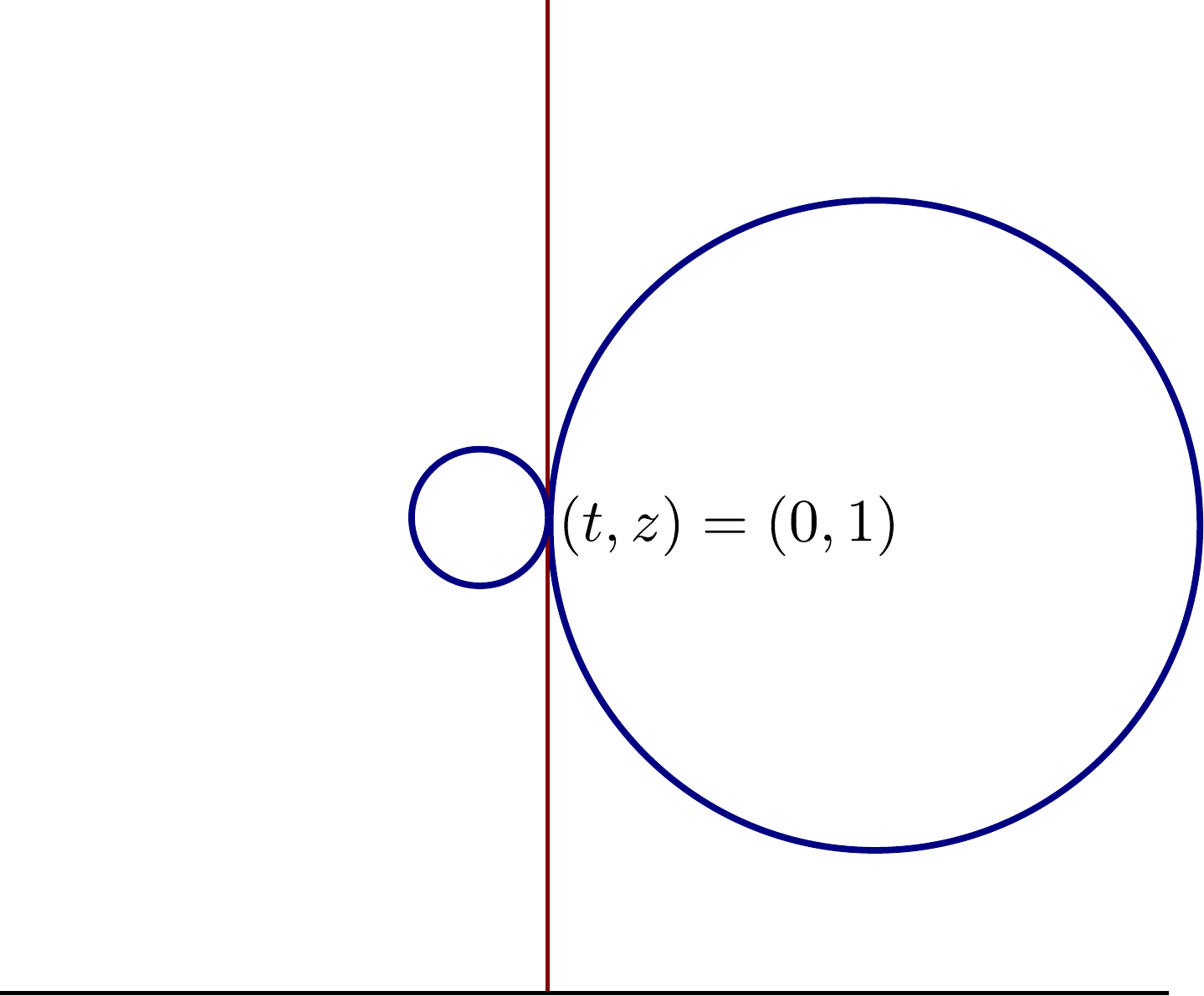}
  \caption{The geodesic length can go to zero when one of the distances $u_{12}$ becomes small.
    This dictates the maximal value of the two-point function $G \sim 1/\gamma^{2m}$.
    In this situation, invariance under inversion fixes $y=1$. 
    Thus we can treat $y=1$ as an ultraviolet cutoff.}
  \label{fig:UV}
\end{figure}

In this Section, we look at how the precise answer for the two-point function coming from (\ref{u_wy}) differs from the small $\epsilon$ approximation we discussed in Section~\ref{sec:2pt_gen}.
The small $\epsilon$ approximation relies on the boundary dilaton $\phi_b$ being large compared to all the components of the dilaton charge vector.
The ``distance'' to the boundary $\epsilon$ also tells how far we are from the pure $AdS$ setup.
We expect the precise two-point function to converge to the approximation when $\epsilon \to 0$.

However, generally $\epsilon$ is finite, and the boundary dilaton is also finitely large.
We see that it makes a difference in the ultraviolet, and in particular that the ultraviolet limit is not conformal.
It should be emphasized that the semiclassical description eventually breaks down in the UV and our analysis no longer works there.
We also show that at finite distances the small $\epsilon$ approximation is very close to the full answer.

As we mentioned before, the integral in (\ref{u_wy}) makes sense only when $y>1$.
Thus we can think of $y \sim 1$ as a condition on the UV cutoff, coming from our semiclassical approximation of the gravitational theory.
On fig.~\ref{fig:UV}, we have schematically drawn a setup with $y=1$. 
The vertical coordinate of one of the operators is always greater than 1, thus it happens when both the operators are at the same point.
This is our cutoff and at the same time the smallest value of $\gamma$ possible.

In the small $\epsilon$ approximation, $y\sim 1$ belongs to the conformal regime, with $\gamma \sim u_{12}$.
From this, we find that the small $\epsilon$ approximation breaks down roughly at distances:
\begin{equation}
  u_{12} \sim \epsilon.
  \label{u_UV}
\end{equation}
This is where our semiclassical approximation stops being reliable.
Together with it, the OPE also breaks down.
Therefore we should consider OPE carefully, taking into account the order of limits we are taking.
For every distance between the two operators, we can find $\epsilon$ small enough that the OPE holds.
However, after we fix $\epsilon$, we can bring operators close enough to ensure that OPE no longer works.

\begin{figure}
  \centering
  \includegraphics[width=.6\textwidth]{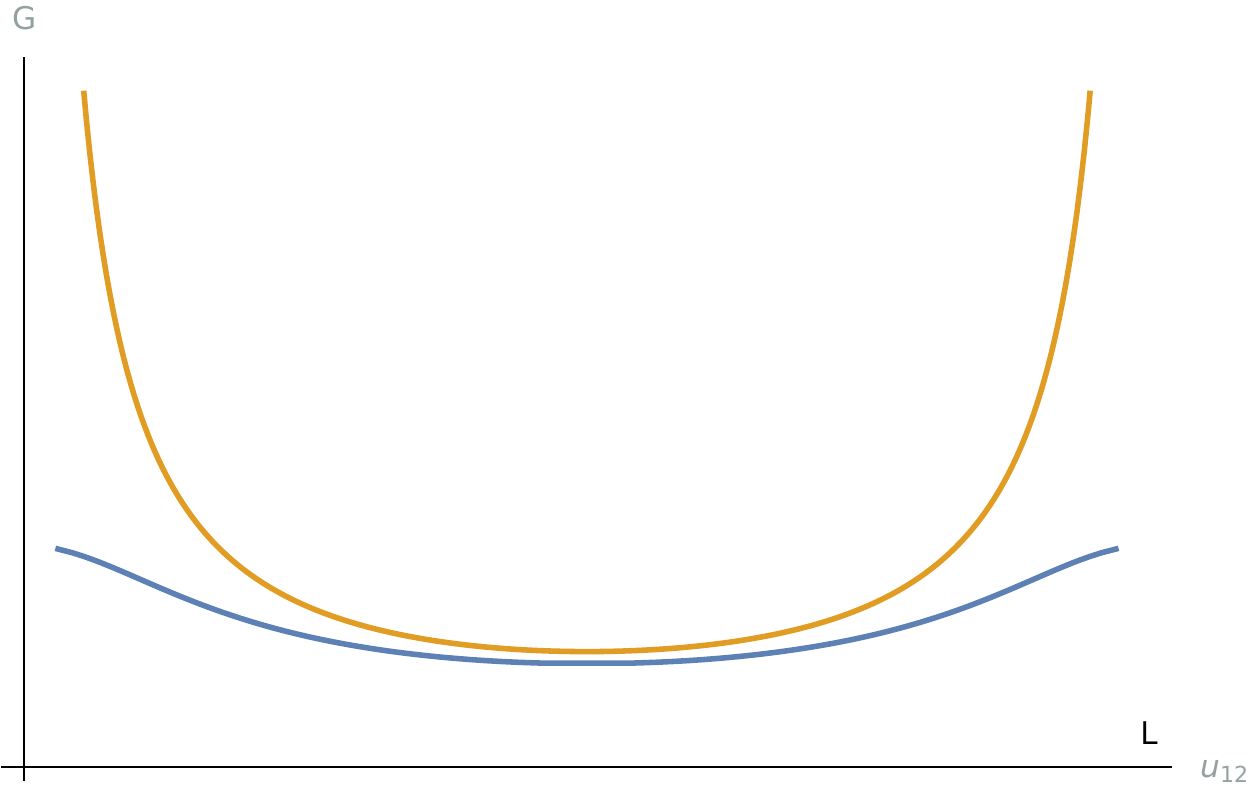}
  \caption{The two-point function in the small $\epsilon$ approximation (yellow) versus the precise one (blue).}
  \label{fig:ang_corr}
\end{figure}

We can think of the small $\epsilon$ approximation as a way to continue our theory in the UV consistently with conformal symmetry.
We should keep in mind that this way may not be physical, and this behavior is not typical for one-dimensional quantum mechanics.
In the well-studied example of the SYK model, the two-point function looks conformal only at large distances, and in the ultraviolet the theory is essentially free.
Thus our full answer (\ref{u_wy}) behaves more like a conventional SYK model, and the approximation (\ref{u12_y_large}) is more like the conformal part of the SYK, or the cSYK of \cite{Gross:2017vhb}.
On fig.~\ref{fig:ang_corr}, we draw the precise result for the two-point function and the small $\epsilon$ approximation to it, and it reminds us of the way the precise two-point function of the SYK compares to the conformal approximation (see for example \cite{Maldacena:2016hyu}).

\begin{figure}
  \centering
  \includegraphics[width=.6\textwidth]{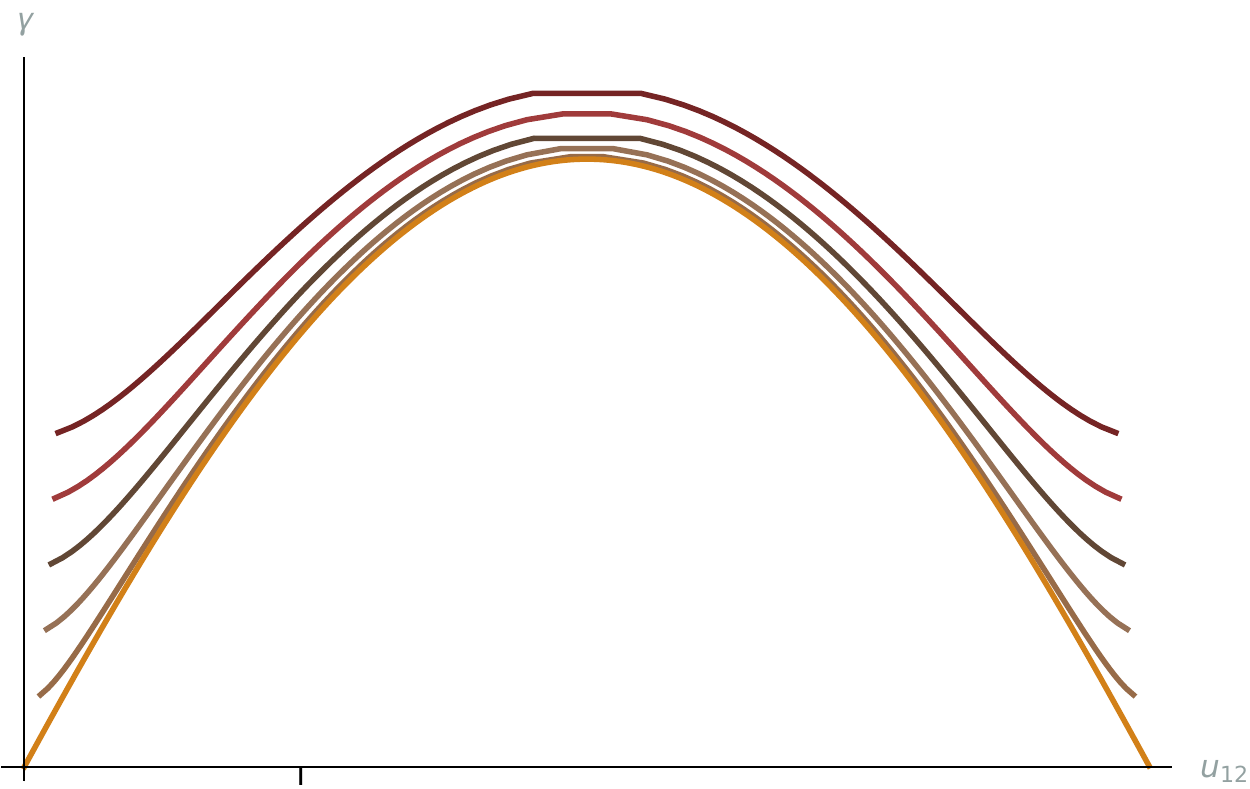}
  \caption{The precise answer for $\gamma\left( u_{12} \right)$ (from red to orange) gets closer to the small $\epsilon$ approximation (light orange) as $\epsilon$ decreases.}
  \label{fig:2pt_UV}
\end{figure}

For large distances the precise answer and the approximation get very close.
On fig.~\ref{fig:2pt_UV}, we fix the length of the boundary and solve for $\gamma\left( u_{12} \right)$ numerically for various values of $\epsilon$.
(It should be said that the approximation is much friendlier to numerical methods.)
We see that as $\epsilon$ decreases, the answer gets closer to the small $\epsilon$ approximation, as expected.

Thus we see that the precise solution has two parameters: $\epsilon$, which says how far in the ultraviolet we can extend the conformal symmetry and therefore how close it is to the small $\epsilon$ approximation, and the ratio $\phi_r/m$, which governs how close we are to the conformal answer, including in the infrared.
It may be instructive to draw a further analogy with the SYK model.
In the large $N$ limit, the SYK model is approximately conformal at large distances.
Since $\phi_r \sim N$, we expect $\phi_r$ to be large, and as it grows the two-point function gets closer to the conformal one.
In the UV, the SYK model is effectively free and the two-point function approaches a limit.
To get the same result in our theory, we keep $\epsilon$ fixed (but small).
Since our results depend only on the ratio $\phi_r/m \sim N/\Delta$, we expect the two-point functions of heavier operators to get farther away from conformal in the SYK as well.

It should be said that this analogy relies on our description of the UV region, where the semiclassical approximation we are working in stops being applicable.
So we think of the SYK as an (approximation to an) effective theory, rather than the precise holographic dual of the gravitational theory in the bulk.

\begin{figure}
  \centering
  \begin{tabular}{lr}
    \includegraphics[width=.5\textwidth]{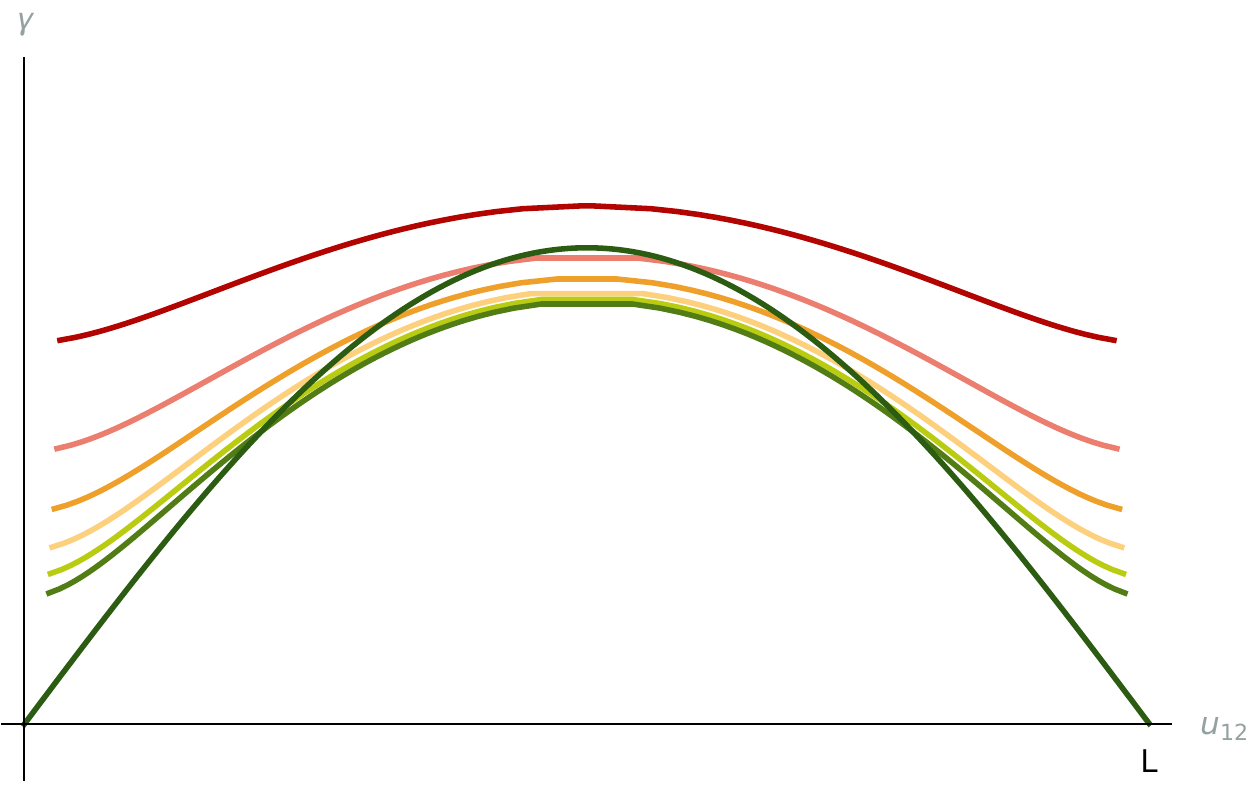}
  &
    \includegraphics[width=.5\textwidth]{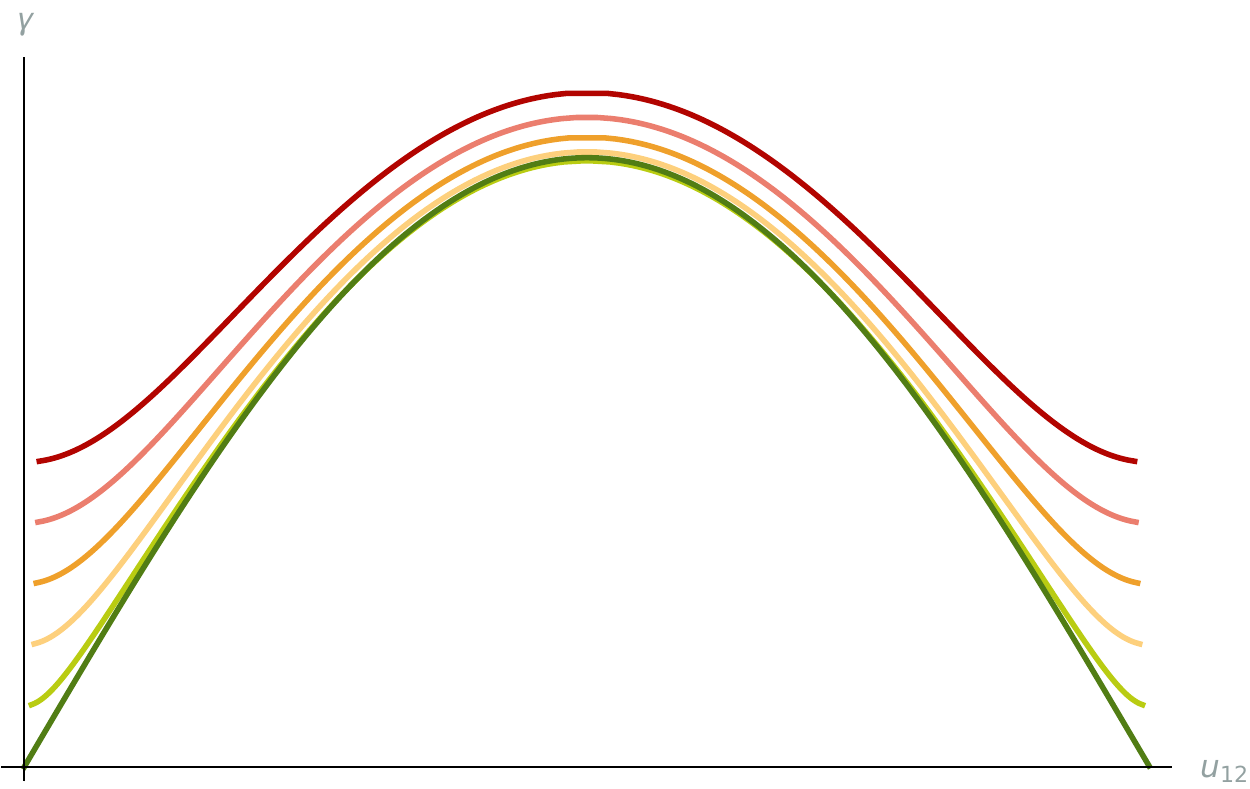}
  \end{tabular}
  \caption{$\gamma\left( u_{12} \right)$ for various values of the boundary dilaton compared to the conformal answer. The curves become greener as $\epsilon$ decreases. The renormalized dilaton $\phi_r$ on the right is larger than on the left, and the result is accordingly close to the conformal.}
  \label{fig:phi_grows}
\end{figure}

Finally, we can look at how the precise answer for $\gamma\left( u_{12} \right)$ approaches the conformal answer when $\epsilon$ changes.
The numerical results are on fig.~\ref{fig:phi_grows}.
When mass is large (or $\phi_r$ small), there is little resemblance of the conformal answer both in the ultraviolet and the infrared.
When mass becomes smaller (or $\phi_r$ larger), the $\gamma\left( u_{12} \right)$ function approaches the conformal answer from above, unlike in the small $\epsilon$ approximation (see fig.~\ref{fig:yu+}).

\subsection{Extrinsic curvature}
\label{sec:K}

Up to this point, when finding the two-point function, we have considered only the action of massive particle.
The full gravitational action (\ref{I_JT}) also contains a term with extrinsic curvature: 
\begin{equation}
  I = - \phi_b \int_\partial \sqrt{h} K.
  \label{I_K}
\end{equation}
Since the massive particle shifts the boundaries of the $NAdS$ space, this term creates a non-zero correction to the two-point function.
To find this correction, we subtract from (\ref{I_K}) the action for $NAdS$ without the massive particle:
\begin{equation}
  I-I_0 = \phi_b \left( \int \sqrt{h} K_0 - \int \sqrt{h} K \right).
  \label{I0}
\end{equation}
Here the second term denotes the curvature of the empty $NAdS$ space with the boundary of length $L_{AdS}$, and the first term is the space  with the same boundary length, but with a massive particle inside.
$\phi_b$ in (\ref{I0}) is a large number, so this difference is potentially large.
The corrected two-point function becomes:
\begin{equation}
  G =\frac{1}{\gamma^{2m}} e^{-\left( I-I_0 \right)}.
  \label{G_K}
\end{equation}
In this Section, we discuss this correction and find it small for small distance, and finite constant when the distance gets large.

In the hyperbolic half-plane, the extrinsic curvature is:
\begin{equation}
  K = \frac{t' \left( t'^2+z'^2+zz'' \right)-t'' zz'}{\left( t'^2+z'^2 \right)^{\frac{3}{2}}}.
  \label{K_tz}
\end{equation}
For a circle of radius $R$ and with a center at vertical coordinate $z=y$ the extrinsic curvature is:
\begin{equation}
  K = -\frac{y}{R}.
  \label{K_yR}
\end{equation}
In Section~\ref{sec:conf}, we have found the boundary length of the empty $NAdS$ (in $AdS$ units) as:
\begin{equation}
  L_{AdS} = \frac{2\pi R}{\sqrt{y^2-R^2}}.
  \label{L_empty}
\end{equation}
Therefore the extrinsic curvature action of the empty $NAdS$ space is:
\begin{equation}
  I_0 = \phi_b \frac{y}{R} \cdot L_{AdS} = \phi_b \sqrt{ {L_{AdS}^2}+\left( 2\pi \right)^2}. 
  \label{I0_ans}
\end{equation}

The extrinsic curvature action for the space with a massive particle is a sum of two parts.
One comes from the finite segments of the boundary, and the other from the cusps where the particle meets the boundary:
\begin{equation}
  I = I_{\text{seg}} + 2 I_{\text{cusp}}.
  \label{I_sum}
\end{equation}
The first part depends on the boundary length and have roughly the same structure as (\ref{I0_ans}).
More precisely, it is:
\begin{equation}
  I_{\text{seg}} = \phi_b \frac{y}{\epsilon}\left( \frac{u_{12}}{R_1} + \frac{u_{21}}{R_2}  \right).
  \label{I_seg_u}
\end{equation}
Here we cannot use the small $\epsilon$ approximation (since it requires $y \to R$), and the distances are as in (\ref{u_wy}, \ref{u21_wy}).
They are chosen so as to match the distance in (\ref{I0_ans}):
\begin{equation}
  u_{12} + u_{21} = \epsilon \cdot L_{AdS}.
  \label{u+u}
\end{equation}

The second term in (\ref{I_sum}) is largely universal and independent of distances.
To find it, we replace a cusp with a segment of a small circle of radius $r$. 
Since the circle is small, we can think that the metric is constant, $g \sim \frac{1}{y^2}$.
If the cusp angle is $\theta$, the extrinsic curvature action is:
\begin{equation}
  I_{\text{cusp}} = \phi_b \lim_{r \to 0} \left( \frac{y}{r} \frac{r \theta}{y} \right) = \phi_b \cdot \theta.
  \label{I_theta}
\end{equation}
The angle is found as:
\begin{equation}
  \theta = \arccos \frac{w_1 y}{R_1} -  \arccos \frac{w_2 y}{R_2} = \arcsin \left( \frac{m}{\phi_b} \frac{y \sqrt{y^2-1}}{R_1 R_2} \right). 
  \label{theta_ans}
\end{equation}
If we take $\phi_b$ to be large, the full cusp action becomes:
\begin{equation}
  I_{\text{cusp}} = m \frac{y \sqrt{y^2 - 1}}{R_1 R_2}.
  \label{cusp_ans}
\end{equation}
Away from the UV cutoff at $y\sim1$, this action changes little with distance. 
When distance is large, $y \sim R$ and the cusp action becomes:
\begin{equation}
  I_{\text{cusp}} \sim m.
  \label{I_m}
\end{equation}

The precise answer for the correction is hard to find analytically.
In the symmetric case, $u_{12} = u_{21}$, the numerical method gives us the answer very close to:
\begin{equation}
  I - I_0 \sim \frac{2m}{1+ \frac{3 \pi^2 \phi_b}{mL}}.
  \label{I_sym}
\end{equation}
When $L$ is small, this action is negligible.
When $L$ is large, it comes mostly from the contribution of the cusps, $I \sim 2m$.
In particular, it changes the value of the two-point at large distances (\ref{G_a_pi}) to:
\begin{equation}
  G \to \left( \frac{4 m}{\phi_r \cdot e} \right)^{2m}.
  \label{G_pi_corr}
\end{equation}

\begin{figure}
  \centering
  \includegraphics[width=.6\textwidth]{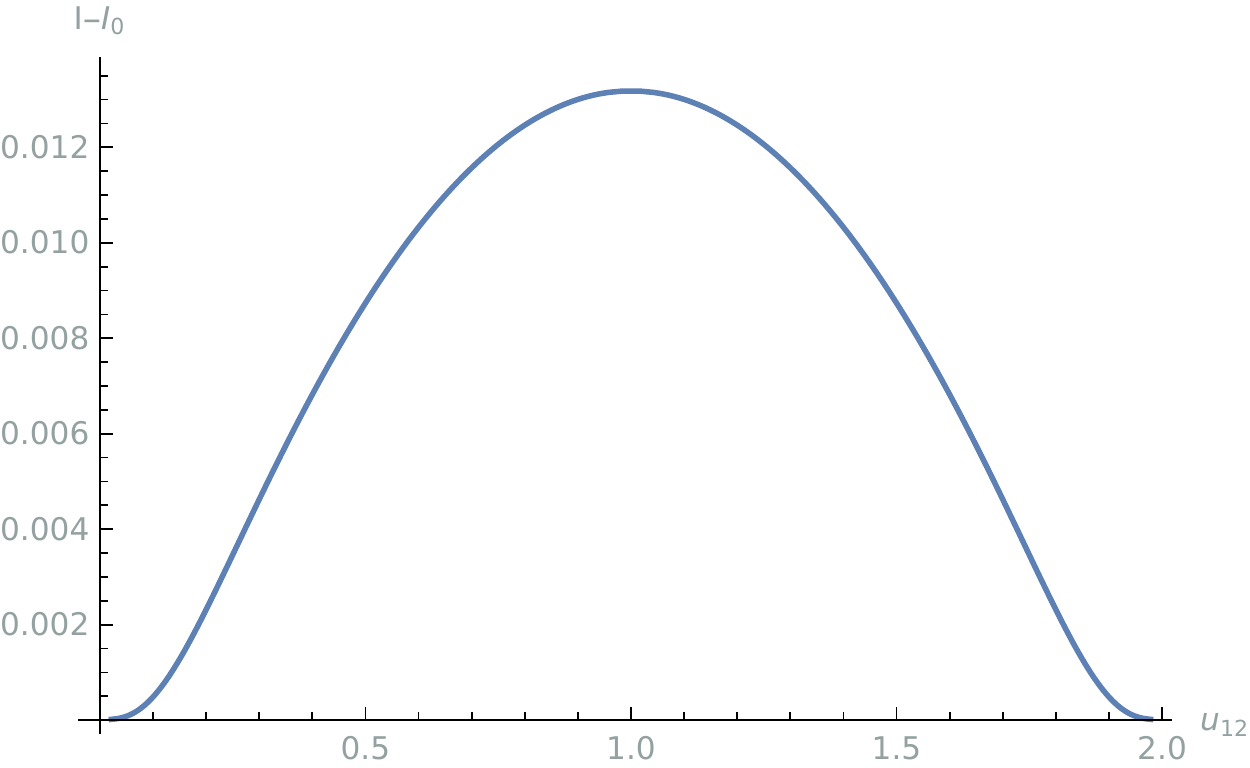}
  \caption{$\left( I-I_0 \right)$ as a function of $u_{12}$. The parameters are: $\phi_r=8, m=1, \epsilon=.1$.}
  \label{fig:SK}
\end{figure}

We can also fix $L$ and find the action as a function of $u_{12}$. 
The numerical result for some fixed values of parameters is plotted on fig.~\ref{fig:SK}.
When the $u_{12}$ distance is small and  $y \to 1$, the angle $\theta$ goes to zero, and the action from the cusps and the $u_{12}$ segment also vanishes.
The contribution of the $u_{21}$ segment exactly coincides with the action of the empty space, and the whole difference in action $\left( I-I_0 \right)$ becomes zero, as can be seen on fig.~\ref{fig:SK}.
Overall, the action the extrinsic curvature is numerically small for finite distances.
However, we see later that this holds only for the Euclidean time, and the real-time correlators receive a correction from the extrinsic curvature action at long time. This corrections makes the correlators exponentially small. 

\subsection{Schwarzian limit}
\label{sec:small_m}

A limit we can use for a reality check is taking $\phi_r/m$ to be large.
Zero mass corresponds to the conformal two-point function, and expansion in $m/\phi_r$ gives a correction to the conformal answer.
The same correction can be found from Schwarzian theory (see \cite{Sarosi:2017ykf}), corresponding to a limit of large $\Delta$.

In the conformal case, the cusp at the insertion of the boundary operator is absent, and the $\alpha_1, \alpha_2$ angles are equal.
We relax this condition and take the angles to be:
\begin{equation}
 \begin{aligned}
  \alpha_1 &= \alpha - \delta, \\
  \alpha_2 &= \alpha + \delta.
  \end{aligned}
  \label{alpha_eps_def}
\end{equation}

As $\delta$ goes to zero, we want to come back to the conformal case discussed in Section \ref{sec:conf}.
The same happens as mass goes to zero.
Therefore we take:
\begin{equation}
  \delta = c \cdot \frac{m}{\phi_r}. 
  \label{eps_l0}
\end{equation}
with $c$ being some constant to be fixed later.

Our goal is to find the two-point function in this limit, up to the second order in $\delta$.
Since $\delta$ is proportional to $m$, it means that we find $\gamma$ in terms of $u_{12}, u_{21}$ to the first order in $\delta$:
\begin{equation}
  \gamma = \frac{2 \phi_r}{m}\left( \cos \alpha_2 - \cos \alpha_1 \right) = \frac{4 \phi_b}{m} \sin \delta \sin \alpha.
  \label{y_eps}
\end{equation}
Expanding $u_{12}$ in $\delta$, we find:
\begin{equation}
  u_{12} = \frac{\phi_r}{m}\cdot 4 \delta  \left( \alpha - \delta \cdot \eta \left( \alpha \right) \right)+O\left( \delta^2 \right),
  \label{u_eps}
\end{equation}
where we have defined:
\begin{equation}
  \eta\left( \alpha \right) \equiv 1-\alpha \cot \alpha.
  \label{eta_def}
\end{equation}

The full length of the boundary then is:
\begin{equation}
  L = u_{12} + u_{21} = \frac{\phi_r}{m} \cdot 4 \pi  \delta \left( 1 - \frac{\delta}{\pi} \left( 1 + \left( \pi-2\alpha \right) \eta \left( \alpha \right) \right) \right)+O\left( \delta^2 \right) .
  \label{L_eps}
\end{equation}
Since we expand to the first order in $\delta$, this equation allows us to fix the constant in (\ref{eps_l0}):
\begin{equation}
  \delta \sim  \frac{mL}{4\pi \phi_r} + O\left( \delta^2 \right).
  \label{eps_l}
\end{equation}

The fraction of the boundary belonging to the first segment is:
\begin{equation}
  \frac{\pi u_{12}}{L} = \alpha + \delta \left( 1 + \left( \frac{2\alpha}{\pi} - 1 \right)\eta \left( \alpha \right) \right)+O\left( \delta^2 \right).
  \label{phase_eps}
\end{equation}
From here, we see that in this near-conformal case, $\alpha$ is roughly the arc angle for $u_{12}$, with a correction of order $\epsilon$. 

Putting everything together, from (\ref{L_eps}) and (\ref{phase_eps}) we find: 
\begin{equation}
  \gamma = \frac{L}{\pi} \sin \frac{\pi u_{12}}{L} \left( 1-\frac{2\delta}{\pi} \eta \left( \alpha \right) \eta\left( \pi-\alpha \right) \right)+O\left( \delta^2 \right).
  \label{y_eps_ans}
\end{equation}
Raising this to a power, we find a correction for the two-point function:
\begin{equation}
  G = \frac{1}{\left( \frac{L}{\pi} \sin \frac{\pi u_{12}}{L} \right)^{2m}} \left(1+\frac{m^2}{ \phi_r} \frac{L}{\pi^2}  \eta \left( \frac{\pi u_{12}}{L} \right) \eta\left( \frac{\pi u_{21}}{L} \right)  \right).
  \label{G_corr}
\end{equation}

This is the correction that was found in \cite{Sarosi:2017ykf} from the Schwarzian propagator.
Since $\phi_r \sim N$, it corresponds to a $1/N$ correction in the SYK model.
\section{Two-point function in real time}
\label{sec:2pt_cont}

Our analysis allows us to extend our discussion to real time and consider a thermal correlator:
\begin{equation}
  G_\beta\left(t \right) = \left \langle \mathcal O\left( \frac{\beta}{2}+it \right) \mathcal O\left( \frac{\beta}{2}-it \right) \right \rangle.
  \label{G_cont_def}
\end{equation}
In quantum mechanics, this correlator looks like a sum over energy eigenstates:
\begin{equation}
  G_\beta \left( t \right) = \frac{1}{Z\left( \beta \right)} \sum_{m,n} e^{-\beta \left( E_m+E_n \right)/2} e^{it \left( E_m-E_n \right)} \left |\left \langle m | \mathcal O| n \right \rangle \right|^2.
  \label{G_E}
\end{equation}
In a chaotic system, it is believed that for very large $t$, the off-diagonal terms have large and essentially random phases.  In this case, after some averaging
over $t$, the off-diagonal terms do not contribute.   If so, one expects that at least in an averaged sense, the large real time behavior can be approximated by:
\begin{equation}
  \left. G_\beta \right|_{t \to \infty} \sim \frac{1}{Z\left( \beta \right)} \sum_{n} e^{-\beta E_n } \left |\left \langle n | \mathcal O| n \right \rangle \right|^2.
  \label{G_t_large}
\end{equation}

If  the non-diagonal elements can be neglected in some sense, and the diagonal elements are of order one, this correlator is a close cousin to the spectral form factor:
\begin{equation}
  \text{SFF}(t) = \frac{\left|Z\left( \frac{\beta}{2}+it \right)\right|^2}{\left|Z\left( \frac{\beta}{2} \right)\right|^2}, 
  \label{SFF_def}
\end{equation}
which has been discussed at length in the context of SYK in \cite{Cotler:2016fpe} and other studies.
At large time, and if the spectrum of the system has no degeneracies, the averaged spectral form factor becomes:
\begin{equation}
  \lim_{T \to \infty} \frac{1}{T} \int_0^T  \text{SFF}(t) dt = \frac{Z(2\beta)}{Z\left( \beta \right)^2}.
  \label{sff_lim}
\end{equation}

The partition function generally scales as $Z \sim \exp\left( -c S \right)$, therefore the long-time value of the spectral form factor (and of the real-time two-point function) is exponential in $S$ \cite{Dyson:2002pf}.
On the gravitational side, the entropy is $S \sim 1/G_N$, and in SYK, the entropy is $S \sim N$.
In our setup, it means that $S \sim \phi_r$, and the two-point function should have an exponentially small limit at long times.

The existence of this limit for the two-point function is a non-perturbative effect, both in the SYK model and in gravity.
We find that our two-point function also approaches an exponentially small number, when the extrinsic curvature term is taken into account.
However, in our case the two-point function is averaged, that is insensitive to small oscillations which have been observed for SYK in \cite{Cotler:2016fpe} and are anticipated on general grounds.
Also, we find that in our case the long time limit is approached much faster than expected for a spectral form factor in SYK.

We first find the two-point function as the exponentiated geodesic length and then take into account extrinsic curvature.
We once again use the small $\epsilon$ approximation and translate the distances between operators to (now complex) angles:
\begin{equation}
  \begin{aligned}
  \frac{\beta}{2}+it =& \frac{2 \phi_r}{m} \frac{\alpha}{\sin \alpha} \left( \cos \alpha + \cos \bar{\alpha} \right), \\
  \frac{\beta}{2}-it =& \frac{2 \phi_r}{m} \frac{\bar{\alpha}}{\sin \bar{\alpha}} \left( \cos \alpha + \cos \bar{\alpha} \right).
  \end{aligned}
  \label{beta_alpha}
\end{equation}
Here we have defined:
\begin{equation}
  \alpha = \alpha_1, \qquad \bar{\alpha} = \pi-\alpha_2.
  \label{alpha_bar_def}
\end{equation}

The two-point function depends, as before, on the $\gamma$ parameter:
\begin{equation}
  G\left( \beta, t \right) = \frac{1}{\gamma^{2\Delta}}, \qquad \gamma = \frac{2 \phi_r}{m} \left( \cos \alpha + \cos \bar{\alpha} \right).
  \label{G_y_cpx}
\end{equation}

To simplify the discussion, we take mass to be relatively small, in particular:
\begin{equation}
  \beta m/\phi_r \ll 1. 
  \label{beta_small}
\end{equation}
In this limit, we can find the two-point function at $t=0$:
\begin{equation}
  \alpha_0 = \frac{\pi}{2}+\frac{\beta m}{4\pi \phi_r} \qquad \Rightarrow \qquad \gamma_0 = \frac{\beta}{\pi}. 
  \label{alpha_0}
\end{equation}
We normalize the thermal two-point function (\ref{G_cont_def}) by the two-point function at $t=0$, imitating the spectral form factor (\ref{SFF_def}):
\begin{equation}
  \frac{G_\beta\left( t \right)}{G_\beta\left( 0 \right)} = \left( \frac{\gamma_0}{\gamma} \right)^{2\Delta}.
  \label{G_norm}
\end{equation}

\begin{figure}
  \centering
  \includegraphics[width = .7 \textwidth]{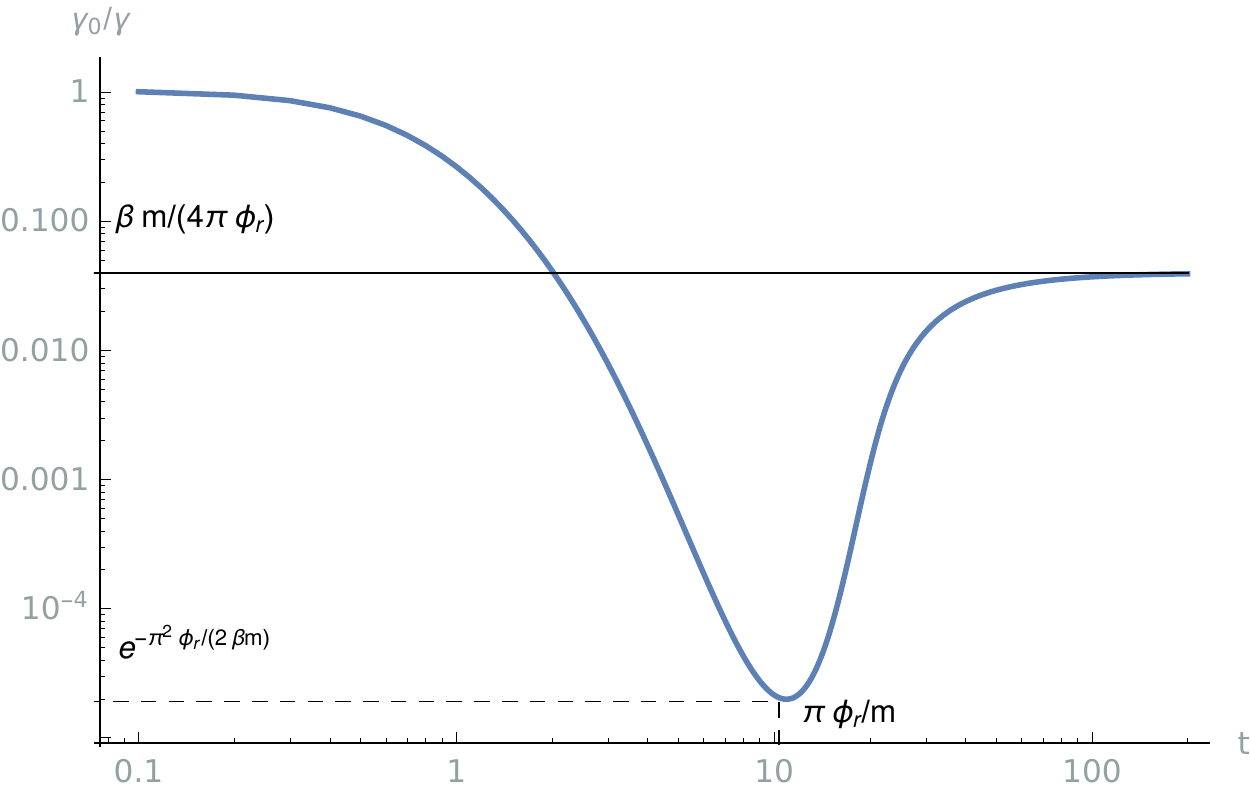}
  \caption{Two-point function in real time without the extrinsic curvature correction. }
  \label{fig:sff}
\end{figure}

On fig.~\ref{fig:sff}, we plot the $\gamma_0/\gamma$ ratio as a function of time.
We cannot solve (\ref{beta_alpha}) analytically, but can see how the correlator behaves in various limits.
Let the angle $\alpha$ be complex:
\begin{equation}
  \alpha = \xi + ix.
  \label{alpha_x}
\end{equation}
Then the two-point function is:
\begin{equation}
  G = \gamma^{-2\Delta} = \left( -\frac{4 \phi_r}{m} \cos \xi \cosh x\right)^{-2\Delta}.
  \label{G_x}
\end{equation}
With time, the real part $\xi$ grows from $\alpha_0$ to $\pi$.
When $\xi$ is small, (\ref{beta_alpha}) gives:
\begin{equation}
  t = \frac{\beta}{\pi} x, \qquad (\text{small }t),
  \label{t=0}
\end{equation}
and $y$ goes as:
\begin{equation}
  \gamma_0/\gamma \sim 1/\cosh \left( \frac{\pi t}{\beta} \right).
  \label{y_t=0}
\end{equation}
The two-point function then decays exponentially, as can be expected from a thermal correlator.
This can also be seen from the Schwarzian theory.
However, this decay eventually stops.
The imaginary part $x$ grows faster than $\xi$, and the minimum on fig.~\ref{fig:sff} occurs when $x$ is large and $\xi$ still close to $\pi/2$.
Plugging this information as an approximation into (\ref{beta_alpha}), we find the minimum at:
\begin{equation}
  \xi_{\text{min}} =\frac{\pi}{2}+ \frac{\beta m}{2\pi }, \qquad t_{\text{min}} = \frac{\pi \phi_r}{m}.
  \label{t_min}
\end{equation}
The minimum of the correlator, consistently with the initial exponential decay, occurs at:
\begin{equation}
  \gamma_0/\gamma_{\text{max}} \sim \exp\left( -\frac{\pi^2 \phi_r}{2\beta m} \right).
  \label{yy_max}
\end{equation}
The minimal value of the normalized two-point function then is:
\begin{equation}
  \frac{G_{\text{min}}}{G_0} \sim  \exp\left( -\frac{\pi^2 \phi_r}{\beta} \right).
  \label{G_min}
\end{equation}
After the minimum, the imaginary part $x$ decreases, and at the same time $\xi$ covers most of the distance to $\pi$.
The two-point function grows exponentially with roughly the same speed as it decreased.
Finally, the correlator approaches a plateau, where both $x$ and $\xi$ are small.

The similarity of the real-time correlator to the form-factor of the SYK \cite{Saad:2018bqo} is striking, but likely accidental.
The finite value at which the two-point function saturates is also exponential in $N$ in the SYK, and is $\sim 1/N$ in our problem.
It is also worth noting that the large real time limit is the same as the  large Euclidean time limit (\ref{G_a_pi}).
From (\ref{G_t_large}), we expect the large $t$ limit of the thermal two-point function to be the square of the thermal one-point function of the operator.
This is a consequence of the eigenstate thermalization hypothesis \cite{deutsch1991quantum}, \cite{srednicki1994chaos}.
Without action of the extrinsic curvature, the thermal one-point function appears to be the same as the one-point function in Euclidean time.

However, when we take into account extrinsic curvature, the picture changes drastically.
In our discussion so far, we worked in the small $\epsilon$ approximation when $y=R$, and the extrinsic curvature is equal to one.
The full answer for $u_{12}$, (\ref{u_wy}), is hard to continue to the complex plane, but for us it is sufficient to find the extrinsic curvature to the first order in $1/y^2$. 
When $\epsilon$ is small, $y$ is a large parameter, so all other corrections will be subleading in $\epsilon$.

\begin{figure}
  \centering
  \includegraphics[width=.6 \textwidth]{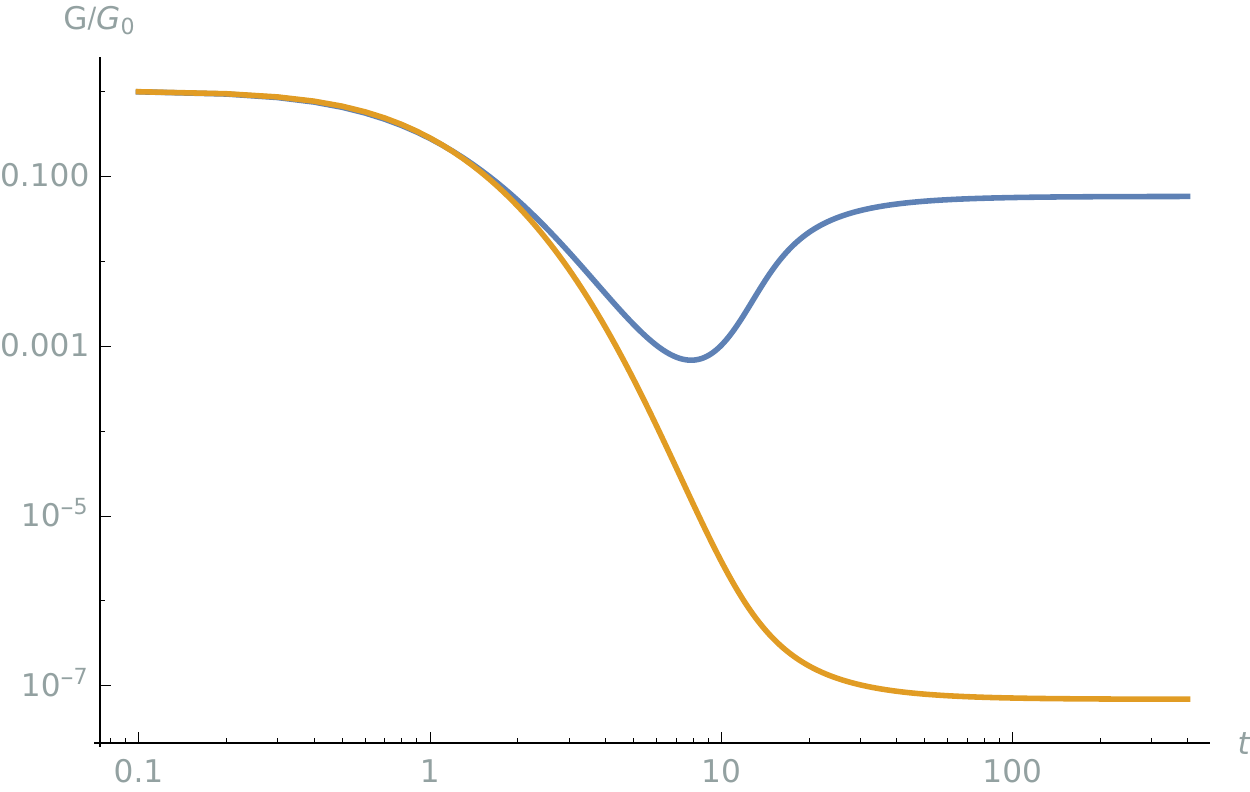}
  \caption{The two-point function including the extrinsic curvature term (yellow) compared to the two-point function as exponentiated geodesic (blue). The long-time value of the two-point function is $\sim \exp(-N)$.}
  \label{fig:real_2pt_K}
\end{figure}

The extrinsic curvature is:
\begin{equation}
  K = -\frac{y}{R} = -\frac{y}{\sqrt{y^2 - \sin^2\alpha}} = -1-\frac{1}{2 y^2} \sin^2 \alpha + O\left( y^{-4} \right)
  \label{K_expand}
\end{equation}
In particular, we see that at large time $y=R$ with good accuracy, so the ``segment'' part of action of the extrinsic curvature is:
\begin{equation}
  \left.I_{\text{seg}}\right|_{t \to \infty } = - \phi_b L/\epsilon = - \phi_r \beta. 
  \label{I_t_inf}
\end{equation}
This is a large number. 
When we normalize to the action of the empty $AdS$ space, it becomes:
\begin{equation}
  \left. \left( I-I_0 \right)\right|_{t \to\infty } =\phi_b\left(\sqrt{\left( \frac{\beta}{\epsilon} \right)^2 + \left( 2\pi \right)^2} - \frac{\beta}{\epsilon}  \right) = \frac{2\pi^2 \phi_r}{\beta} + O\left( \epsilon^2 \right). 
  \label{dI_t_inf}
\end{equation}
This creates a large correction to the two-point function at long times:
\begin{equation}
  \left. \frac{G}{G_0} \right|_{t \to \infty} \sim  \exp\left( -\frac{2\pi^2 \phi_r}{\beta} \right).
  \label{G_t_inf}
\end{equation}
This correction is similar to the minimal value of the two-point function (\ref{G_min}).
Therefore, when the extrinsic curvature is taken into account, the two-point function decays to a value $\sim \exp( - \phi_r)$, and therefore exponential in $N$.
This can be expected in quantum mechanics on general grounds, as an average of oscillations with a large phase \cite{Saad:2018bqo}. 
We plot the numerical result (the first correction to the small $\epsilon$ approximation) on fig.~\ref{fig:real_2pt_K}.
We see that the ``ramp'' is gone, and the two-point function decays monotonically.
Roughly at $t \sim \phi_r/m \sim N$ this decay slows down, and the final value is $\sim \exp(-N)$.

It would seem that our calculation reproduces the plateau in the SYK two-point function on the $NAdS$ side.
However, the time at which this plateau is reached seems much shorter in our case.
In \cite{Cotler:2016fpe}, the ``plateau time'' has been found to be exponential in $N$ from random matrix considerations.
In our case the plateau starts at roughly the time when the two-point function (\ref{G_norm}) reaches its minimum:
\begin{equation}
  t_{\text{min}}\sim \phi_r\sim N.
  \label{t_N}
\end{equation}
This time is linear in $N$. 
The reason for this behavior is not clear, but it points out that the plateau we find in JT gravity may be governed by different physics than the plateau in SYK.

\section{Euclidean four-point function}
\label{sec:4pt}

\begin{figure}
  \centering
  \includegraphics[width=1.0\textwidth]{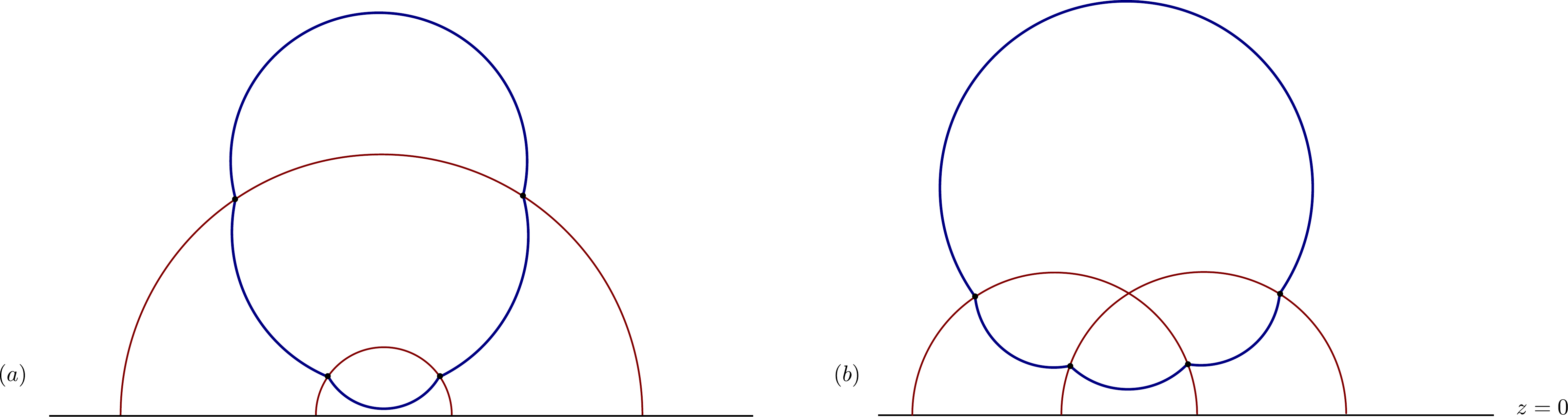}
  \caption{Four-point function for $(a)$ time-ordered and $(b)$ out of time ordered operators.}
  \label{fig:4pt}
\end{figure}

Following the same logic, we can find the four-point function in the semi-classical approximation. 
This time, we consider trajectories of two massive particles, both intersecting the boundary of the $NAdS$ space.
In a stable theory, when particles have positive masses, they create inward cusps (see fig.~\ref{fig:4pt}).

The four-point function is defined by the lengths of the trajectories, lying inside the $NAdS$ space.
As before, we are going to express this four-point function in terms of the lengths of the boundary segments.
For simplicity, we take the segments to be pairwise equal.
This means that the picture on fig.~\ref{fig:4pt} is both left-right symmetric, and invariant under inversion.

In this picture, there are two significantly different cases.
If the trajectories of the particles do not intersect, it corresponds to the time-ordered four-point function.
However if the trajectories do intersect, we get an out of time ordered correlator, and expect to see exponential growth after analytical continuation.
The presence or absence of an intersection is conformally invariant, and can be defined from the cross product of the charge vectors for the massive particles.
Let one trajectory be defined by the three-vector $A$ and the other by $B$.
We will be working in the convention where:

\begin{equation}
  \begin{aligned}
  \left( A \times B \right)^2 >0  \Rightarrow & \text{intersection}, \\
  \left( A \times B \right)^2 <0  \Rightarrow & \text{no intersection}.
  \end{aligned}
  \label{intersect_inv}
\end{equation}
There is a boundary case when both trajectories are straight vertical lines and the cross product is exactly zero, but we will not be considering it.

In this Section, we work in the small $\epsilon$ approximation.
We do not discuss the precise answer with a UV cutoff, as we did for the two-point function.
We expect the approximation to work in a similar way for a four-point function, removing a cutoff with an approximately conformal region in the ultraviolet.
This approximately conformal region allows us to find a small mass correction to the four-point function in Section~\ref{sec:4pt_sch}.

We impose a significant amount of symmetry, making the pictures on fig.~\ref{fig:4pt} both left-right and inversion symmetric.
This allows us to parameterize the boundary distance in a relatively simple way, using the angular variables.
In particular, the answer for the time-ordered four-point function is strikingly similar to the answer for the two-point function, compare (\ref{ans_a_angles}, \ref{gamma_a_angles}) to (\ref{u_12_21}, \ref{y_alpha}).
For the out of time ordered four-point function, we can also use a similar parameterization, with the result being (\ref{u_b_angles}, \ref{gamma_b_psi}), however in this case there is an extra condition (\ref{b_e1}).

Our parameterization helps us to analytically continue the four-point function to real time.
However, here the imposed symmetry appears to be restrictive and does not allow to find the time-ordered correlator.
For the out of time ordered four-point function, we find that it first decays exponentially and then stabilizes at a small ($\sim \exp \left( -N \right)$) value.
We find this after taking into account the action of extrinsic curvature, which is small in Euclidean signature but is significant for real-time correlators.

\subsection{Four-point function: time-ordered}
First, we start with the picture of the type $(a)$ on fig.~\ref{fig:4pt} with non-intersecting trajectories.
We want to make it left-right symmetric and inversion invariant.
These two symmetries act on the embedding coordinates $Y$ as reflections:

\begin{equation}
  \begin{aligned}
  \text{Left-right}: & Y_1 \to -Y_1, \\
  \text{Inversion}: &   Y_2 \to -Y_2.
\end{aligned}
  \label{sym_Y}
\end{equation}
Let us denote $Z$ the charge vector of the dilaton inside the ``smaller'' circle, $A$ the vector for the trajectory on the ``smaller'' circle, and $B$ the vector for the trajectory of the ``larger'' circle.
To make the picture invariant under the left-right reflection, we make the first component of each of these vectors vanish:
\begin{equation}
  Z_1 = A_1 = B_1 = 0.
  \label{Z_sym_a}
\end{equation}
To make them inversion-invariant, we take:
\begin{equation}
  A_2 = B_2 = -Z_2.
  \label{Z_inv_a}
\end{equation}

\begin{figure}
  \centering
  \includegraphics[width=.6\textwidth]{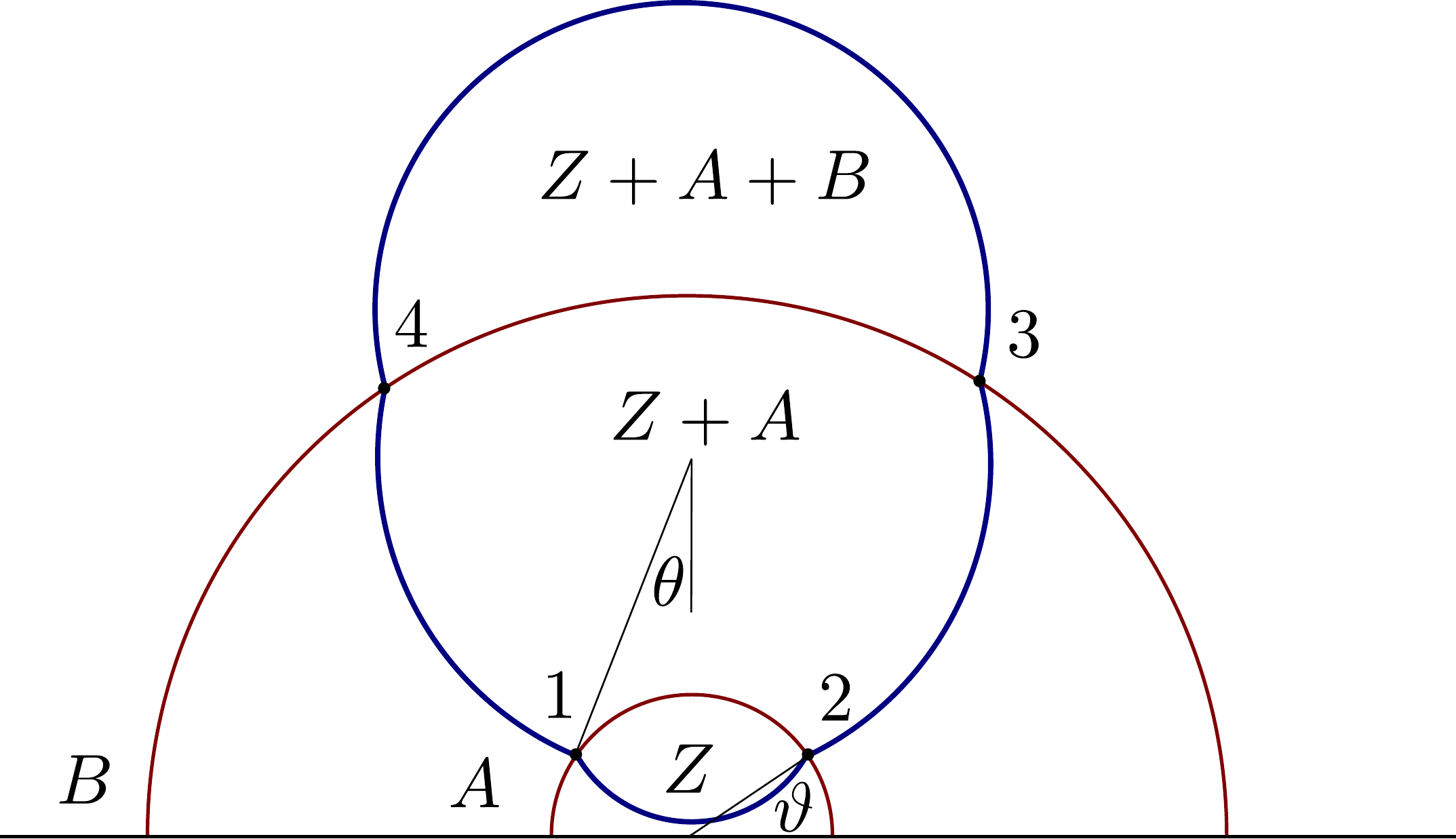}
  \caption{Bulk diagram for the time ordered four-point function.}
  \label{fig:4pta}
\end{figure}

The individual components of the vectors prove not to be the convenient parameters for our calculations.
Instead, we introduce two angle parameters, as we did previously for the two-point function.
Using the condition on masses,
\begin{equation}
  A^2 = B^2 = -m^2,
  \label{A2m}
\end{equation}
together with (\ref{Z_sym_a}), we parameterize $A$ and $B$ vectors as:
\begin{equation}
  \begin{aligned}
  A &= \left( -m \sinh \alpha, 0, m \cosh \alpha \right), \\
  B &= \left( m \sinh \alpha, 0, m \cosh \alpha \right), \qquad \alpha>0.
  \end{aligned}
  \label{AB_alpha}
\end{equation}
In these variables, the radius of the ``larger'' circle is:
\begin{equation}
  r \equiv e^{\alpha},
  \label{r_def}
\end{equation}
and the radius of the ``smaller'' circle is $1/r$, since they are exchanged by inversion.

By the same logic, we parameterize the dilaton vector $Z$ as:
\begin{equation}
  Z = \left( Z \cosh \zeta, 0, -Z \sinh \zeta \right).
  \label{Z_zeta}
\end{equation}
From inversion invariance (\ref{Z_inv_a}), we find the $Z$ constant:
\begin{equation}
  Z = \frac{m \cosh \alpha}{\sinh \zeta}, \qquad \zeta>0.
  \label{Z_def}
\end{equation}
Thus we are left with two angles $\alpha, \zeta$, which we are going to determine from two boundary lengths, which we call $u_{12}$, $u_{23}$.

The boundary length of the lower segment is given by an integral:
\begin{equation}
  u_{12} =2 \epsilon \cdot \int_0^{\theta_*} \frac{R d\theta}{y - R \sin \theta}
  \label{u12_int}
\end{equation}
We have rescaled from the $AdS$ units to the quantum mechanicals ones, so that the distance stays finite as $\epsilon$ becomes small.
Here $R$ is the radius of the boundary segment and $y$ is the vertical position of its center.
Both are taken to be large, and are connected by:
\begin{equation}
  y^2-R^2 = e^{-2\zeta}.
  \label{y2R2}
\end{equation}
Hence $\zeta$ is the measure of how ``close'' the $NAdS$ is to the real boundary (coordinate-wise, since the real distance to the boundary is infinite).
$y$ is given by:
\begin{equation}
  y =\frac{\phi_b}{m}  \cdot \frac{1 - e^{-2\zeta}}{2 \cosh \alpha} \sim \phi_b/m,
  \label{y_a_I}
\end{equation}
We take $\phi_b/m$ to be large, and find the answer in the leading order in $1/y$.
This is the condition of the small $\epsilon$ approximation.

In this limit, the $\theta_*$ angle in (\ref{u12_int}) is small and is found to be:
\begin{equation}
  \theta_* = \frac{r^{-1}}{y}.
  \label{theta_a_I}
\end{equation}

Expressing everything in terms of the $(\alpha, \zeta)$ angles, we take the integral (\ref{u12_int}) and find the boundary distance:
\begin{equation}
  u_{12} =\frac{2 \phi_r}{m} \cdot\frac{\sinh \zeta}{\cosh \alpha} \arctan\left(\frac{1}{\sinh\left( \alpha-\zeta \right)} \right),
  \label{u12_a_ans}
\end{equation}
A quick check shows that $u_{34}=u_{12}$.

In the same way, we can find the $u_{23}$ distance.
We use the same formula (\ref{u12_int}) for the integral, except with different parameters of the circle.
The angles between which we integrate are found from (\ref{theta_a_I}) and inversion invariance,
\begin{equation}
  \theta_1 = \frac{r^{-1}}{y}, \qquad \theta_2 = \frac{r}{y}.
  \label{theta_a_II}
\end{equation}
Here $y$ is different from before and is equal to:
\begin{equation}
  y = \frac{\phi_b}{m} \cdot \frac{\sinh \zeta}{ \cosh \left( \alpha-\zeta \right)}.
  \label{y_a_II}
\end{equation}
Bringing everything together, we get for the second distance:
\begin{equation}
  u_{23} =\frac{2 \phi_r}{m} \cdot \frac{\sinh \zeta}{\cosh \left( \alpha-\zeta \right)} \arctan\left( \sinh \alpha \right).
  \label{u23_a_ans}
\end{equation}

The four-point function in the semiclassical approximation is determined by the geodesic distances between the operators.
We rescale the four-point function, so that it is consistent with our definition of the two-point function (\ref{G_rescaled}):
\begin{equation}
  W = \epsilon^{-4\Delta} \exp\left(- 2 \ell \cdot 2\Delta \right) = \frac{1}{\gamma^{4m}}, \qquad \gamma \equiv \epsilon \cdot e^\ell.
  \label{W_ell_a}
\end{equation}
The geodesic length of the trajectory of one particle is given by an integral:
\begin{equation}
  \ell = \int_{\vartheta_*}^{\frac{\pi}{2}} \frac{r d\vartheta}{r \sin \vartheta} = -\ln \tan \frac{\vartheta}{2},
  \label{geo_a}
\end{equation}
with $\vartheta$ defined on fig.~\ref{fig:4pta}.
$\vartheta_*$ is a small angle, determined by the geometry of fig.~\ref{fig:4pta} to be:
\begin{equation}
  \vartheta_* = \frac{1+r^2}{2 y r}.
  \label{varq_a}
\end{equation}
Plugging it in the integral (\ref{geo_a}) and using the definition of the $\left( \alpha,\zeta \right)$ angles, we find the exponentiated geodesic length $\gamma$:
\begin{equation}
  \gamma =  \frac{2 \phi_r}{m} \cdot \frac{\sinh \zeta}{\cosh \alpha \cosh \left( \alpha-\zeta \right)}.
  \label{ell_a_ans}
\end{equation}
Another quick check shows that the lengths of the both segments of geodesics inside $NAdS$ are the same.

\begin{figure}
  \centering
  \includegraphics[width = .6\textwidth]{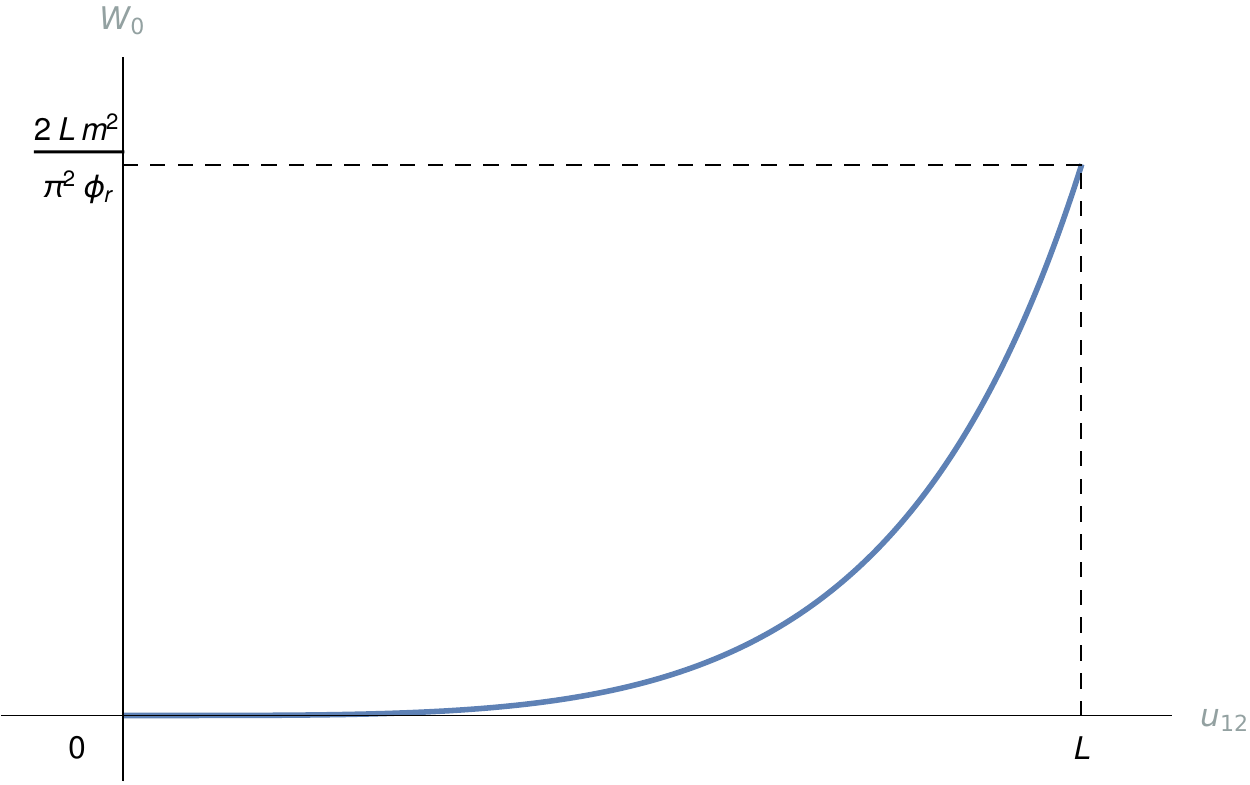}
  \caption{The connected part of the in-order four-point function.}
  \label{fig:W0_a}
\end{figure}

We can find the four-point function using directly (\ref{u12_a_ans}) and (\ref{u23_a_ans}), but find it convenient to change variables once again.
Let us denote:
\begin{equation}
  \begin{aligned}
  \sigma & \equiv  \frac{\pi}{2} - \arctan \left( \sinh \left( \alpha-\zeta \right) \right), \qquad 0 \le \sigma \le \pi,\\
  \psi &\equiv  \frac{\pi}{2} - \arctan \left( \sinh \alpha \right), \qquad 0 \le \psi \le \pi/2.
  \end{aligned}
  \label{angles_def}
\end{equation}
Then using various identities for hyperbolic and trigonometric functions, we find:
\begin{equation}
  \begin{aligned}
  u_{12} &= \frac{2 \phi_r}{m} \cdot \frac{\sigma}{\sin \sigma} \left( \cos \psi - \cos \sigma \right), \\
  u_{23}&= \frac{2 \phi_r}{m} \cdot \frac{\pi/2-\psi}{\sin \psi}\left( \cos \psi - \cos \sigma \right),
\end{aligned}
  \label{ans_a_angles}
\end{equation}
and the exponentiated geodesic length is now:
\begin{equation}
  \gamma =  \frac{2 \phi_r}{m} \cdot \left( \cos \psi - \cos \sigma\right).
  \label{gamma_a_angles}
\end{equation}
These expressions are much easier to analyze.
They are also very similar to the answer for the two-point function (\ref{u_12_21}, \ref{y_alpha}).
As was the case for the two-point function, we cannot find an analytical solution for $\gamma\left( u_{12}, u_{23} \right)$ in a closed form.
Nevertheless, we can find a numerical solution relatively easily.

It is convenient to focus on the connected part of the four-point function:
\begin{equation}
  W_0 \equiv \frac{\left \langle \mathcal O_1\left( u_1 \right) \mathcal O_1\left( u_2 \right) \mathcal O_2\left( u_3 \right) \mathcal O_2\left( u_4 \right) \right \rangle }{\left \langle \mathcal O_1\left( u_1 \right) \mathcal O_1\left( u_2 \right)  \right \rangle \left \langle\mathcal O_2\left( u_3 \right) \mathcal O_2\left( u_4 \right)  \right \rangle } - 1 = \frac{W\left( u_{12}, u_{23} \right)}{G^2\left( u_{12} \right)}-1.
  \label{W0_def}
\end{equation}
In terms of $\gamma$, this is:
\begin{equation}
  W_0 = \left( \frac{\gamma_{\text{2pt}}}{\gamma_{\text{4pt}}} \right)^{4m} -1.
  \label{W0_gamma}
\end{equation}

We plot the numerical solution for $W_0$ on fig.~\ref{fig:W0_a}.
When the distance between operators $u_{12}$ is small, the connected part is close to zero.
It grows monotonically and reaches a maximum when $u_{12} = L/2$.
If the boundary length $L$ is relatively small, $L \ll \phi_r/m$, the maximum value of the four-point function is:
\begin{equation}
  W_0\left( u_{12} = L/2 \right) \sim \frac{2m^2 L}{\pi^2 \phi_r}.
  \label{W0_max}
\end{equation}
We see that in general, the in-order four-point function is relatively close to zero.
In the next Section, we find the out-of-order four-point function numerically and see that it is also closer to zero for lighter particles.

\subsection{Four-point function: out-of-time ordered}
\label{sec:otoc}
Next we turn to the out-of-time ordered four-point function.
We consider the worldlines of the particles intersecting in the $NAdS$ space.
As before, we use the symmetries of the problem to simplify the discussion.
We pick the parameters in such a way that the picture is both left-right and inversion symmetric.

\begin{figure}
  \centering
  \includegraphics[width = .6\textwidth]{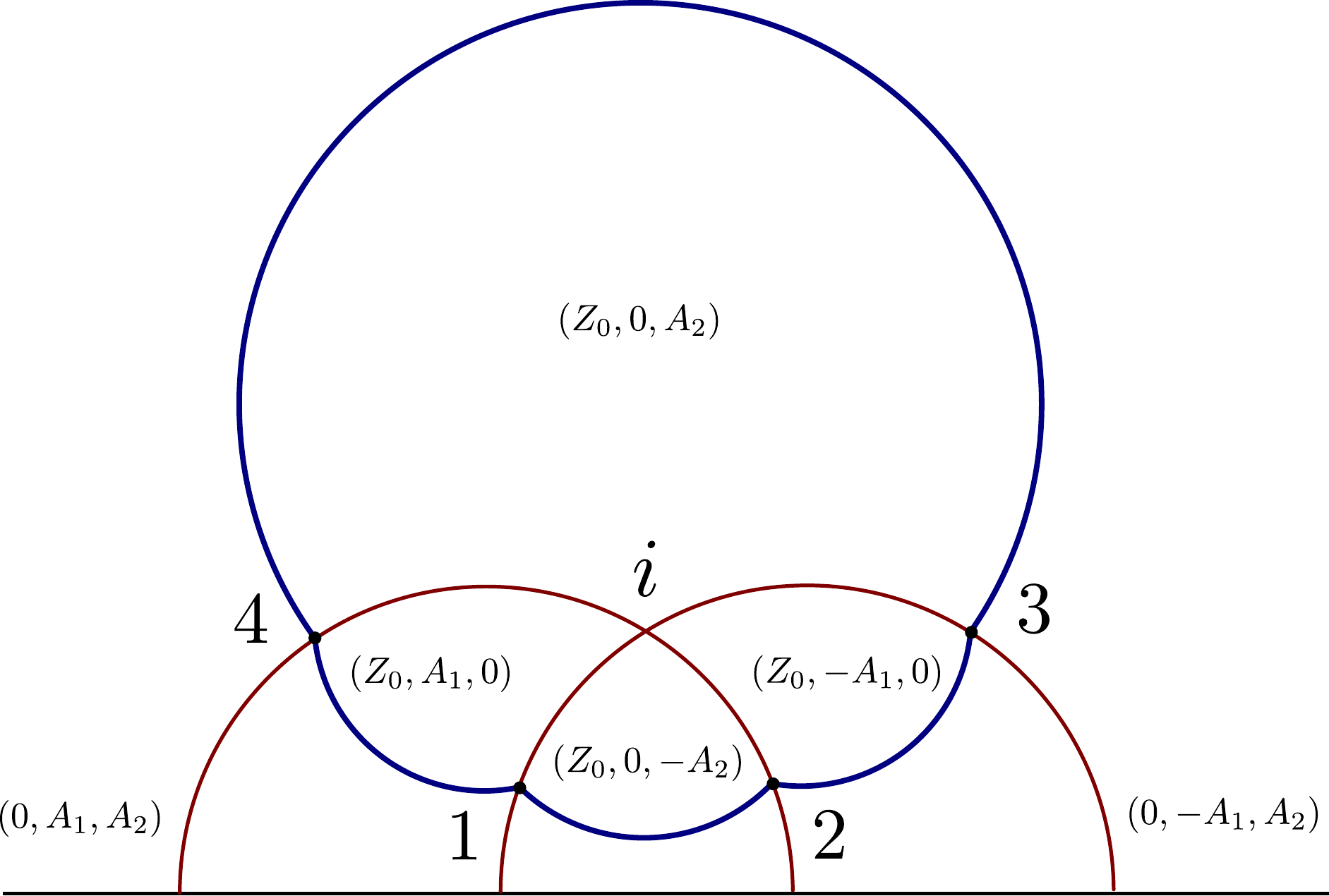}
  \caption{Out-of-time ordered four-point function}
  \label{fig:4pt_b}
\end{figure}

The left-right symmetry requires that the first component of the dilaton on the left-hand side of the picture was the opposite of the one on the right-hand side:
\begin{equation}
  \left.Z_1\right|_{\text{left}}=\left.-Z_1\right|_{\text{right}}.
  \label{4pt_b_lr}
\end{equation}
In particular, it means that the first component of the dilaton in the 12 and 34 segments is zero (see fig.~\ref{fig:4pt_b}).

The invariance under inversion states that the trajectories of the particles intersect at the point $(t,z) = (0,1)$, or $i$ in the complex plane.
It implies that the zeroth component of the particles' charge vector vanishes:
\begin{equation}
  A_0 = 0.
  \label{A0=0}
\end{equation}
In addition, it requires that the second component of the dilaton vector be the opposite on the top and bottom parts of the picture:
\begin{equation}
  \left.Z_2\right|_{\text{left}}=\left.-Z_2\right|_{\text{right}}.
  \label{4pt_b_inv}
\end{equation}
It implies that the second component of the dilaton vanishes in the 23 and 14 segments.
It also implies that the second component of the dilaton in the 12 segment is:
\begin{equation}
  Z_2 = -A_2.
  \label{Z2A2}
\end{equation}
Bringing all this together, we arrive at the setup on fig.~\ref{fig:4pt_b}.

These symmetries guarantee that the lengths of the boundary segments are pairwise equal:
\begin{equation}
  \begin{aligned}
  u_{12} &= u_{34},\\
  u_{23} &= u_{14}.
  \end{aligned}
  \label{u_b_sym}
\end{equation}
Also, from fig.~\ref{fig:4pt_b} we notice that there is yet another symmetry.
Unlike in the time-ordered case, here there is no topological difference between segments 12 and 23, so after an exchange:
\begin{equation}
  \begin{aligned}
  A_1 & \leftrightarrow  A_2, \\
  u_{12} & \leftrightarrow  u_{23},
  \end{aligned}
  \label{A1A2}
\end{equation}
the picture goes back to itself.
This means that when determining how the four-point function depends on the distance, we only have to find $u_{12}$ and the other distance can be recovered from this symmetry.

The square of the vector $A$ is fixed by the mass of the particle, and for convenience we introduce a parameter $\alpha$ such that:
\begin{equation}
  \begin{aligned}
  A_1 &= m \sin \alpha, \\
  A_2 &= m \cos \alpha.
  \end{aligned}
  \label{alpha_b}
\end{equation}
Therefore we have two parameters, $Z_0$ and $\alpha$.
Our goal is to find the two boundary distances, $u_{12}$ and $u_{23}$, and the exponentiated geodesic length $\gamma$ in terms of these parameters.

We focus on the 12 segment.
The boundary distance is, as before, given by an integral:
\begin{equation}
  u_{12} =  2 \epsilon \cdot \int_0^{\theta_{*}} \frac{R d\theta}{y - R \cos \theta}.
  \label{u12_b}
\end{equation}
Here we have rescaled from $AdS$ to quantum mechanical length.
As before, $y$ and $R$ are the vertical coordinate and the radius of the circle describing the $NAdS$ boundary, and they are given by:
\begin{equation}
  \begin{aligned}
  y &= \frac{\phi_b}{Z_0+m \cos \alpha}, \\
  R^2 &= y^2 -\frac{Z_0 - m \cos \alpha}{Z_0+m \cos \alpha}.
  \end{aligned}
  \label{yR_b}
\end{equation}
We treat $y$ as a large parameter.
By the same logic as in Section~\ref{sec:2pt}, this allows us to simplify the expressions for distances and at the same time gives a conformal limit in the ultraviolet.

The radius of the trajectory of a particle is:
\begin{equation}
  r = \frac{m}{A_2} = \frac{1}{\cos \alpha}, 
  \label{r_b_def}
\end{equation}
and the center of the right half-circle has the horizontal coordinate of:
\begin{equation}
  v = \frac{A_1}{A_2}  = \tan \alpha.
  \label{v_def}
\end{equation}
The angle $\theta_{*}$ in \ref{u12_b} is found from the intersection of the two circles as:
\begin{equation}
  \theta_{*} = \frac{r-v}{y} + O\left( y^{-2} \right). 
  \label{theta*}
\end{equation}
Bringing all the parameters together and taking the integral, we find (in the $y \gg 1$ approximation):
\begin{equation}
  u_{12} = \frac{4 \phi_r}{\sqrt{Z_0^2 - m^2 \cos^2 \alpha}} \arctan \left( \sqrt{\frac{Z_0 + m \cos \alpha}{ Z_0 - m \cos \alpha}}\cdot \frac{1-\sin \alpha}{\cos \alpha} \right).
  \label{u12_b_ans}
\end{equation}
Using the symmetry (\ref{A1A2}), we immediately find the $u_{23}$ distance as well:
\begin{equation}
  u_{23} = \frac{4 \phi_r}{\sqrt{Z_0^2 - m^2 \sin^2 \alpha}} \arctan \left( \sqrt{\frac{Z_0 + m \sin \alpha}{ Z_0 - m \sin \alpha}}\cdot \frac{1-\cos \alpha}{\sin \alpha} \right).
  \label{u23_b_ans}
\end{equation}
To make these expressions more manageable, we take the inverse tangents to be the new angular variables $\psi/2$, $\sigma/2$.
In terms of these variables, the boundary distances become:
\begin{equation}
  \begin{aligned}
  u_{12} =& \frac{2 \phi_r}{m} \frac{\cos \alpha - \cos \psi}{\sin \psi \sin^2 \alpha} \cdot \psi, \\
  u_{23} =& \frac{2 \phi_r}{m}  \frac{\sin \alpha - \cos \sigma}{\sin \sigma \cos^2 \alpha} \cdot \sigma,
\end{aligned}
  \label{u_b_angles}
\end{equation}
together with a constraint:
\begin{equation}
  \left( \cos \alpha - \cos \psi \right) \cos^3 \alpha = \left( \sin \alpha - \cos \sigma \right) \sin^3 \alpha.
  \label{b_e1}
\end{equation}
The conformal limit is reached when:
\begin{equation}
  \begin{aligned}
  \psi  \quad \to & \quad \alpha, \\
  \sigma \quad  \to & \quad \pi/2 - \alpha.
\end{aligned}
  \label{sigma_psi_conf}
\end{equation}

The exponentiated length $\gamma$ of a geodesic is:
\begin{equation}
  \gamma =\epsilon \cdot e^{\ell} = \frac{2 \phi_r}{Z_0 - m \sin \alpha \cos \alpha}.
  \label{gamma_b_Z}
\end{equation}
In terms of the angles, it becomes:
\begin{equation}
  \gamma = \frac{2 \phi_r}{m} \frac{\cos \alpha - \cos \psi}{\sin^3 \alpha}.
  \label{gamma_b_psi}
\end{equation}
This answer might appear not symmetrical in $(\psi, \sigma)$, which is an effect of the constraint~(\ref{b_e1}).

\begin{figure}
  \centering
  \includegraphics[width=.6\textwidth]{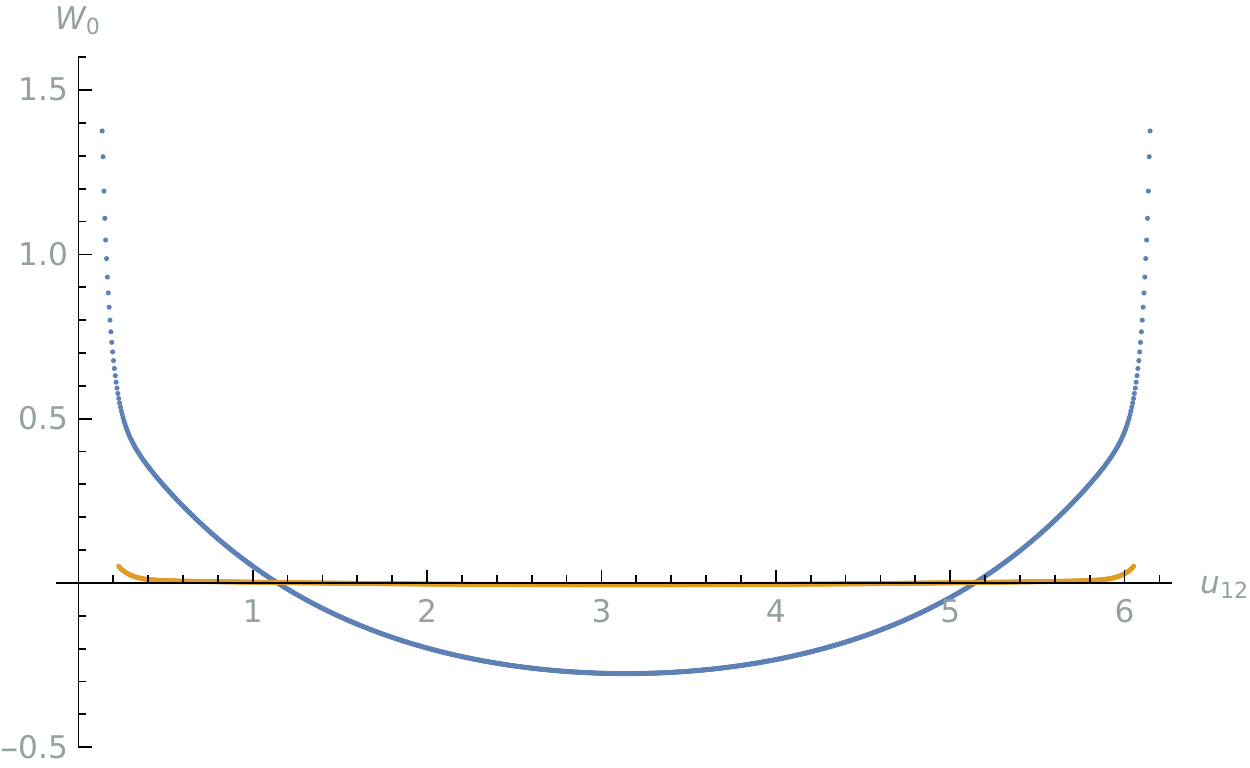}
  \caption{The connected part of the out-of-time order four-point function. 
  The yellow graph has a larger $\phi_r$ and is closer to zero.}
  \label{fig:W0_b}
\end{figure}

We can find $\gamma$ as a function of $u_{12}$ numerically.
As before, we are interested in the connected part of the four-point function:
\begin{equation}
  W_0 = \left( \frac{\gamma_{\text{2pt}} \left( L/2 \right)}{ \gamma_{\text{4pt}}} \right)^{4m}-1.
  \label{W0_def_2}
\end{equation}
The solution for the connected part of the four-point function is plotted on fig.~\ref{fig:W0_b}.

The UV limit, when $u_{12} \sim 0$ of the four-point function is reached when $\psi \sim 0$, $\sigma \sim \pi$, $\alpha \sim 0$.
In this limit $\gamma$ grows linearly with distance.
When mass is relatively small, $mL \ll 1$, the four-point function is:
\begin{equation}
  \left.W_0 \right|_{u_{12} \to 0} \sim \left( \frac{\gamma_{\text{2pt}}}{\gamma_{\text{4pt}}} \right)^{4m} - 1 \sim 0.
  \label{W0_UV_1}
\end{equation}
When mass is relatively large, $mL \gg 1$, the four-point function becomes:
\begin{equation}
  \left.W_0 \right|_{u_{12} \to 0} \sim \left( \frac{\gamma_{\text{2pt}}}{\gamma_{\text{4pt}}} \right)^{4m} - 1 \sim 2^{4m}-1.
  \label{W0_UV_2}
\end{equation}

In the symmetric case, when $u_{12} = u_{23} = L/4$, the angles are $\sigma=\psi$, $\alpha = \pi/4$.
When $mL \ll 1$, then $\gamma \sim L/\pi$ and the four-point function is once again close to zero.
When $mL \gg 1$, the exponentiated length is constant:
\begin{equation}
  \gamma \sim \frac{4 \phi_r}{m} \left( 1+\sqrt{2} \right),
  \label{gamma_sym}
\end{equation}
and the four-point function becomes:
\begin{equation}
  W_0 \sim \left( \frac{1}{1+\sqrt{2}} \right)^{4m}-1.
  \label{gamma_lim}
\end{equation}

So we see that the connected four-point function for the particles of small mass always stays close to zero.
The four-point function for heavier particles grows to a potentially large value in the ultraviolet, and becomes negative in the ``infrared'' when both distances are macroscopic.
This is exactly what we see on fig.~\ref{fig:W0_b}.

\subsection{Schwarzian limit}
\label{sec:4pt_sch}

As a reality check, we find a correction to the four-point function when mass is small.
This should reproduce the result found in \cite{Sarosi:2017ykf} and \cite{Maldacena:2016upp}.
We proceed in the same way we did for the two-point function in Section \ref{sec:small_m}.

The time-ordered four-point function is conformal when $\psi=\sigma$.
We relax the condition and take:
\begin{equation}
  \begin{aligned}
  \sigma =& \alpha + \delta,\\
  \psi =& \alpha - \delta.
\end{aligned}
  \label{sigma_psi_sch}
\end{equation}
The $\delta$ parameter measures how far we are from the conformal limit and should be proportional to mass:
\begin{equation}
  \delta = c \cdot m.
  \label{eps_l0_4}
\end{equation}

Then we find for the segment lengths:
\begin{equation}
  \begin{aligned}
  u_{12} =& \frac{4 \phi_r}{m} \delta \left( \alpha + \delta \left( 1 - \alpha \cot \alpha \right) \right)  + O\left( \delta^2 \right), \\ 
  u_{23} =& \frac{4 \phi_r}{m} \delta \left( \pi/2 - \alpha + \delta \left( 1+ \left( \pi/2 - \alpha \right) \cot \alpha \right) \right)  + O\left( \delta^2 \right).
\end{aligned}
  \label{u_a_eps}
\end{equation}
We see that $\sigma, \psi$ are roughly the segment angles for $u_{12}, u_{23}$.

The full boundary length then is:
\begin{equation}
  L = 2\left( u_{12} + u_{23} \right)= \frac{4\phi_r}{m} \delta \left( \pi + 4 \delta \left( 1+\left( \pi/4 -\alpha \right) \cot \alpha \right) \right)  +O\left( \delta^2 \right),
  \label{L_4_a}
\end{equation}
which allows us to fix $c$:
\begin{equation}
  c = \frac{L}{4\pi \phi_r}.
  \label{c_4_a}
\end{equation}
The exponentiated geodesic length is:
\begin{equation}
  \gamma = \frac{4 \phi_r}{m} \delta \sin \alpha + O\left( \delta^2 \right).
  \label{gamma_a_sch}
\end{equation}
Bringing everything together, we can find $\gamma$ in terms of the segment length:
\begin{equation}
  \gamma = \frac{L}{\pi} \sin \frac{\pi u_{12}}{L}\left( 1  - \frac{4 \delta}{\pi} \left( 1-\alpha \cot \alpha \right) \left( 1+\left( \pi/2 - \alpha \right) \cot \alpha \right) \right)+ O\left( \delta^2 \right).
  \label{gamma_a_ans1}
\end{equation}
The four-point function  $W = 1/{\gamma^{4m}}$.
To study the corrections, we extract the connected part of the four-point function:
\begin{equation}
  W_0  = \frac{W\left( u_{12}, u_{23} \right)}{G^2\left( u_{12} \right)}-1.
  \label{W0_Sch}
\end{equation}
In the Schwarzian limit, the two-point function is given by (\ref{G_corr}), and the connected part of the four-point function becomes:
\begin{equation}
  W_0^{\text{(in-order)}} = \left( \frac{\gamma_{\text{2pt}}}{\gamma_{\text{4pt}}} \right)^{4m} -1 = \frac{2m^2L}{\pi^2 \phi_r} \cdot \eta\left( \frac{\pi u_{12}}{L} \right)^2 + O\left( \delta^3 \right).
  \label{W0_a_ans}
\end{equation}
where as before $\eta\left( \alpha \right) \equiv 1-\alpha \cot \alpha$.
This is the same answer as in \cite{Maldacena:2016upp}.

In the same way, we find the first correction to the out-of-time ordered four-point function.
To do that, we relax the condition (\ref{sigma_psi_conf}).
In doing so, we need to ensure that the constraint (\ref{b_e1}) is satisfied.
Then the angles become:
\begin{equation}
  \begin{aligned}
  \psi =& \alpha + \delta \sin^2 \alpha, \\
  \sigma =& \pi/2 - \alpha + \delta \cos^2 \alpha.
  \end{aligned}
  \label{psi_sigma_schw}
\end{equation}
The small parameter $\delta$ is again proportional to the mass:
\begin{equation}
  \delta = c \cdot m.
  \label{eps_b}
\end{equation}
Expanding (\ref{u_b_angles}) in $\delta$, we find:
\begin{equation}
  \begin{aligned}
  u_{12} =& \frac{2 \phi_r}{m} \delta \left( \alpha + \delta \sin^2 \alpha \left( 1-\frac{\alpha}{2} \cot \alpha \right) \right) + O\left( \delta^2 \right), \\
  u_{23} =& \frac{2 \phi_r}{m} \delta \left(\pi/2 -  \alpha + \delta \cos^2 \alpha \left( 1-\left( \pi/4 - \frac{\alpha}{2} \right) \tan \alpha \right) \right)+ O\left( \delta^2 \right).
\end{aligned}
  \label{u_b_schw}
\end{equation}
We see that in this case as well, $(\sigma, \psi)$ are approximately the segment angles.
The full length of the boundary is:
\begin{equation}
  L = \frac{2\pi \phi_r}{m} \cdot \delta \left( 1+ \delta \left( \frac{2}{\pi} - \frac{1}{4} \sin 2\alpha \right) \right)+ O\left( \delta^2 \right).
  \label{L_b_schw}
\end{equation}
From here we fix the coefficient in (\ref{eps_b}):
\begin{equation}
  \delta \sim \frac{m L}{2\pi \phi_r}.
  \label{eps_fix}
\end{equation}
The exponentiated geodesic length becomes:
\begin{equation}
  \gamma = \frac{2 \phi_r}{m} \delta \left( 1 + \frac{1}{4} \delta \sin 2\alpha \right) + O\left( \delta^2 \right).
  \label{gamma_b_schw}
\end{equation}
Plugging in the boundary length (\ref{L_b_schw}), we find:
\begin{equation}
  \gamma = \frac{L}{\pi}\left( 1+\frac{mL}{2\pi \phi_r}\left( \frac{1}{2} \sin 2\alpha - \frac{2}{\pi}\right) \right).
  \label{gamma_b_ans}
\end{equation}
The connected part of the four-point function then becomes:
\begin{equation}
  W_0^{\text{(out-of-order)}} =  - \frac{2m^2 L}{\pi^2 \phi_r} \left( \frac{\pi}{2}\sin \frac{2 \pi u_{23}}{L}+ 1\right)+O\left( \delta^3 \right).
  \label{W0_otoc}
\end{equation}


After analytic continuation, the sine in (\ref{W0_otoc}) becomes exponentially decaying.
The real-time correlation function corresponds to $u_{23} = \beta/4 +it$.
Then the connected four-point function found from the Schwarzian limit becomes:
\begin{equation}
  W_0^{\text{(out-of-order)}} \sim -\frac{1}{N}\cosh\left( \frac{2 \pi}{\beta}t \right).  
  \label{W0_schw_real}
\end{equation}
This demonstrates the chaotic behavior of the four-point function.
In the next Section, we see that the Schwarzian limit describes well the out-of-order four-point function at early times.

\section{Four-point function in real time}
\label{sec:4pt_cont}

A useful measure of chaotic behavior of theory is the out-of-time ordered four-point function \cite{larkin1969quasiclassical}, \cite{Maldacena:2015waa}.
To construct it, we place our operators equidistantly on the thermal circle as follows:
\begin{equation}
  W^{\text{(out-of-order)}} = \left \langle \mathcal O_1 \left( -it/2 \right) \mathcal O_2 \left( \beta/4+it/2 \right) \mathcal O_1\left( \beta/2-it/2 \right) \mathcal O_2 \left( 3\beta/4+it/2 \right)\right \rangle.
  \label{otoc_def}
\end{equation}
It is convenient to divide this four-point function by the product of the two-point functions $\left \langle \mathcal O_1 \mathcal O_1 \right \rangle $, $\left \langle \mathcal  O_2 \mathcal O_2 \right \rangle $:
\begin{equation}
  W_0^{\text{(out-of-order)}} = \frac{W^{\text{(out-of-order)}} }{ G\left( \beta/2 \right)^2}.
  \label{W0_otoc_def}
\end{equation}
In terms of the geodesic lengths, the normalized four-point function becomes:
\begin{equation}
  W_0^{\text{(out-of-order)}}=\left| \frac{\gamma_{\text{2pt}} \left( \beta/2 \right)}{\gamma_{\text{4pt}} \left( \beta/2-it \right)}\right|^{4m}. 
  \label{W0_otoc_gamma}
\end{equation}

The out of time ordered four-point function is connected to the thermal average of a double commutator:
\begin{equation}
  C(t) = \left \langle - \left[ \mathcal O_2\left( t \right), \mathcal O_1\left( 0 \right) \right]^2 \right \rangle_\beta = 2\left( 1-W_0^{\text{(out-of-order)}} \right).
  \label{Ct_def}
\end{equation}
In a theory dual to a black hole \cite{polchinski2015chaos}, one expects the double commutator to grow exponentially at first, $C(t) \sim \exp\left( \lambda_L t \right)$.
This growth is referred to as  Lyapunov behavior.
The exponent of the growth is bounded above, $\lambda_L \le 2\pi/\beta$.
This growth does not continue indefinitely, and at about scrambling time $t_{*} \sim \beta$ the double commutator saturates.
It approaches a constant value exponentially slowly.
This is similar to what is expected at the beginning of the Ruelle region.

\begin{figure}
  \centering
  \includegraphics[width=.6\textwidth]{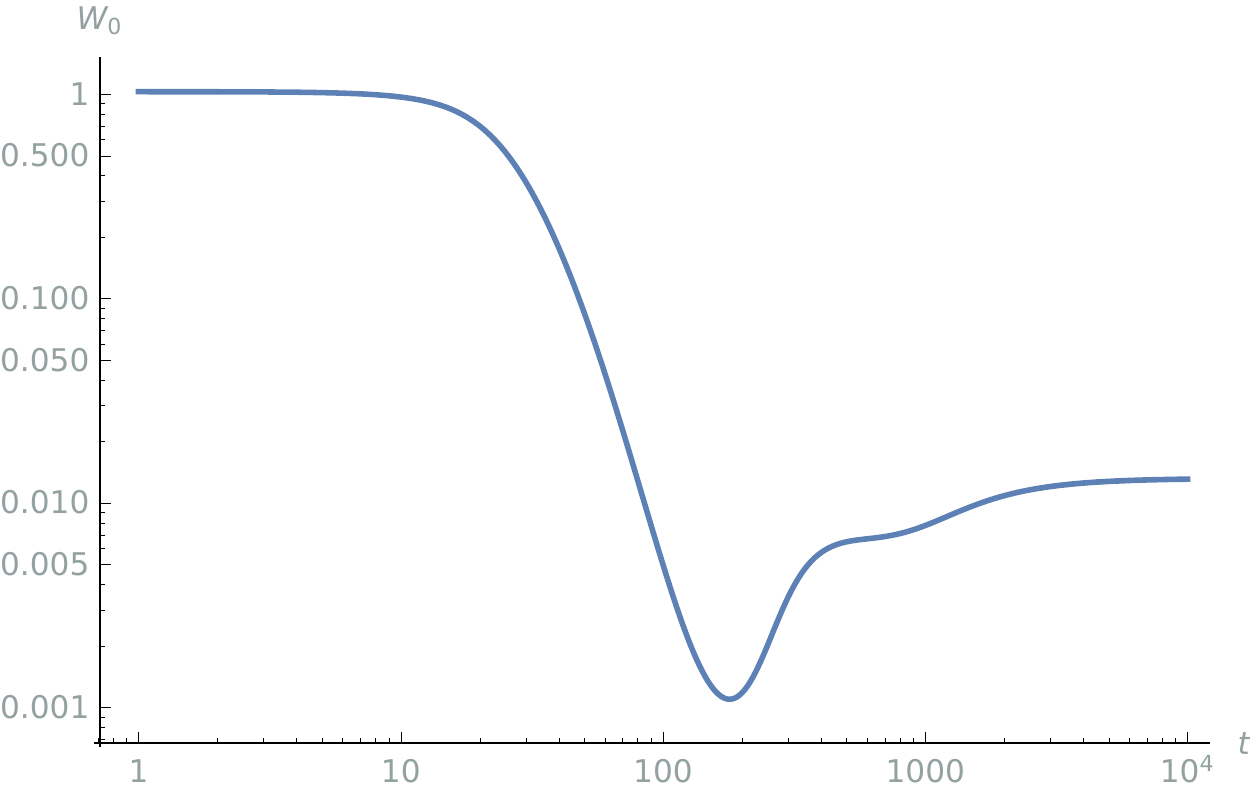}
  \caption{The out-of-time ordered four-point function in the small $\epsilon$ approximation.}
  \label{fig:otoc_vanilla}
\end{figure}

In our notation, the arrangement of operators in (\ref{otoc_def}) corresponds to the distances being complex conjugates:
\begin{equation}
  \begin{aligned}
  u_{12} =& \beta/4+it, \\
  u_{23} =& \beta/4-it.
  \end{aligned}
  \label{u_otoc}
\end{equation}
In the small $\epsilon$ approximation, the boundary distances $u_{12}, u_{23}$ and the exponentiated geodesic distance $\gamma$ are analytic functions of the angles and therefore can be relatively easily continued to complex plane.
The angles become subject to the conditions:
\begin{equation}
  \begin{aligned}
  \sigma=&\bar{\psi}, \\
  \alpha+\bar{\alpha}=&\pi/2.
  \end{aligned}
  \label{angles_otoc}
\end{equation}
Using these two conditions (which improve the convergence of the numerical method), the definitions of the boundary distances (\ref{u_b_angles}) and the exponentiated geodesic distance (\ref{gamma_b_psi}), we can find the out-of-time ordered four-point function numerically.
The result is on fig.~\ref{fig:otoc_vanilla}.
The overall structure of this four-point function is very similar to the real-time two-point function on fig.~\ref{fig:sff}.
We see that the normalized four-point function starts close to 1, and exponentially decays to a minimum at $t \sim \phi_r/m$. 
After that, it grows to a plateau.
At long times, the four-point function is:
\begin{equation}
  \left.W_0^{\text{(out-of-order)}}\right|_{t \to \infty} \sim \frac{m}{\phi_r} \sim 1/N.
  \label{W0_t_inf}
\end{equation}

\begin{figure}
  \centering
  \includegraphics[width=.6\textwidth]{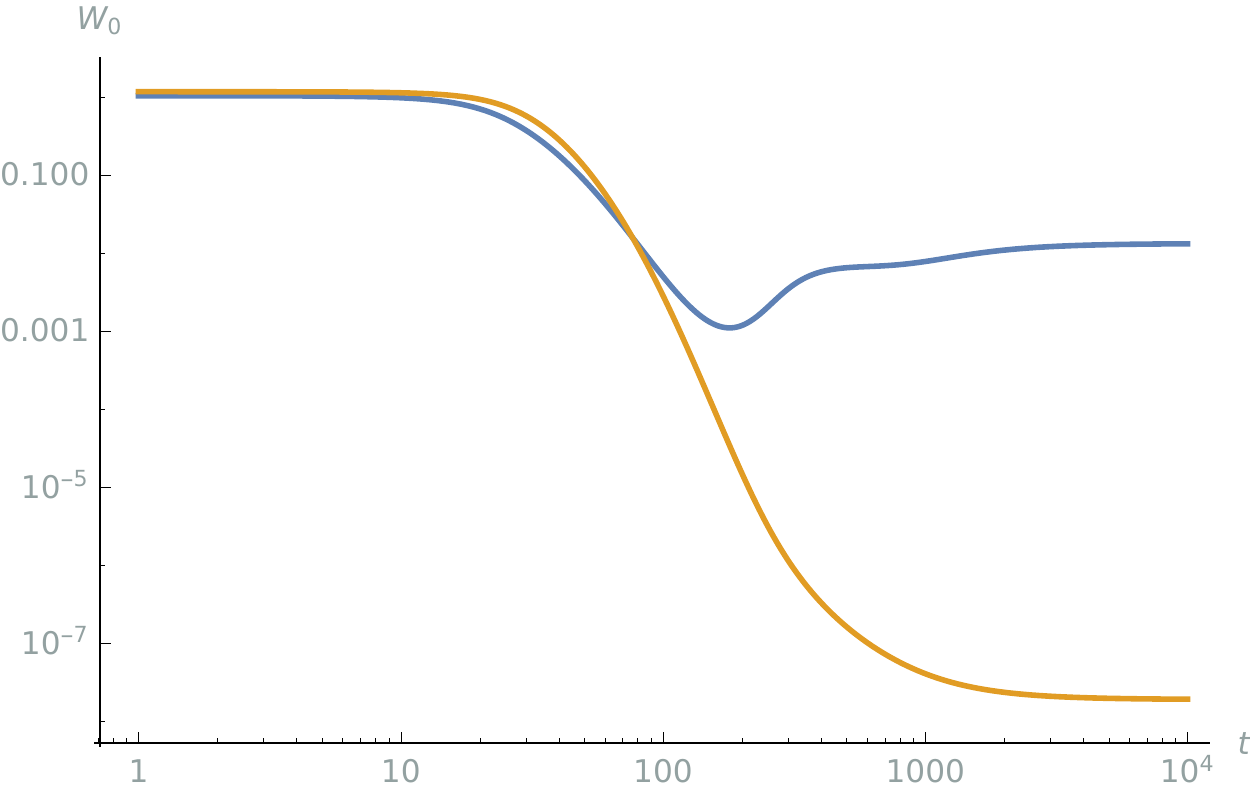}
  \caption{The out-of-time ordered four-point function with the extrinsic curvature correction (blue) compared to the two-point function found as the exponentiated geodesic distance (yellow).}
  \label{fig:otoc_K}
\end{figure}

As we have seen for a real-time two-point function, addition of the extrinsic curvature term changes the picture quite a bit.
The two-point function in (\ref{W0_otoc_def}) is in Euclidean time, and the extrinsic curvature correction for it is small.
Therefore, to find the full correction, we consider only the four-point function.
We do it numerically, but first let us consider the long time limit.

At long times, the curvature is very close to 1:
\begin{equation}
  K_{t \to \infty} = -\frac{y}{R} = -1.
  \label{K_otoc_t_inf}
\end{equation}
The curvature of the empty $AdS$ space is never very close to 1.
Hence the action of extrinsic curvature (normalized by the action of the empty $AdS$ space) is:
\begin{equation}
  \left( I-I_0 \right)_{t\to\infty}=\left( 4I_{\text{seg}} + 4 I_{\text{cusp}} -I_0 \right)_{t\to \infty} = -\phi_b \left( \frac{\beta}{\epsilon} - \sqrt{\left( \frac{\beta}{\epsilon} \right)^2 + \left( 2\pi \right)^2}  \right) - 4m = \frac{2 \pi^2 \phi_r}{\beta} - 4m.
  \label{I_otoc}
\end{equation}
Therefore, the plateau on fig.~\ref{fig:otoc_vanilla} becomes exponentially lower:
\begin{equation}
  \left.W_0^{\text{(out-of-order)}}\right|_{t \to \infty} \sim \exp\left( -\frac{2 \pi^2 \phi_r}{\beta} + 4m \right) \sim \exp\left( -N \right).
  \label{W0_t_inf_K}
\end{equation}

Expanding the extrinsic curvature to first order in $\epsilon^2$, we can find the correction numerically. 
(We do not use the full answer for the four-point function, since it has branch points and is hard to numerically continue to complex plane.)
The result is on fig.~\ref{fig:otoc_K}.
At small time, the correction is small, and the four-point function demonstrates the same exponential decay.
However, at time $t \sim \beta$ the decay slows down, and the four-point function approaches a small but nonzero value.
This is the beginning of the Ruelle region, describing thermalization of a black hole \cite{polchinski2015chaos}.
However, the onset of the Ruelle behavior is expected to be at roughly the scrambling time $t_* \sim \log N$, and in our case the exponentially decaying four-point function reaches the value (\ref{W0_t_inf_K}) at time $t \sim \phi_r \sim N$.
This seems puzzling to us.

Note that the four-point function never approaches zero, and its value at long times is exponentially small in $N$.
If we look at the four-point function in the energy basis, we find a very similar structure to what we have seen in Section~\ref{sec:2pt_cont}.
Disregarding the off-diagonal terms in the $\mathcal O_1$, $\mathcal O_2$ operators, we can write the four-point function as:
\begin{equation}
  W \sim \sum_{n,m} \left|\left \langle n | \mathcal O_1 | n \right \rangle \right|^2 \left|\left \langle m | \mathcal O_2 | m \right \rangle \right|^2 e^{-\beta/2  \cdot  E_n - \beta \cdot E_m} e^{i\left( E_n-E_m \right)t}.
  \label{W_basis}
\end{equation}
If the diagonal terms in the operators are close to 1, then by the same logic as before the four-point function approaches a finite value at long times.
Since we normalize by the two-point functions which are independent of real time $t$, this should hold after normalization as well.
As was the case for the two-point function, our four-point function does not capture the rapid fluctuations in (\ref{W_basis}) and therefore represents the quantum mechanical four-point function only in the averaged sense.

To our knowledge, this effect has not been tested in the SYK model.
Although it is non-perturbative in $N$, it should be visible in a numerical simulation.

\appendix
\section{Two-point function for particles with negative mass}
\label{app:m_neg}

Our setup allows us to also study the two-point function for particles with negative mass.
It corresponds to a picture like fig.~\ref{fig:ears}$(a)$, with cusps in the $NAdS$ boundary pointing outward.
We use the same angular parameterization as before:
\begin{equation}
  w_{1,2} = \cos \alpha_{1,2},
  \label{w_def_neg}
\end{equation}
with the $w_1>w_2$ to make the cusps point outward.

For a positive $w$, there is no upper bound on its value.
It can be greater than one, making the formal parameter $\alpha$ imaginary.
We will see that the point where $\alpha$ becomes imaginary does not introduce any singularities, and it is special only for our choice of parameters.

\begin{figure}
  \centering
  \includegraphics[width = .6\textwidth]{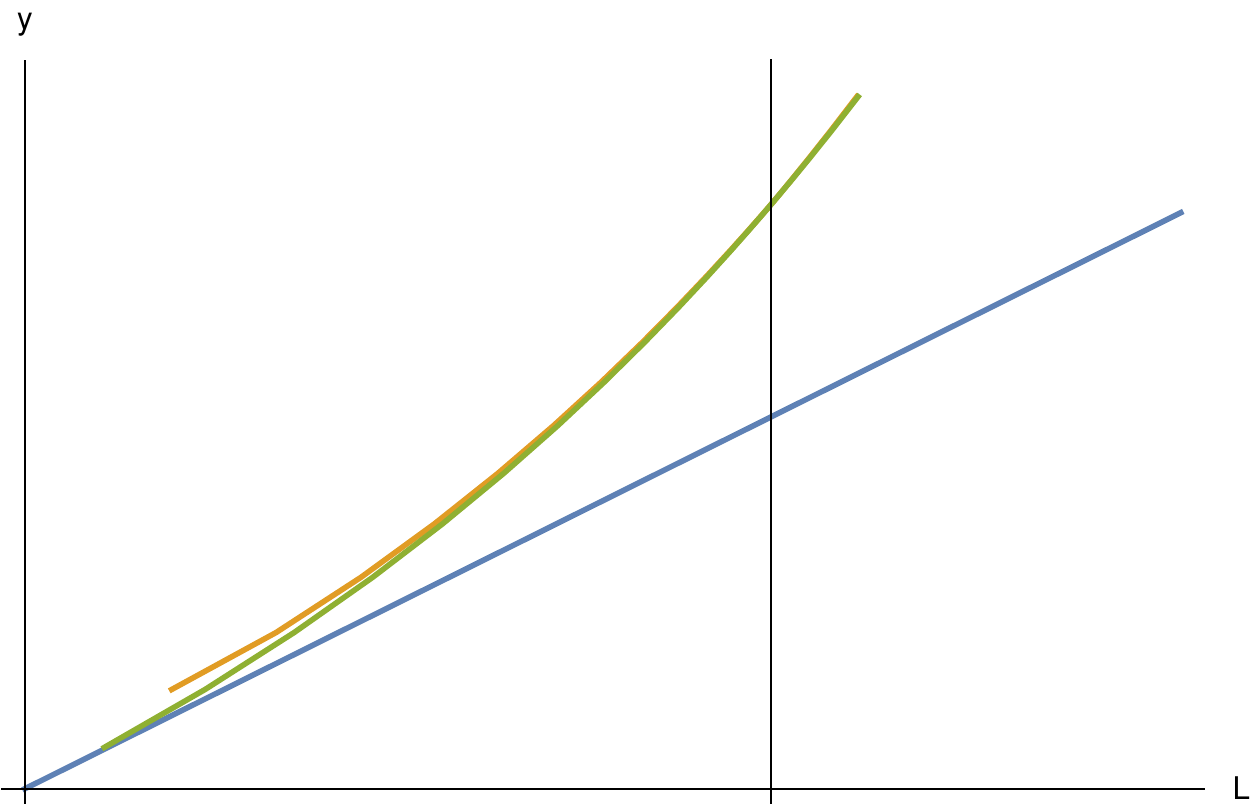}
  \caption{$y$ as a function of boundary length $L$ for a theory with negative mass.
    The blue line is the conformal result, the blue line is the result with $y \sim R$, or very large $l_0$ compared to all the other parameters.
  The vertical line divides the region with real $\alpha$ from the region with imaginary $\alpha$, and one can see that there is no cusp or discontinuity at this point.}
  \label{fig:y(L)}
\end{figure}

We start with finding the symmetric correlation function, that is, consider $\alpha_1 = \pi-\alpha_2$.
The length of the boundary $L$ and the exponentiated geodesic length $\gamma$ are as before, see (\ref{L_sym}) and (\ref{y_sym_alpha}).
$\gamma(L)$ cannot be in general solved analytically, but large and small distances are accessible to us.
First, let us take $\alpha$ close to $\pi/2$:
\begin{equation}
  \alpha = \frac{\pi}{2} - \delta, \qquad \delta \gg \frac{m}{2\phi_b}.
  \label{alpha_pi2}
\end{equation}
We need the second condition in (\ref{alpha_pi2}) to keep $y$ large.
Then:
\begin{equation}
  \gamma = \frac{4}{m} \delta, \qquad L = \pi \gamma.
  \label{y_alpha_pi2}
\end{equation}
From this, we see that the two-point function has the conformal form:
\begin{equation}
  G \sim \left( \frac{L}{\pi} \right)^{-2\Delta}.
  \label{G_alpha_pi2}
\end{equation}
Like before, it is a consequence of the approximation we are taking.
The two-point function has to be cut off at $L \sim m \epsilon^2/\phi_r$.
We can see it on fig.~\ref{fig:y(L)}.

\begin{figure}
  \centering
  \includegraphics[width=.6\textwidth]{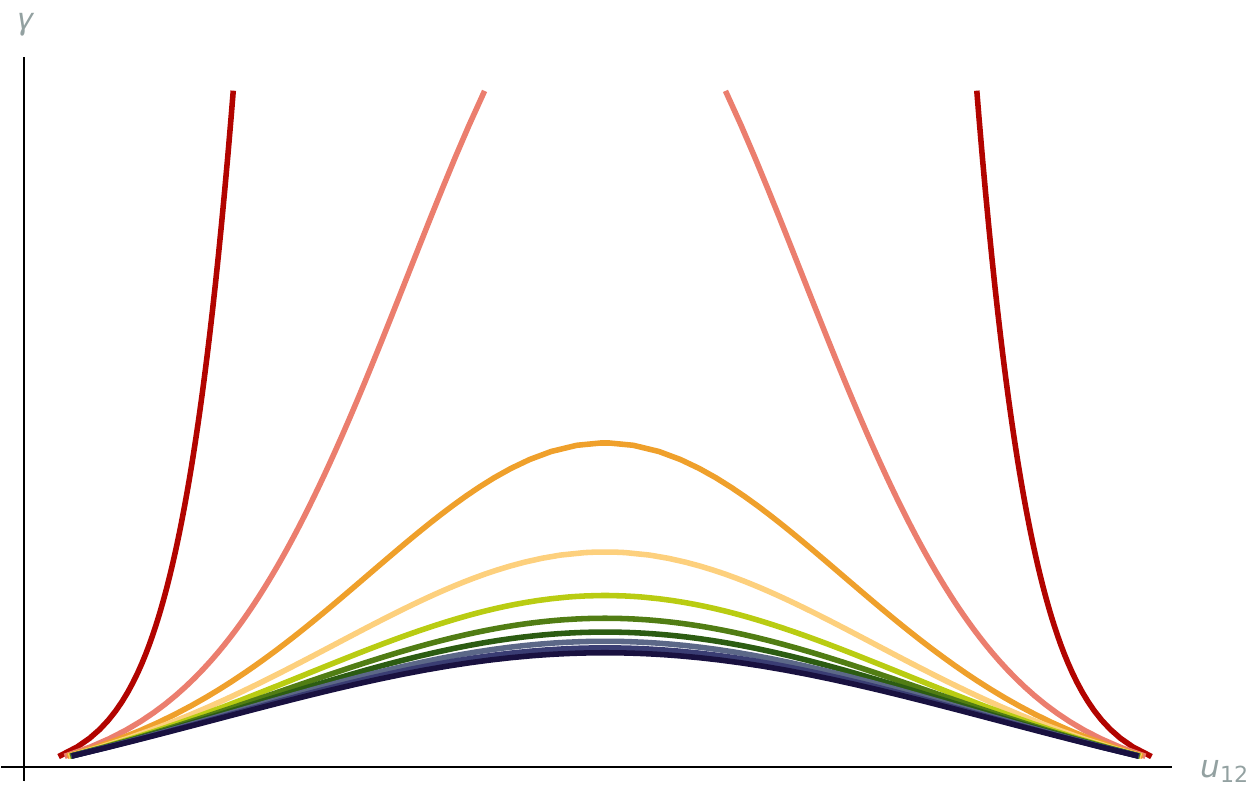}
  \caption{$\gamma\left( u_{12} \right)$ for negative mass. The conformal result $\gamma \sim \sin u_{12}$ is the dark blue line on the bottom.
  The red line corresponds to the largest mass.}
  \label{fig:yu-}
\end{figure}

Next, let us see what happens when $\alpha$ goes to the complex plane:
\begin{equation}
  \alpha = \delta.
  \label{neg_m_a_0}
\end{equation}
The two-point function is finite at that point.
The length parameters are:
\begin{equation}
  \gamma = \frac{4 \phi_r}{m}, \qquad L = 2\gamma.
  \label{y_small_alpha}
\end{equation}
The two-point function is still close to the conformal one (\ref{G_conf_sym}), although it starts to move away from it:
\begin{equation}
  G \sim \left(\frac{L}{2} \right)^{-2\Delta}.
  \label{G_small_alpha}
\end{equation}
We can also look at the expressions for $L$ and $\gamma$,  (\ref{L_sym}, \ref{y_sym_alpha}), and see that they are regular at $\alpha \to 0$.
Therefore we do not encounter any cusp or discontinuity when $\alpha$ becomes imaginary.

Considering large imaginary $\alpha$, we recover the region of large $L$:
\begin{equation}
  \alpha = i\Lambda \qquad \Rightarrow \qquad y = \frac{\phi_b}{m} \cdot e^\Lambda.
  \label{alpha_im_sym}
\end{equation}
Then the radius of the boundary is:
\begin{equation}
  R^2 = y^2 + \sinh^2 \alpha \sim y^2, \qquad \phi_b/m \gg 1,
  \label{R=y_sim}
\end{equation}
and the length of the boundary is:
\begin{equation}
  L = 4y \epsilon \frac{\Lambda}{\sinh \Lambda} \sim  \frac{4 \phi_r}{m} \Lambda.
  \label{L_lambda}
\end{equation}
Then the two-point function is exponentially growing, as one would expect in a theory containing particles with negative mass:
\begin{equation}
  G \sim e^{m^2 L/(2 \phi_r)}.
  \label{G_alpha_im_sym}
\end{equation}
On fig.~\ref{fig:y(L)} we see the beginning of this exponential growth.

\begin{figure}
  \centering
  \includegraphics[width=.4\textwidth]{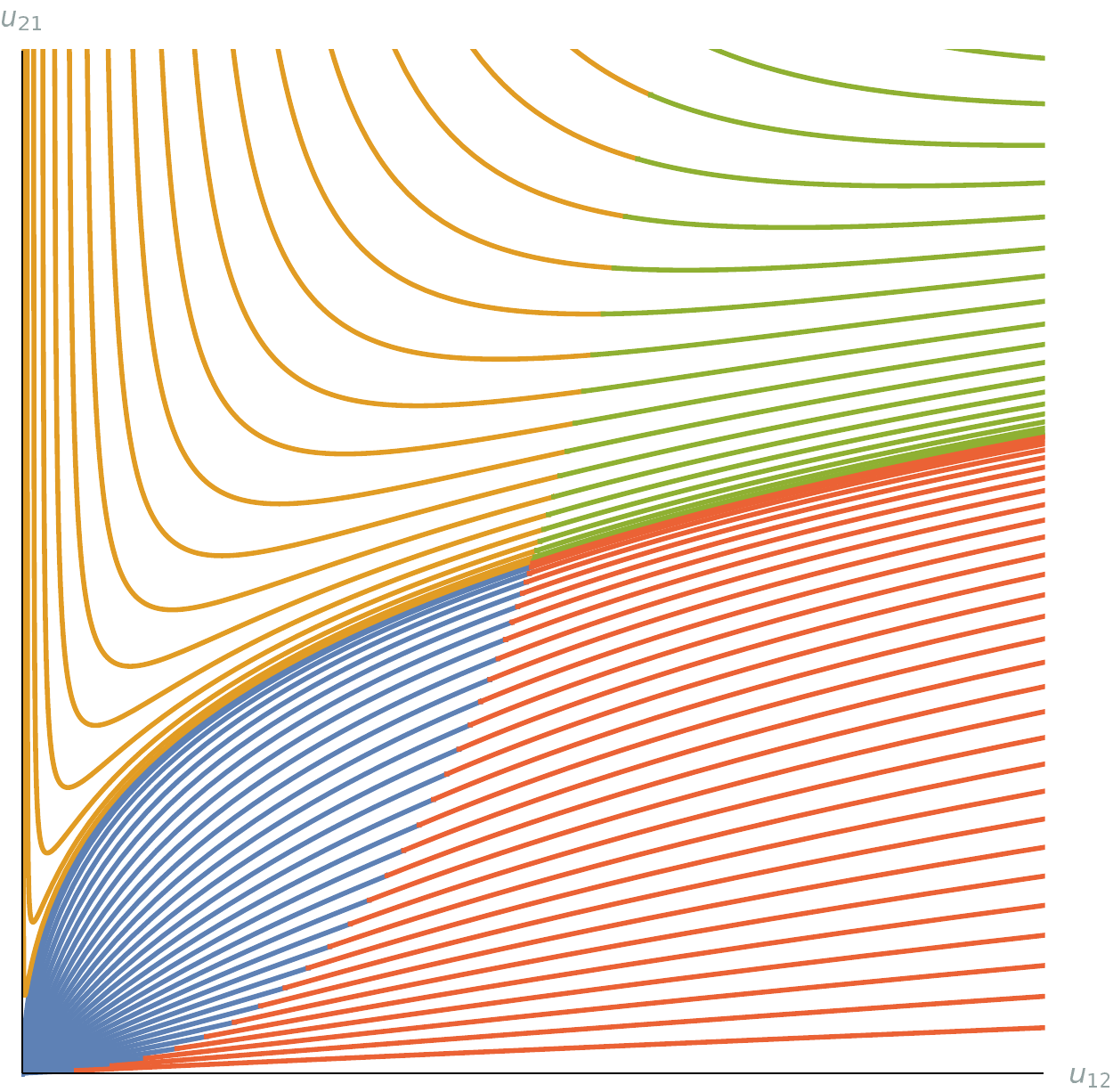}
  \caption{Absolute values of boundary distances $u_{12}$ and $u_{21}$ as functions of angles $\alpha_1, \alpha_2$. 
  Blue region corresponds to real angles, green to imaginary angles, and yellow and red to one angle being real and one imaginary.}
  \label{fig:uu_plus}
\end{figure}

We can also numerically solve for the two-point function with fixed boundary length $L$, while changing mass (see fig.~\ref{fig:yu-}).
In the small $\epsilon$ approximation, the two-point function approaches the conformal $\sin u_{12}$ as mass decreases.
For larger mass, the two-point function grows exponentially for small distances.

Using numerical methods, we can also answer the question of whether the angular parameters are in one-to-one correspondence with the length parameters.
To answer that, we vary $\alpha_{1,2}$ and plot the length on the $u$ plane.
The angles are artificial parameters and can be real or imaginary:
\begin{equation}
  0 \le \alpha \le \pi \qquad \text{or} \qquad \alpha \in i \mathbb R.
  \label{alpha_ranges}
\end{equation}
We can easily check that the distances are either both positive or both negative, depending on the sign of the mass.
So varying angles, we cover two quarters of the $u$ plane.
Right now we are interested in the case when both distances in (\ref{u_generic}) are negative.

The results are on fig.~\ref{fig:uu_plus}.
Notice that varying the $\alpha$ parameters, we cover a quarter of the $u$ plane exactly once.

\bibliographystyle{JHEP}

\bibliography{mass}

\end{document}